\documentclass[conf]{new-aiaa}
\usepackage[utf8]{inputenc}

\usepackage{caption}
\usepackage{subcaption}
\usepackage{placeins}
\usepackage{soul}
\usepackage{color}
\usepackage{comment}

\usepackage{tikz,xcolor}
\usetikzlibrary{calc,intersections,arrows.meta}

\usepackage{graphicx}
\usepackage{amsmath}
\usepackage[version=4]{mhchem}
\usepackage{siunitx}
\usepackage{fancyhdr}
\usepackage{longtable,tabularx}

\title{A Review of the Design of Cone-Cylinder-Flare Geometries for Stability Analyses and Conventional/Quiet Wind Tunnel Tests*}

\author{\textbf{\normalsize Sebastien Esquieu$^\text{(1)}$, Steven P. Schneider$^\text{(2)}$, Elizabeth K. Benitez$^\text{(3)}$, Jean-Philippe Brazier$^\text{(4)}$}
\\{\normalsize\itshape
$^\text{(1)}$CEA, The French Alternative Energies and Atomic Energy Commission, Le Barp, France}
\\{\normalsize\itshape
$^\text{(2)}$School of Aeronautics and Astronautics, Purdue University, West Lafayette, Indiana, USA}
\\{\normalsize\itshape
$^\text{(3)}$US Air Force Research Laboratory, Wright-Patterson Air Force Base, OH, 45433}
\\{\normalsize\itshape
$^\text{(4)}$ONERA / DMPE, Université de Toulouse, F-31055 Toulouse - France}}
\date{}

\begin{document}
\maketitle

\begin{abstract}
In order to study the effects of pressure gradients, flow expansion, and recompression on the stability of hypersonic boundary-layers, axisymmetric cone-cylinder-flare configurations have been specifically designed for wind tunnel experiments.
The objective is to create well-suited geometries for hypersonic laminar-turbulent transition analyses, while also adding the capability to generate attached and separated flows. 
After a thorough review of the aerodynamic flows obtained at Mach 6 and zero degree angle of attack in fully laminar conditions on each configuration, linear stability theory (LST) and linear parabolized stability equations (PSE) are used to predict the amplification rates of the boundary-layer disturbances for the case without flow separation since local stability analysis cannot handle highly non-parallel flows.
For this attached flow case, the numerical stability results are compared to wind tunnel measurements obtained in the BAM6QT (Boeing AFOSR Mach-6 Quiet Tunnel). The semi-empirical $e^N$ method allows to correlate transition with the integrated growth of the linear instability waves.  
%As regards the stability of the separated flows, they are investigated only experimentally, since local stability analysis cannot handle such highly non-parallel flows. 
The influence of tunnel noise is also investigated, using conventional and quiet experiments.\\
\end{abstract}

\noindent \textbf{Nomenclature}
\\
\begin{tabular}{l l}
%\\
$Mach$  & 		Mach number \\
$pi$ &    		stagnation pressure (bar) \\
$Ti$ &    		stagnation temperature (K) \\
$Hi$ &    		total enthalpy (J/kg) \\
$p$& 		static pressure (Pa) \\
$T$& 		static temperature (K) \\
$\rho$& 		density (kg/m$^3$)\\
u,v,w &		velocity in x,y,z directions (m/s) \\
V &			velocity magnitude \\
$Re_{/m}$ &    	freestream unit Reynolds number (1/m) \\
x &			geometric coordinate in streamwise direction (m)\\
y &			geometric coordinate in radial direction (m) \\
$Rn$  & 		nose radius (m) \\
$\delta$ &		boundary-layer thickness (m) \\
$\alpha_i$ & 		amplification rate (1/m) \\
$\psi$  &		wave propagation angle ($^o$) \\\\
\end{tabular}
\\\\
\noindent \textbf{Subscript} 
\\
\begin{tabular}{l l}
%\\
$\infty$  & 		value in freestream \\
$e$  & 		value at the boundary-layer edge \\\\
\end{tabular}
\\\\
\noindent * This review is an extension of the work presented in paper 2019-2115 at the AIAA SciTech Forum (San Diego, CA).

%=======
\section{Introduction}
%=======
\label{sec:intro}
Laminar-to-turbulent transition is a primordial input for trajectory and conceptional design of hypersonic vehicles. Indeed, transition at hypersonic speeds causes large changes in heat transfer, skin friction and boundary-layer separation. Turbulent boundary-layers increase the heating rate and the viscous drag, and offer a greater resistance to the formation of separation bubbles. As a consequence, the aerodynamic lift and drag, the stability and control of the vehicle as well as the thermal protection system are directly affected by the state of the boundary-layer. So the knowledge of stability and transition of high-speed boundary-layers is an important issue for hypersonic flight. But understanding how and where transition occurs on a given configuration during re-entry or high speed flight is one of the long-standing problems in aerodynamics \cite{leyva2017hypflight}.

In flight or in a wind tunnel, disturbances are present in the flow and interact with the boundary-layer that develops along the object. The boundary-layer acts as a selective filter where only certain frequencies and wavelengths can be amplified and eventually induce a laminar-to-turbulent transition. For a low disturbance environment, as explained by \citep{stetson1990hypbl}, four fundamentally different instability mechanisms can produce disturbance growth in a hypersonic boundary-layer: first-mode, second-mode, crossflow instabilities and Görtler vortices. Several paths to transition exist \cite{morkovin1984trans} but, here, only the low disturbance level is considered.

As detailed by Schneider \cite{schneider2015mechanisms}, the first-mode instability is similar to low-speed Tollmien-Schlichting waves and occurs for subsonic and moderate supersonic flows. It is most amplified when the wavefronts are oblique to the stream direction. The second-mode instability concerns hypersonic flows. It is similar to a trapped acoustic wave and is most amplified when the wavefronts are normal to the freestream direction. The cross-flow instability occurs in three-dimensional boundary-layers, and has both traveling and stationary forms. The G\"{o}rtler instability is important for boundary-layers on concave walls, and perhaps in some regions of concave streamline curvature.

Very instructive and thorough studies of the transition development are available for the attached boundary layer mainly on cone shapes. Some examples are presented by \cite{casper2009pressfluct} and \cite{marineau2015BLStab}.

Here, the cone-cylinder-flare configurations add the influence on the boundary-layer stability of pressure gradients, flow expansion, and compression. Different flare angles are used to generate different levels of compression: a limited pressure gradient to keep an attached boundary-layer \cite{esq2019-2115} and a higher adverse pressure gradient to generate a flow separation \cite{benit2020-3072, esq2023-3af, benit2023-jfm}.

The hypersonic flows addressed here are considered in wind tunnel conditions at Mach 6. In these high-speed conditions and with the sharp cone-cylinder shape, the edge Mach number will be greater than Mach 5 all along the configuration. Following Mack's theory \cite{mack1984BLStabl}, for the attached boundary-layer case, the dominant instabilities may be the second-mode waves. Other instabilities such as G\"{o}rtler instabilities could also be present in the concave part of the geometry. For small flare angles, these concave wall instabilities should not be predominant. But for larger flare angles with a separation bubble, G\"{o}rtler instabilities, streaks or even other instability mechanisms should be considered for the study of the transition process. Such studies have been conducted on a hollow-cylinder flare configuration by \cite{ginoux65,bur2009HCF}, and more recently by \cite{lugrin2020TransitionSI,lugrin2022TransSBLI}.

Here, the focus will be mainly on the second-mode waves with the influence of pressure gradients on boundary-layer stability.

%=======
\section{Cone-cylinder-flare geometries: conceptual design and final characteristics}
%=======
\label{sec:geoms}

\subsection{Design: physical objectives}

Many studies are devoted to the stability analysis of conical shapes, but in real flight, the objects are often more complex than these ideal configurations. This implies that several levels of geometric complexity and associated physical phenomena have to be considered. As a first step towards increased physical complexity, in the present study it has been decided to create an innovative axisymmetric shape in order to study:

\begin{itemize}
    \item firstly, the influence of pressure gradients on the boundary-layer stability. Indeed, some abrupt shape changes can generate flow expansions and corner flows. Their influence on boundary-layer transition is significant ;
    \item secondly, the effect of flow separation on laminar-turbulent transition. Indeed, it is well-known that separation bubbles lead to early transition breakdown but little information is available on the physical description of this phenomenon. In real flight, the consequent shock-boundary layer interaction (SBLI) can generate significant heat flux increases at the reattachment point or also, for example, a large decrease of control authority of maneuvering surfaces. So, a generic configuration enabling analysis of these complex SBLI appears of first importance too.
\end{itemize}

%So it has been decided to create a new generic configuration in order to promote transition development for attached and detached boundary-layers with expansion and recompression in high speed flow.

\subsection{Design: wind tunnel requirements}

The cone-cylinder-flare configurations have been designed for experimental tests in Purdue University’s Boeing/Air Force Mach 6 Quiet Tunnel (BAM6QT). This wind tunnel has the capacity to operate with low freestream noise levels, less than 0.02\%, close to the ones observed in real atmospheric flights. This tunnel, very well suited for boundary-layer stability measurements, will be presented later in section \ref{sec:BAM6QT}.

\begin{figure}
	\centering
	\includegraphics[width=0.88\linewidth]{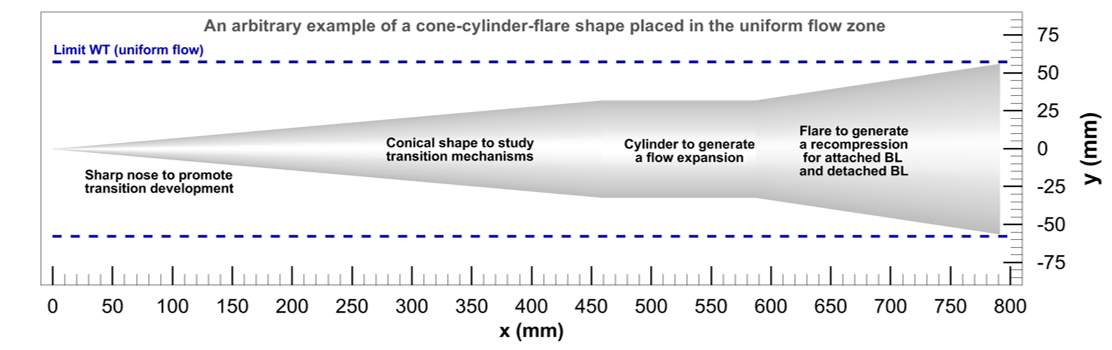}
	\caption{Physical goals of the design / Zone of uniform flow in the wind tunnel}
	\label{fig:CCF_WT_limits}
\end{figure}

The cone-cylinder-flare configurations have been specifically designed to fit in the wind tunnel and mainly in the zone of uniform flow in the test section. The geometric constraints are the following:
\begin{itemize}
    \item model length up to 800 mm ;
    \item model diameter up to 115 mm.
\end{itemize}
These physical boundaries are illustrated in figure \ref{fig:CCF_WT_limits} where an arbitrary shape is placed in the test section. On this example, the flare angle cannot be increased to significant angles. This will lead to a very short flare length that won't be well suited to convective instability-wave development and also to sensor implementation. So a compromise between the length and angle of each part (cone / cylinder / flare), as well as the cylinder diameter, has to be identified to obtain the best configuration for the study of laminar-turbulent  transition mechanisms, with expansion and recompression, for attached and detached boundary-layers in high speed flow.

\subsection{Design: geometrical constraints / computational validation}

To identify the best configurations, a cycle of laminar computations and stability analyses have been realized with the following objectives:
\begin{itemize}
    \item define the length and angle of the conical part in order to generate a zone of highly amplified second-mode waves at the end of the cone before reaching the flow expansion zone at the cone-cylinder junction ;
    \item have a significant cylinder length to analyze the flow expansion effects on the boundary-layer stability and to allow measurements in the recirculation bubble region for the configuration with flow separation. The cylinder length has also to be significant in order to keep the whole separation bubble on the cylinder part without reaching the cone-cylinder junction ;
    \item for the first configuration, the flare angle has to be limited in order to keep an attached boundary-layer while keeping a recompression. For the second configuration, the flare angle has to be important (around 10°) in order to generate a significant adverse pressure gradient leading to a large separation bubble. The flare has to be long enough to be equipped with high frequency sensors, to be visualized by Schlieren technique and the influence of it to be analyzed at the wall by infra-red thermography. 
\end{itemize}
To do so many laminar computations / stability analysis cycles have been realized to identify the best suited geometries. These computational steps are not detailed here, it was decided to proceed directly to the description of the final forms. The physics of the flows that were obtained on these designed configurations will be presented in detail later in the article.

\begin{figure}
	\centering
	\includegraphics[width=0.80\linewidth]{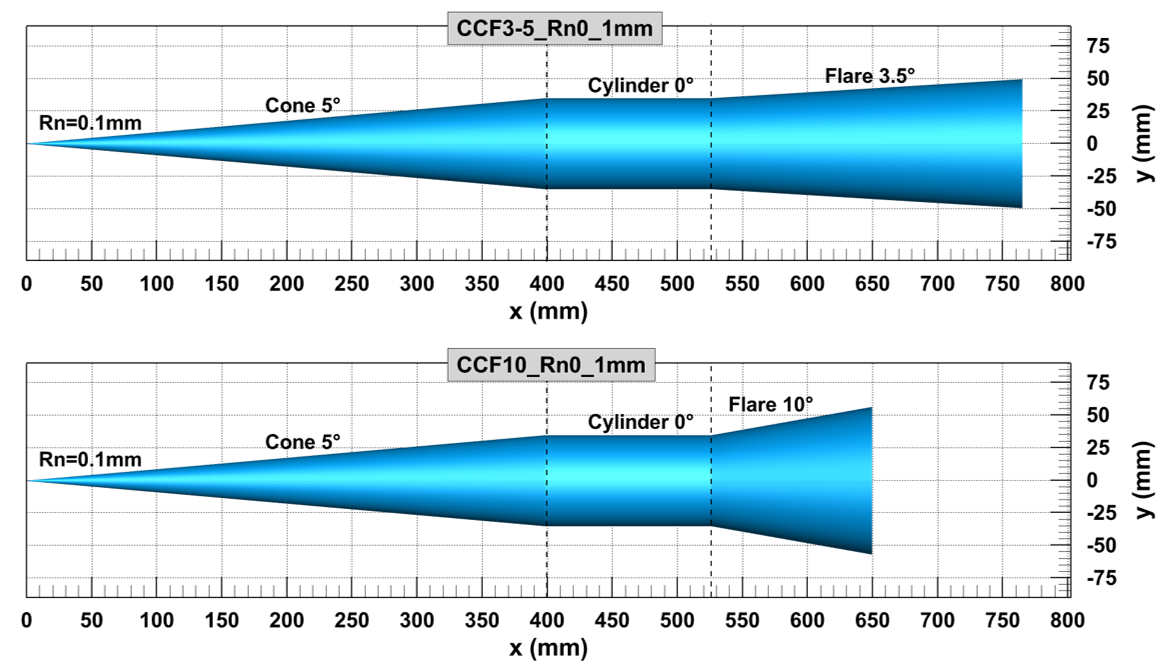}
	\caption{Final design of CCF3-5 and CCF10 cone-cylinder-flare geometries}
	\label{fig:CCF3-5_10geoms}
\end{figure}

\begin{table}
\begin{center}
\def~{\hphantom{0}}
  \begin{tabular}{|l|c|c|c|}
        \hline
		           &  Nose-Cone junction     &  Cone-Cylinder junction  &  Cylinder-Flare junction  \\
        \hline
		x (mm) &        0.0920   &           398.15             &             525.67          \\
        \hline
		y (mm) &        0.0997   &             34.93            &              34.93             \\
        \hline
	\end{tabular}
	\caption{CCF3-5 and CCF10 model coordinates}
	\label{tab:coor} 
	\end{center}
\end{table}

\subsection{The final design: CCF3-5 and CCF10 configurations}

Following all the geometrical constraints and physical goals presented above, the final design results in two slender cone-cylinder-flare configurations.\\

These two configurations are named:
\begin{itemize}
    \item CCF3-5 for the case with attached boundary-layer ;
    \item CCF10 for the case with flow separation.
\end{itemize}

These new axisymmetric cone-cylinder-flare configurations are illustrated in figure \ref{fig:CCF3-5_10geoms}. Here is a summary of the design's main features: 

\begin{itemize}
    \item small nose radius Rn = 0.1 mm ;
    \item half-cone angle equal to 5 degrees ;
    \item cylinder at 0 degree on the central part ;
    \item flare angles:
\end{itemize}
\indent \indent \indent \indent  $\Rightarrow$ 3.5 degrees (CCF3-5 configuration) ;\\
\indent \indent \indent  $\Rightarrow$ 10 degrees (CCF10 configuration).\\
The main dimensions of the cone-cylinder-flare configurations, CCF3-5 and CCF10, are given in table \ref{tab:coor}.

As stated earlier, the following sections will be devoted to a detailed presentation of the aerodynamic flows, as well as numerical stability analyses and experimental measurements.

%=======
\section{CFD computations}
%=======
\label{sec:aeroflows}

%----------
\subsection {Structured grids}
%----------
The grids used for the computations are single-block structured grids.
 
\begin{figure}
 	\centering
 	\includegraphics[width=0.74\linewidth]{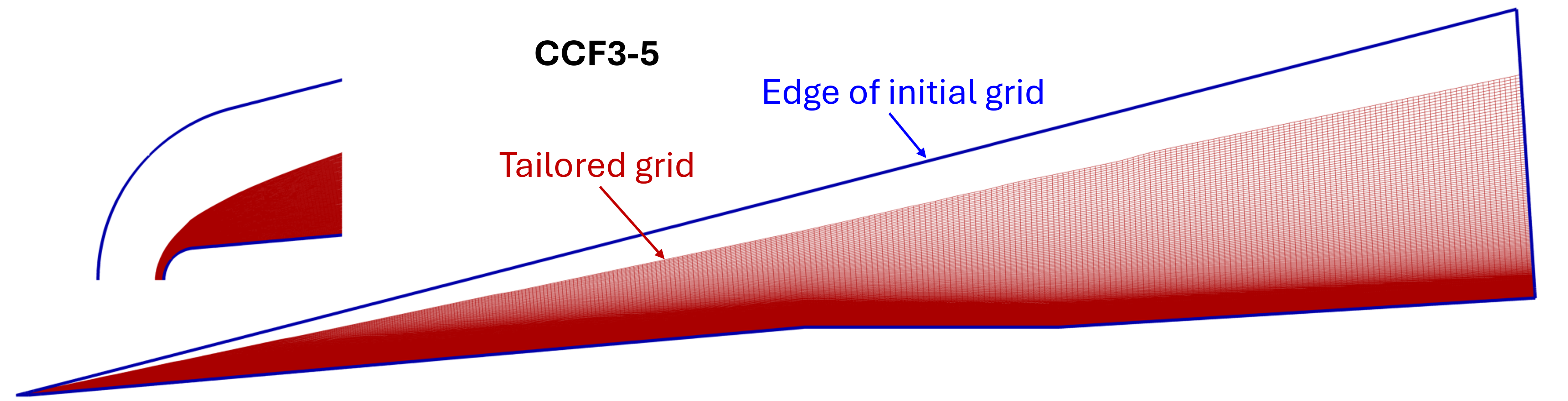}
 	\includegraphics[width=0.74\linewidth]{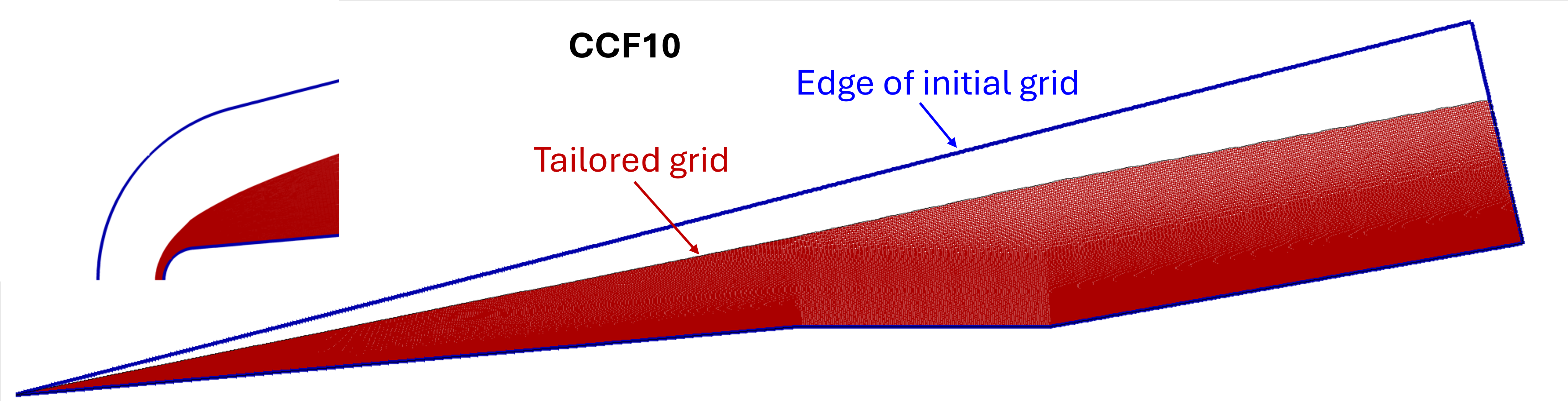}
 	\caption{CCF3-5 / CCF10 - Initial (edge only) and tailored grids at Mach=6}
 	\label{fig:CCF3-5_CCF10_tailgrid}
 \end{figure}

For the CCF3-5 configuration, the initial grid (blue edge in the top picture of figure \ref{fig:CCF3-5_CCF10_tailgrid}) is composed of 1009 points in the longitudinal direction and 301 points in the wall normal direction. The initial grid can be qualified as very good quality  but in order to ensure higher accuracy of the mean flow solution for the stability analysis, a shock tailoring procedure is used. This grid adaptation procedure allows to capture all the gradients precisely by redistributing the grid points based on the initial converged solution. The obtained shock tailored grid is represented in red in figure \ref{fig:CCF3-5_CCF10_tailgrid}.
 
% \begin{figure}
% 	\centering
% 	%\includegraphics[width=0.74\linewidth]{IMG/CCF_STABL_Grid_Ini_1009_301}
% 	\includegraphics[width=0.74\linewidth]{IMG/CCF_STABL_Grid_Tailor_Rem5_57M_Edge_Grid_Ini}
% 	\caption{CCF3-5 configuration - Initial and tailored grids (Mach=6 - Re${/m}$ = 5.6x10$^{6}$)  (1009 x 301)}
% 	\label{fig:CCF35_tailgrid}
% \end{figure}
% 
% \begin{figure}
% 	\centering
% 	\includegraphics[width=0.74\linewidth]{IMG/CCF10_M6_10_34bar_430K_Initial_TG_Grids_2311x401}
% 	\caption{CCF10 configuration - Initial (edge only) and tailored grid (Mach=6 - Re${/m}$ = 11.2x10$^{6}$) (2311 x 401)}
% 	\label{fig:CCF10_tailgrid}
% \end{figure}
 
The CCF10 grids (bottom picture of figure \ref{fig:CCF3-5_CCF10_tailgrid} with edges of initial grid in blue and shock tailored grid in red) are made of a large number of points: 2311 x 401 grid points. Indeed the flow separation region is a complex zone that needs very high refinement level to capture efficiently the complex physics: separation bubble, induced shock as well as reattachment shock among others (see figures \ref{fig:FlowSep1} and \ref{fig:FlowSep2} for SBLI physical details).

%---------- 
\subsection {Laminar mean flows}
%---------- 
 
The laminar mean flow solutions are performed with the axisymmetric CFD code provided with the STABL software suite \cite{johnson2005stabl}. This axisymmetric flow solver DPLR2D uses a finite volume formulation and solves the reacting-flow Navier-Stokes equations. The second-order inviscid fluxes are based on the modified Steger-Warming flux vector splitting method. The viscous fluxes are also second-order accurate. The time integration method is the implicit first-order data parallel line relaxation (DPLR) method.

Since the aerodynamic flows considered here correspond to cold wind tunnel conditions in air, the gas mixture considered for the computation is non-reacting gas "Air" (perfect gas). The effects of chemistry and molecular vibration are omitted for the calculations. The flow is a non-reacting mixture of $N_2$ and $O_2$. 

Free-stream conditions were obtained by applying the isentropic flow relations from the stagnation conditions to the specified free-stream Mach numbers. The main freestream aerodynamic conditions used for the computations are given in table \ref{tab:aero_cond1} where pi and Ti are the stagnation pressure and temperature, Re$/m$ is the unit Reynolds number, p$_{\infty}$, T$_{\infty}$, $\rho_{\infty}$ are the freestream pressure, temperature and density, T$_{wall}$ is the wall temperature and $V_{\infty}$ is the freestream velocity.

The viscosity law used is Sutherland’s law and the heat conductivity is calculated using Eucken’s relation. For the freestream temperature $T_{\infty}=52.4K$, the molecular viscosity value is $\mu$ = 3.400x10$^{-6} kg/(m.s)$ and the heat conductivity value is $k_{cond}$ = 0.004627 $W/(m.K)$.

\begin{table}
\begin{center}
\def~{\hphantom{0}}
	\centering
	\begin{tabular}{|c|c|c|c|c|c|c|c|c|c|}
        \hline
		Case & Mach & pi (bar/psia)& Ti (K)& Re${/m}$ & p$_{\infty}$ (Pa) & T$_{\infty}$ (K)  & $\rho_{\infty}$ (kg/m$^3$) & T$_{wall}$ (K)  & $V_{\infty}$ (m/s) \\
        \hline
		1  &  6.0  & 2.59 /  37.5 &430.  &     2.8x10$^{6}$  &  164.0  &  52.4 &  0.01090 &  300. &  871.0 \\
        \hline
		2  &  6.0  & 5.17 /  75.0 &430.  &     5.6x10$^{6}$  &  327.5  &  52.4 &  0.02175 &  300. &  871.0  \\
        \hline
		3  &  6.0  & 10.34 / 150.0  &430.  &  11.2x10$^{6}$ &  654.9 &  52.4 &  0.04351 &  300. &  871.0 \\
        \hline
	\end{tabular}
	\caption{Main aerodynamic conditions used for the computations}
	\label{tab:aero_cond1} 
	\end{center}
\end{table}

\begin{figure}
	\centering
	\includegraphics[width=0.98\linewidth]{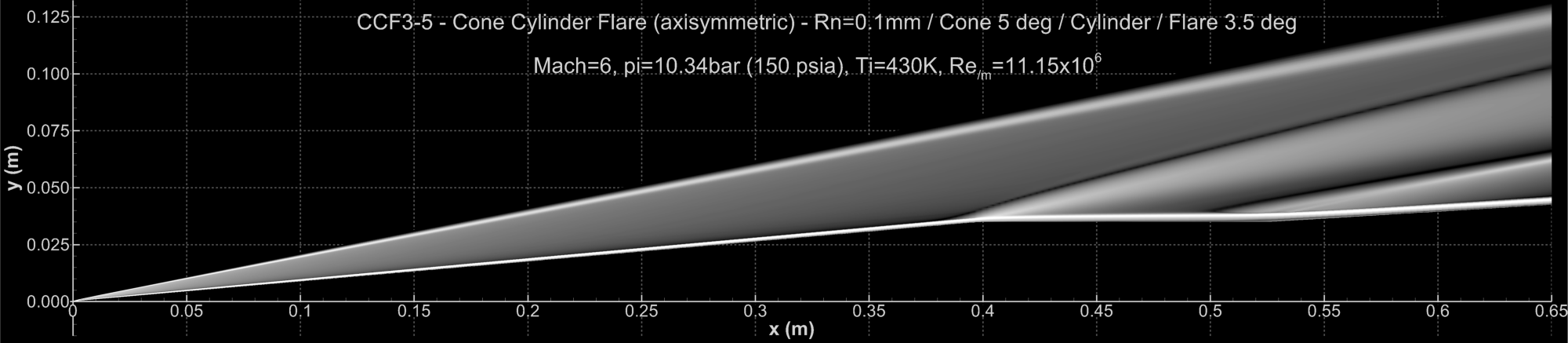}
	\includegraphics[width=0.10\linewidth]{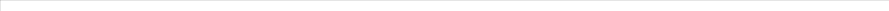}
	\includegraphics[width=0.98\linewidth]{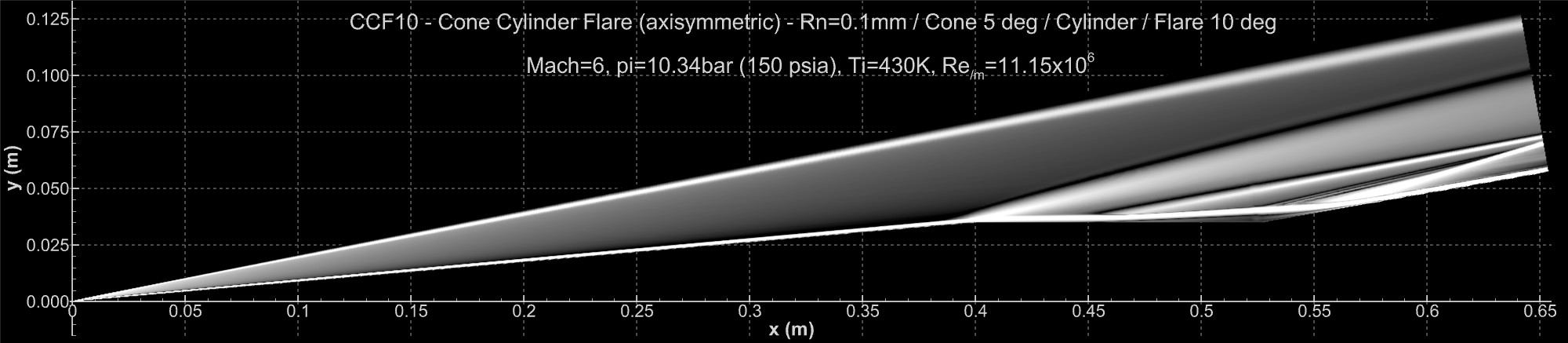}
	\caption{CCF3-5 / CCF10 - Density gradient magnitude at Re${/m}$ = 11.2x10$^{6}$}
	\label{fig:CCF3-5_10_strios}
\end{figure}

The laminar mean flow solutions obtained, respectively, in the CCF3-5 and CCF10 geometries at Re${/m}$ = 11.2x10$^{6}$ are presented in figure \ref{fig:CCF3-5_10_strios} as pseudo-schlieren visualizations where shock, boundary layer along vehicle, as well as flow expansion and recompression can be clearly identified.\\\\

There is a dramatic effect of the flare angle on the boundary-layer:
\begin{itemize}
    \item CCF3-5: the limited recompression generated by the flare can be handled by the incoming boundary-layer on the cylinder. The flow stays attached to the wall on the cylinder and on the flare ;
    \item CCF10: the boundary-layer, still attached just after the cone-cylinder, cannot handle the adverse pressure gradient imposed by the 10 degree flare. The boundary-layer separates from the wall, and a large separation bubble is generated. 
\end{itemize}

 %----------------------------
 \subsection{Flow analysis}
 %----------------------------
 
 %---------------
 \subsubsection{Flow around the nose}
 %---------------
 From a stability analysis point of view, it is known that the increase of the bluntness of the nose (until a given limit to avoid the blunt-body paradox) has a stabilizing effect on the boundary-layer \cite{stetson1987blt}. Indeed, in hypersonic flow, increasing nose bluntness pushes back the point where second-mode disturbances become active. Here, the cone has a quasi-sharp nose ($R_n=0.1mm$) so it is very favorable to develop second-mode instabilities along the cone.
 
\begin{figure}
 	\centering
 	\includegraphics[width=0.43\linewidth]{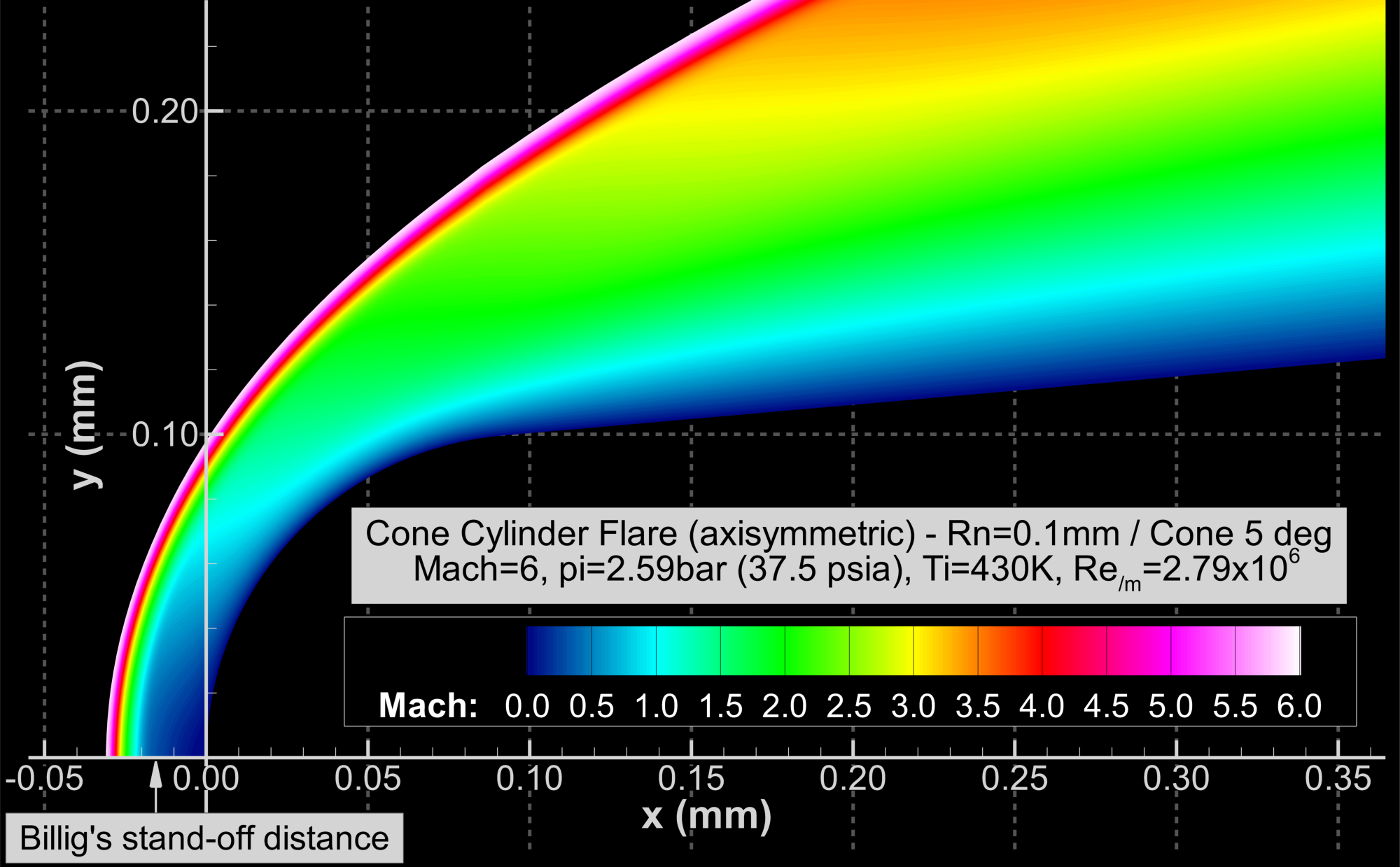}
    \includegraphics[width=0.43\linewidth]{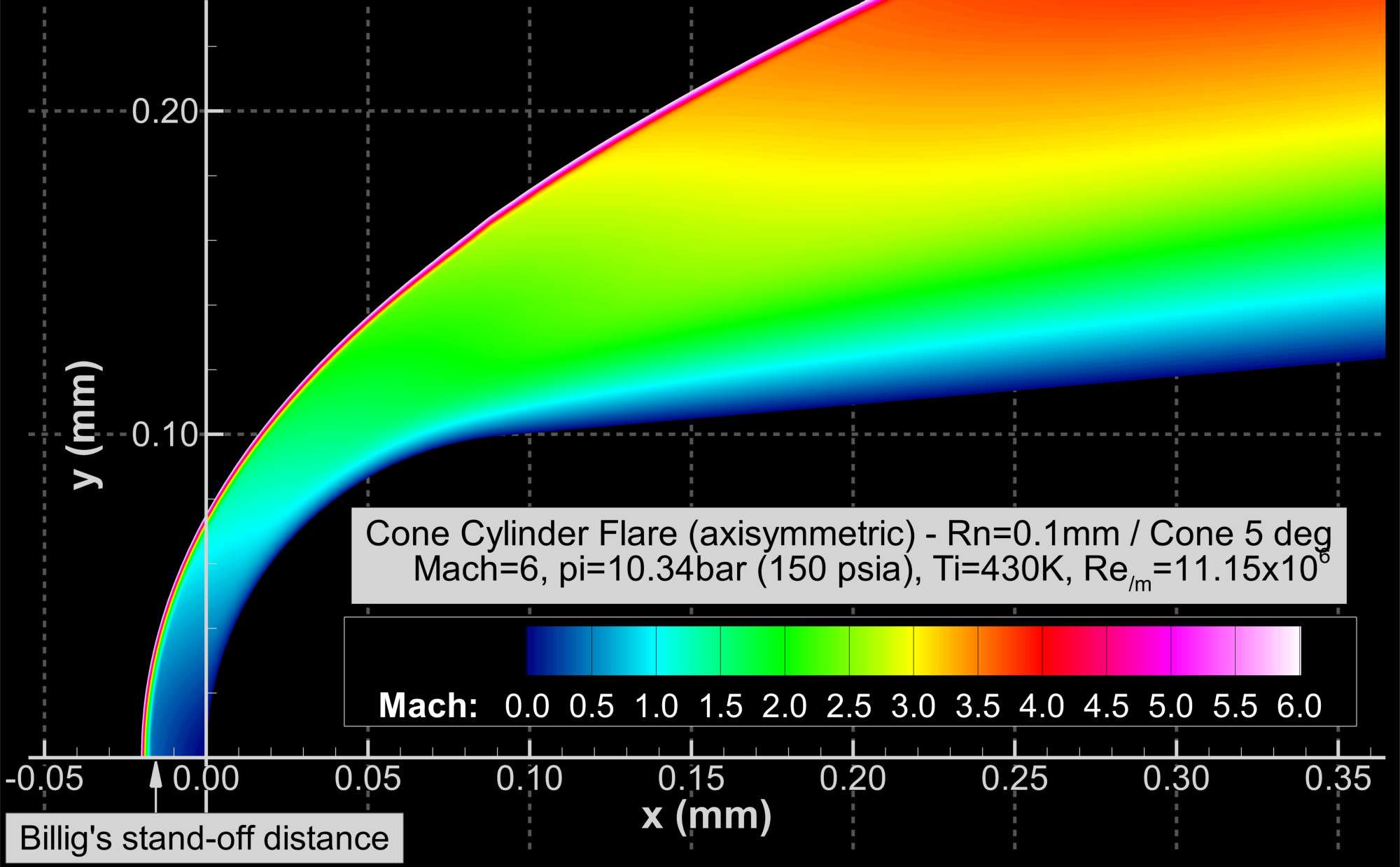}
    \caption{Mach number distribution around the nose at Re${/m}$ = 2.8x10$^{6}$ and 11.2x10$^{6}$}
 	\label{fig:bow_shock_Mach}
 \end{figure}

 For the accuracy of the stability analysis, it was decided to give special attention to the nose region in the computations. For this perfect-gas flow, the Mach number distribution around the stagnation point is presented in figure \ref{fig:bow_shock_Mach} for two Reynolds numbers.

 From a quantitative point of view, the shock stand-off distance $\Delta$ can be compared to the Billig's empirical formula for sphere-cone shapes: % for inviscid flows
 \begin{equation}
  {\Delta}/{R_n} = 0.143\exp({3.24}/{M_{\infty}^2})
  \label{Eq.Delta_Rn}
\end{equation}
 
 Here with $R_n=0.1mm$, the shock detachment evaluated with Billig's formula is equal to $\Delta=0.0156 mm$. This stand-off distance, plotted in figure \ref{fig:bow_shock_Mach}, highlights the physical Reynolds number effect on the shock layer thickness. A decrease in thickness of the shock layer due to the increase of density is clearly visible at Re${/m}$ = 11.2x10$^{6}$ when compared to Re${/m}$ = 2.8x10$^{6}$. As a consequence, a better agreement is logically observed between the higher Reynolds number considered here and the Billig's empirical stand-off distance.
 
\begin{figure}
 	\centering
 	\includegraphics[width=0.43\linewidth]{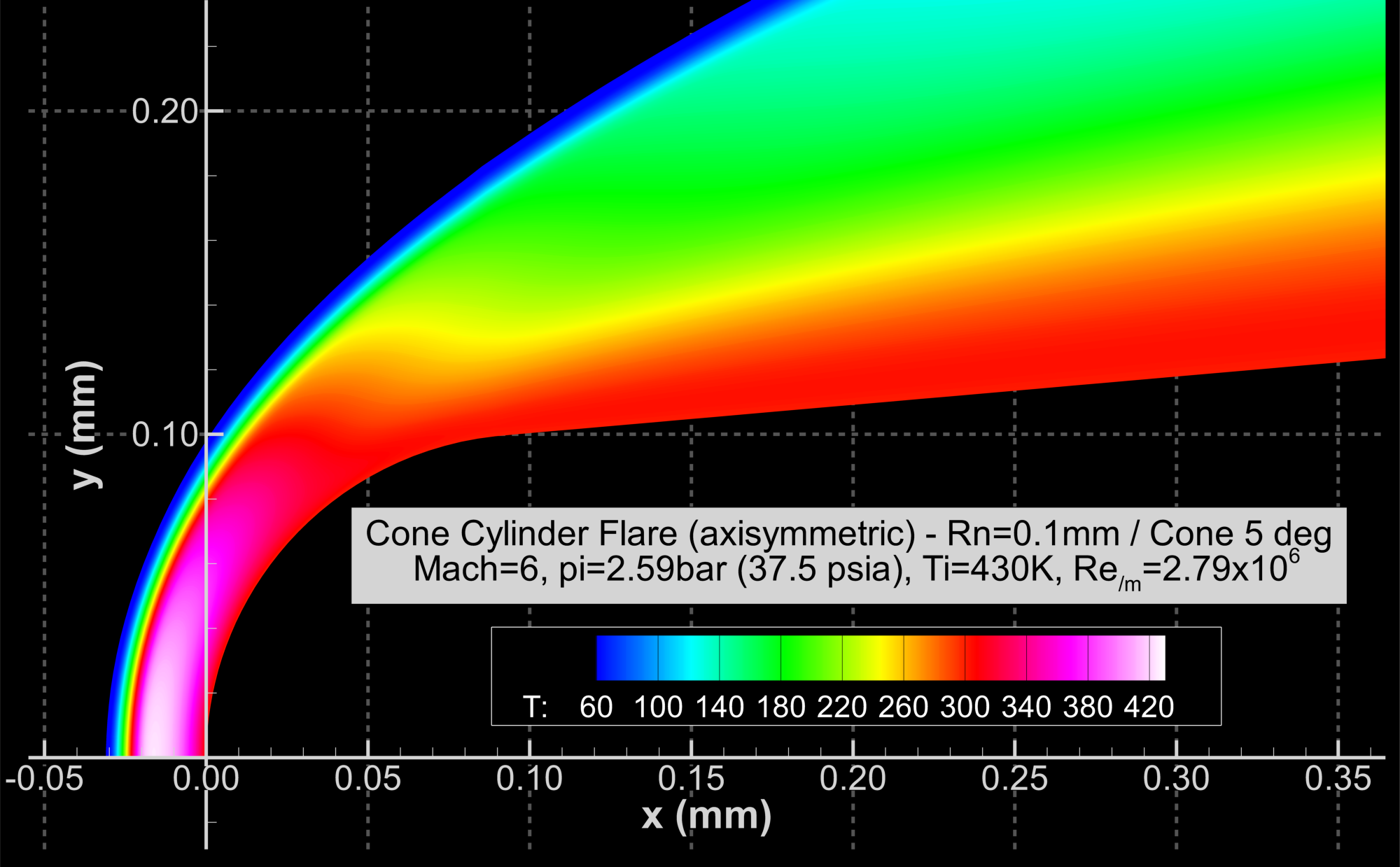}
 	\includegraphics[width=0.43\linewidth]{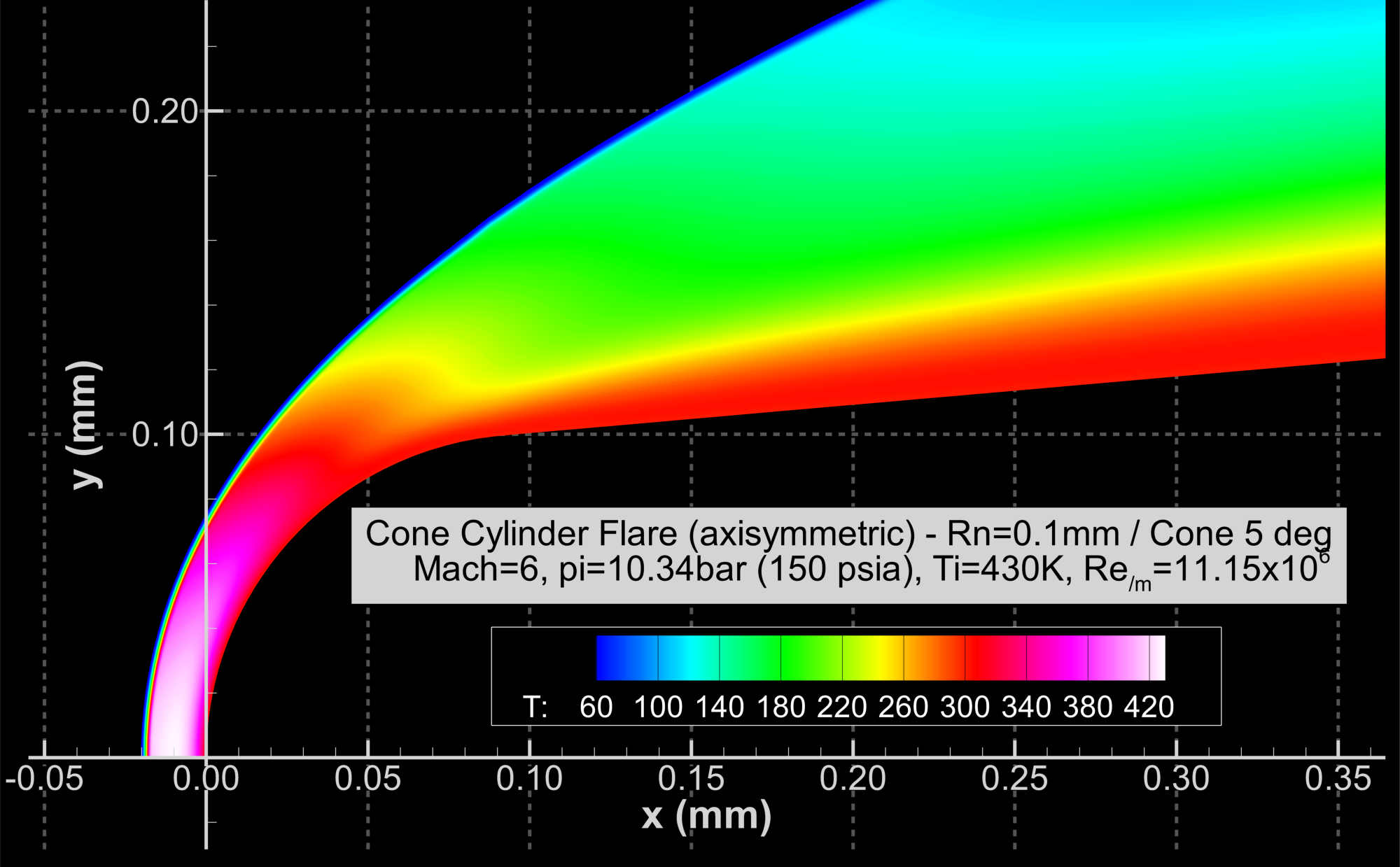}
 	\caption{Temperature distribution around the nose at Re${/m}$ = 2.8x10$^{6}$ and 11.2x10$^{6}$}
 	\label{fig:bow_shock_temp}
 \end{figure}

 The thermal environment around the nose is due to the conversion of the kinetic energy associated with hypersonic flight. As shown in figure \ref{fig:bow_shock_temp}, this conversion, due to the high compression and viscous energy dissipation mechanisms around this stagnation region, leads into increasing temperature around the blunt nose as explained by \cite{bertin1994hyp}. The isothermal wall temperature equal to 300 K explains the decrease of temperature near the wall.

 The high gradients around the nose are very accurately captured by the computations thanks to the fine initial grids used and to the grid tailoring procedure. So the influence of this very small bluntness is thoroughly defined for stability analysis purposes.

 %---------------
 \subsubsection{Flow separation}
 %---------------
 
\begin{figure}
 	\centering
 	\includegraphics[width=0.88\linewidth]{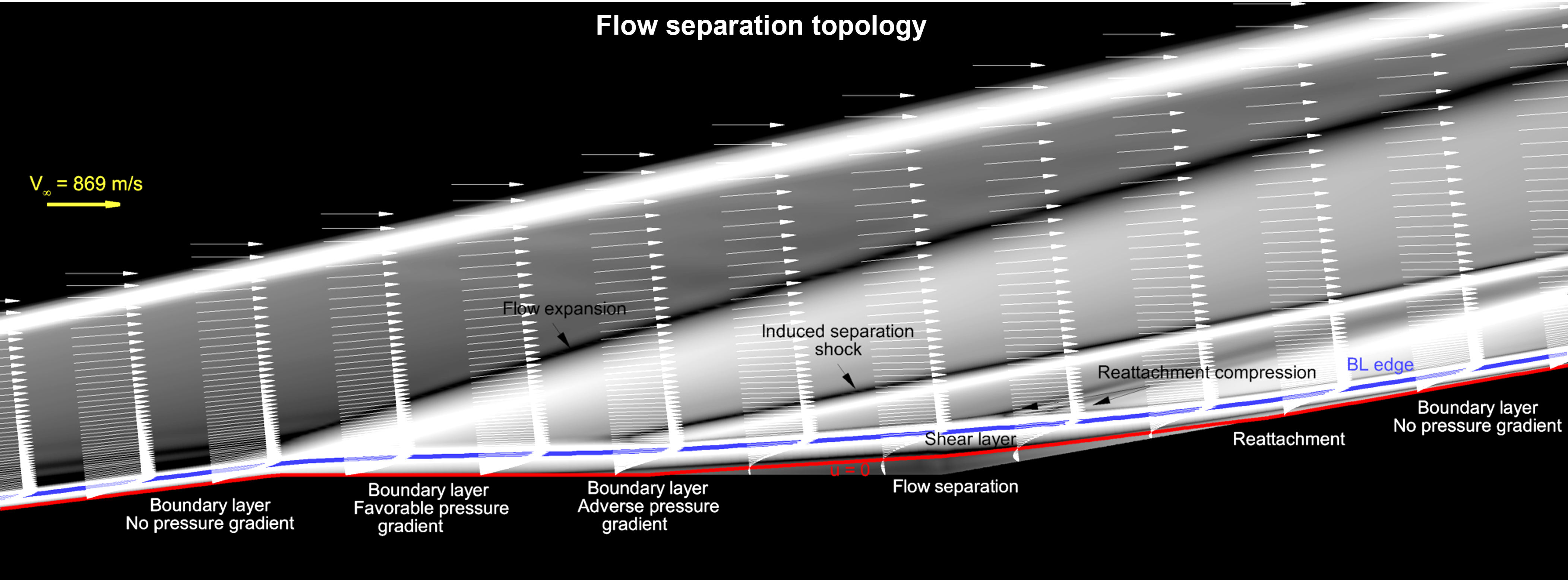}
 	\caption{Physics of the flow separation region}
 	\label{fig:FlowSep1}
 \end{figure}

\begin{figure}
 	\centering
 	\includegraphics[width=0.74\linewidth]{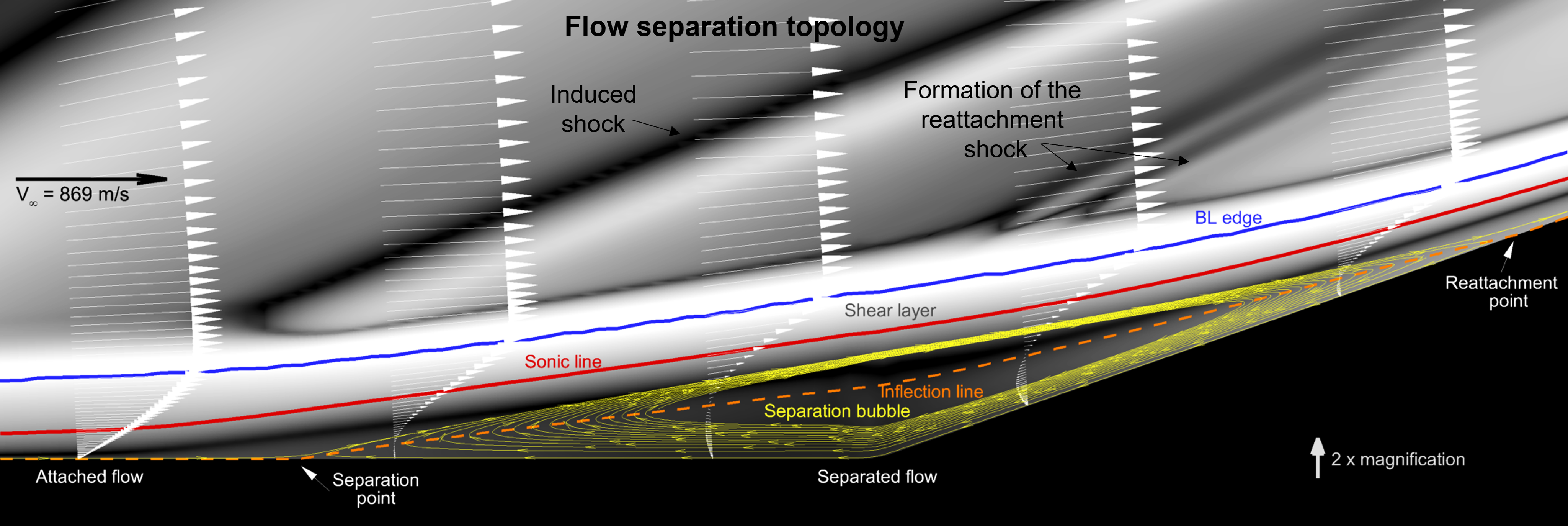}
 	\caption{Physics of the flow separation region: enhanced view of the separation bubble}
 	\label{fig:FlowSep2}
 \end{figure}

 As indicated previously, the 10-degree flare angle generates a flow separation at the junction of the cylinder and the flare as illustrated in figures \ref{fig:FlowSep1} and \ref{fig:FlowSep2}. 
 
Listed below is a short description of the flow separation zone:
\begin{itemize}
    \item Because of the presence of the separation bubble, an induced shock is generated near the separation point, leading to a local pressure increase ;
    \item Inside the bubble, the velocity profiles are strongly modified and negative values appear in the lower part of the bubble below the inflection line ;
    \item Upstream of the reattachment point, a recompression shock is gradually forming leading to local high pressure gradient on the flare ;
    \item The rapid change of the boundary-layer properties in this region leads to high heat flux values at the reattachment point where the boundary-layer thickness reaches a minimum ;
    \item Above the bubble, the mixing layer is known to be a region of important instability amplification leading to abrupt laminar-turbulent transition breakdown under some conditions.
 \end{itemize}
 The complexity of the separation region gives an idea why laminar-turbulent transition prediction in the presence of hypersonic flow separation is one of the long-standing problems for high speed flows.\\

%---------------
\subsubsection{Pressure and skin friction}
%---------------

\begin{figure}
	\centering
        \includegraphics[width=0.46\linewidth]{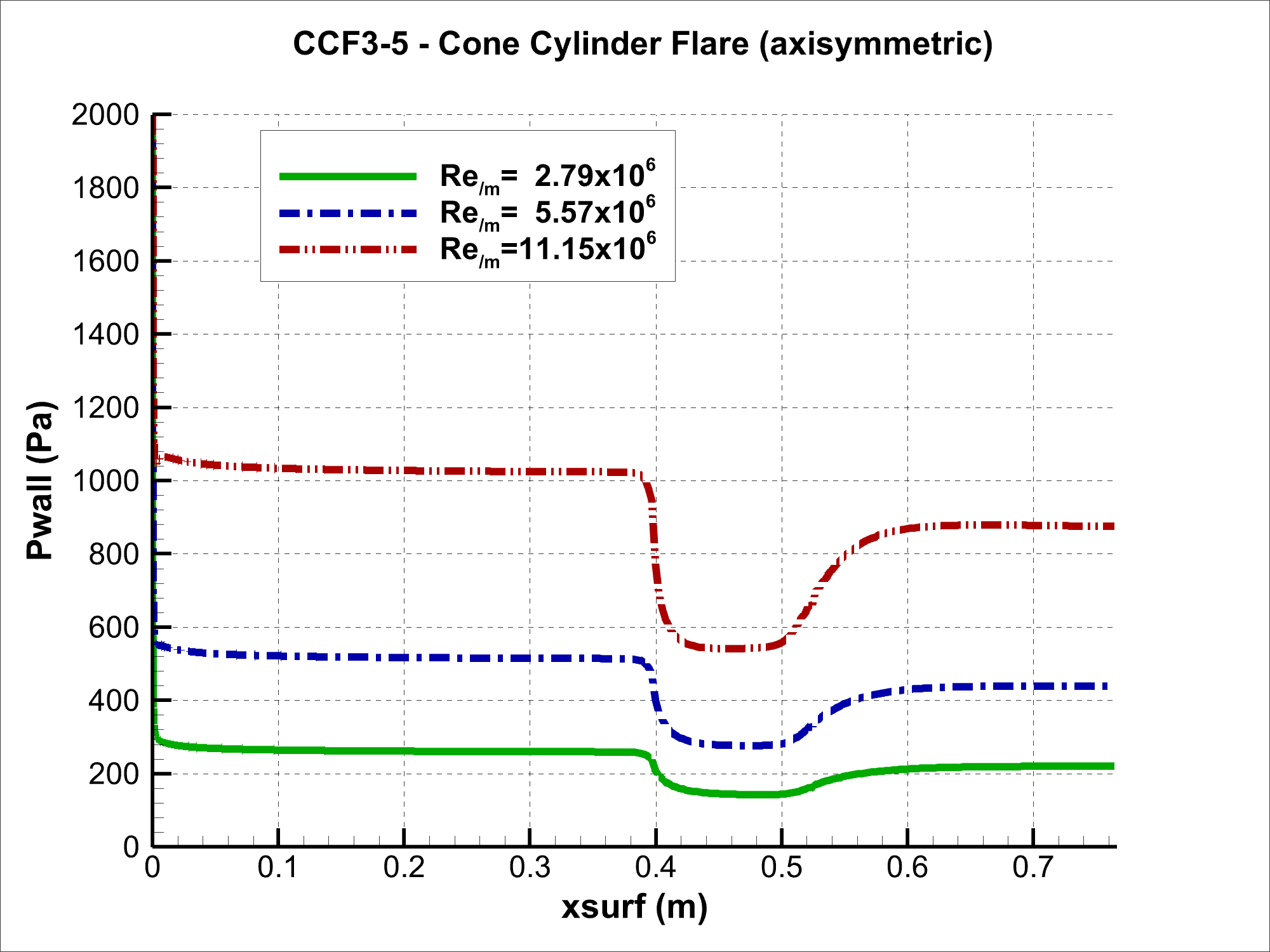}
        \includegraphics[width=0.46\linewidth]{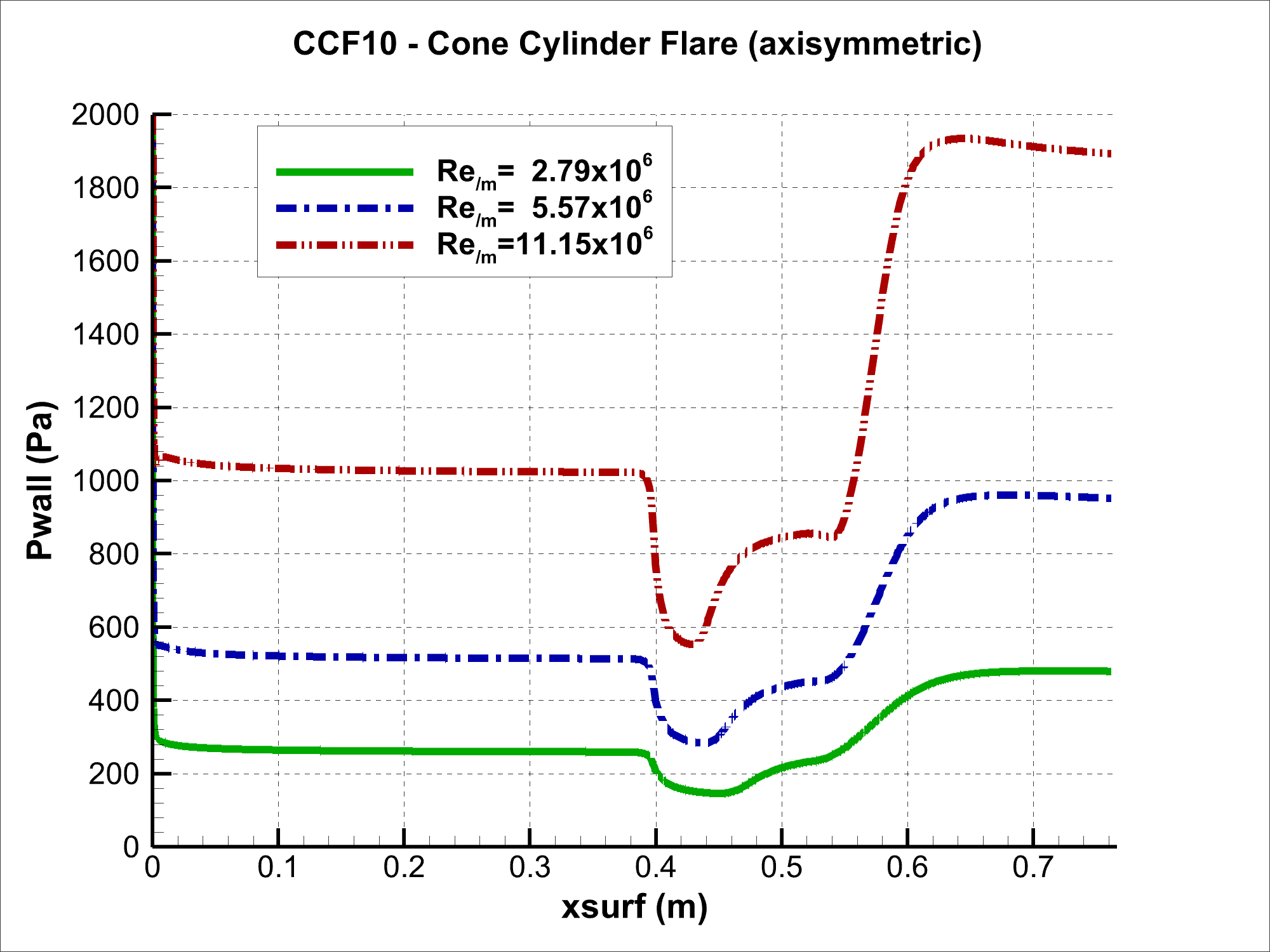}
	\caption{CCF3-5 / CCF10 - Pressure as a function of Reynolds number}
	\label{fig:CCF3-5press_cf}
\end{figure}

\begin{figure}
	\centering
		\includegraphics[width=0.46\linewidth]{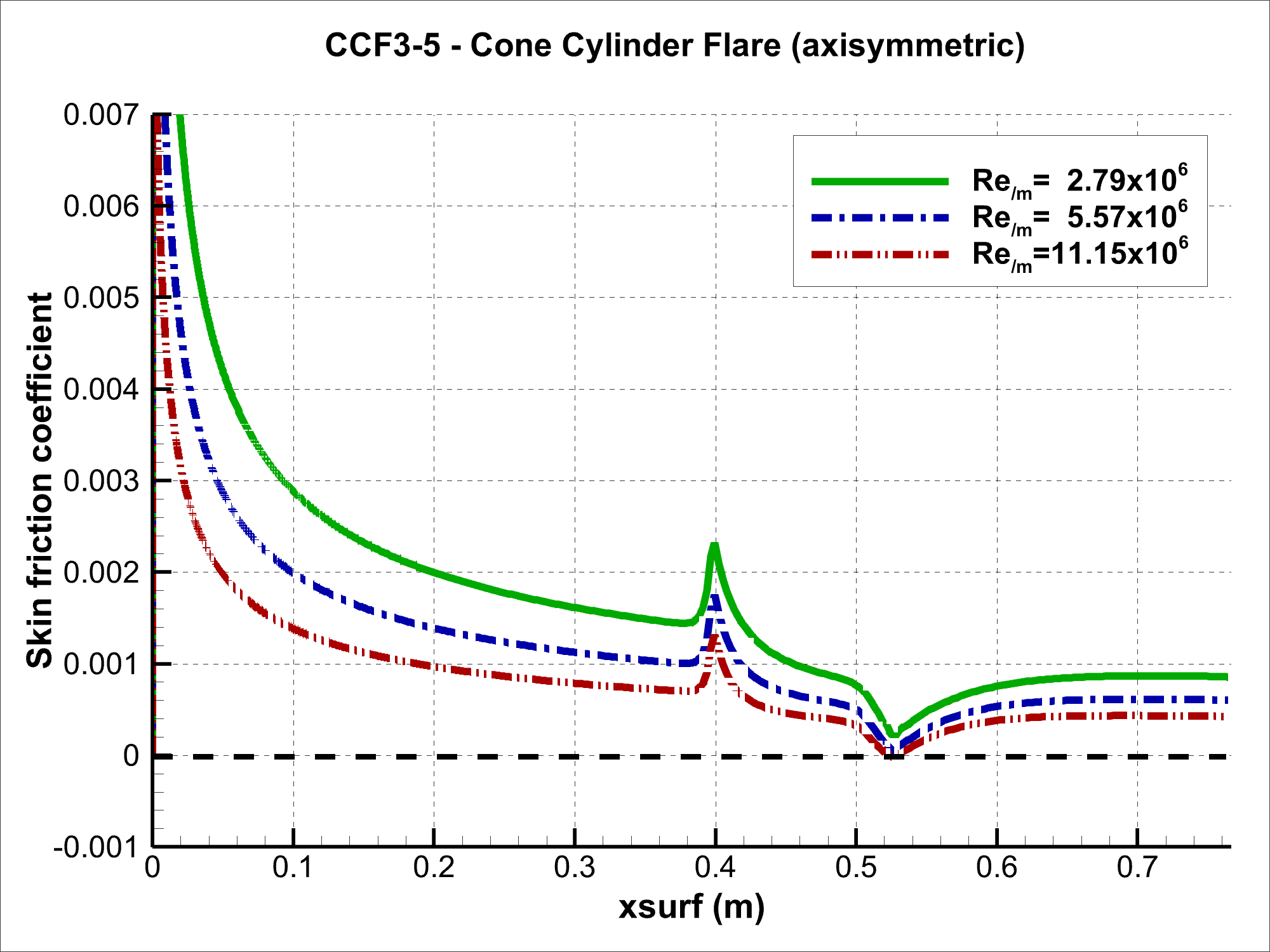}
		\includegraphics[width=0.46\linewidth]{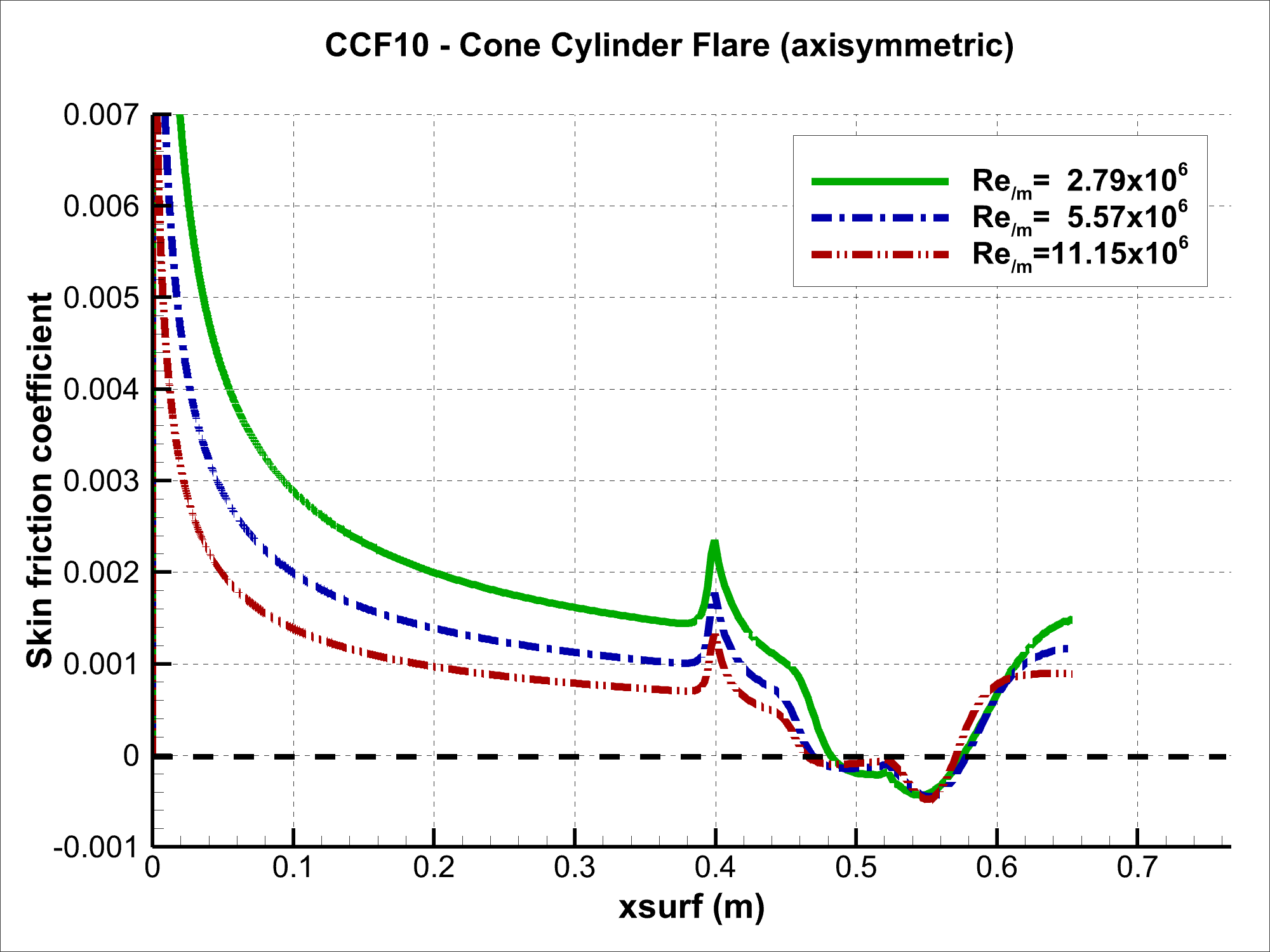}
	\caption{CCF3-5 / CCF10 - Skin friction as a function of Reynolds number}
	\label{fig:CCF10press_cf}
\end{figure}

%---------------
\noindent {\it{\textbf{Attached boundary-layer case (CCF3-5)}}}\\
%---------------
On cone-cylinder-flare geometry, the flow encounters not only the classical quasi-constant pressure level on the cone but also experiences pressure gradients further downstream: first a favorable pressure gradient on the cylinder and second an adverse pressure gradient on the flare as shown in figure \ref{fig:CCF3-5press_cf}. The skin friction constantly diminishes on the cone and, after a narrow peak at cone-cylinder junction, experiences a new significant decrease on the cylinder (see figure \ref{fig:CCF10press_cf}). The minimum skin friction value at the cone-flare junction is followed by a slight and progressive increase on the flare.\\

As stated previously, near the cylinder-flare junction the flow faces an adverse pressure gradient due to the presence of the flare. When considering a significant flare angle, the flow of incoming fluid near the wall might not cross the recompression region and might not remain attached to the wall. In this situation, the pressure gradient will lead to a flow separation \cite{wuerer1965sep}.

\begin{figure}
	\centering
        \includegraphics[width=0.46\linewidth]{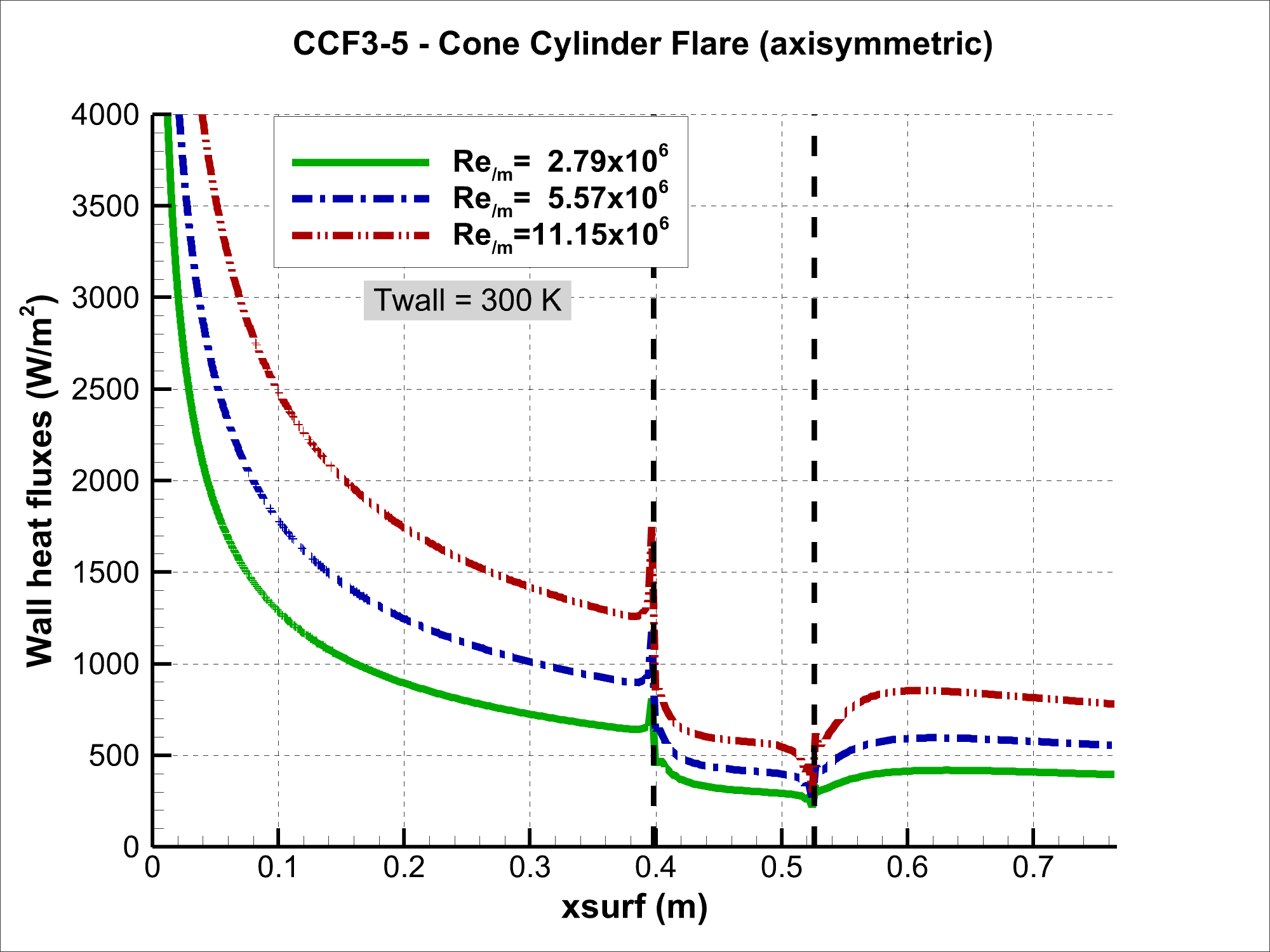}
        \includegraphics[width=0.46\linewidth]{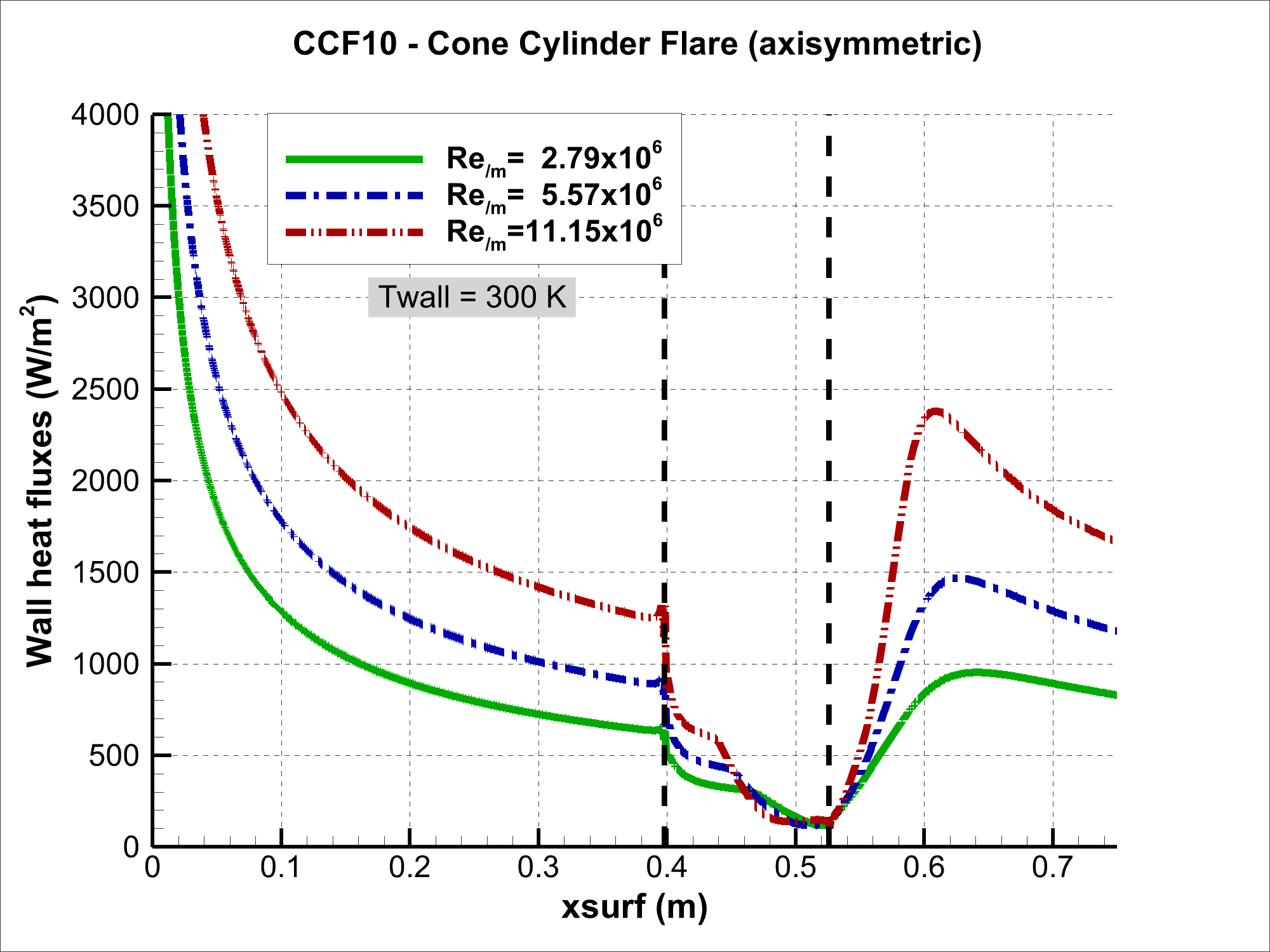}
	\caption{CCF3-5 / CCF10 - Wall heat-fluxes as a function of Reynolds number}
	\label{fig:CCF3-5_CCF10_heat_fluxes}
\end{figure}

Here the CCF3-5 shape has been designed with a small flare angle in order to generate a limited adverse pressure gradient so that no separation bubble appears. It is confirmed that the boundary-layer is still attached. It can be noticed nevertheless that the skin friction coefficient is close to zero at the cylinder-flare junction. The addition of the wall shear stress and the positive pressure increase is not sufficient to generate a back-flow region but it can be stated that any increase of the flare angle will lead to the formation of a separation bubble. This is exactly the case when increasing the flare angle to 10 degrees on the CCF10 configuration.\\

%---------------
\noindent {\it{\textbf{Flow separation case (CCF10)}}}\\
%---------------
The effect of flow separation on the pressure and skin friction is illustrated on figures \ref{fig:CCF3-5press_cf} and \ref{fig:CCF10press_cf}. The shock induced by the separation leads to a pressure increase followed by a pressure plateau. The recompression moves downstream near the reattachment point and is visible as a strong pressure gradient on the flare. The skin friction is now negative over the extent of the separation bubble.

%---------------
\subsubsection{Wall heat-fluxes}
%---------------
The wall heat fluxes are obviously very similar on the conical part knowing, nevertheless, that the grid refinement is higher for CCF10 in the boundary-layer direction (401 points for CCF10 compared to 301 points for CCF3-5).\\

For CCF3-5, the heat fluxes are low on the cylinder and on the flare (see figure \ref{fig:CCF3-5_CCF10_heat_fluxes}). On CCF10, the flow separation has a dramatic consequence: the heat flux is very low below the separation bubble and reach significant levels (in fully laminar conditions) at reattachment where the boundary-layer thickness is minimal.

%---------------
\subsubsection{Boundary-layer properties and velocity profiles}
%---------------
The boundary-layer edge detection is never a trivial calculation for hypersonic flows. Here the detection is based on the ``Return from enthalpy overshoot`` criterion. After detecting the total enthalpy peak in the boundary-layer profile, the algorithm moves away from the wall until the total enthalpy has returned to a value near the freestream one.\\

For this criterion, the total enthalpy at the boundary-layer edge satisfies:

\begin{equation}
  Hi_{e} < \frac{Hi_{\infty}-Hi_{wall}}{0.995}+Hi_{wall}
  \label{Eq.Hi_Criterion}
\end{equation}

A detailed analysis of the STABL edge detection algorithms is available in \cite{chynoweth2018thesis}.\\

\begin{figure}
	\centering
	\includegraphics[width=0.46\linewidth]{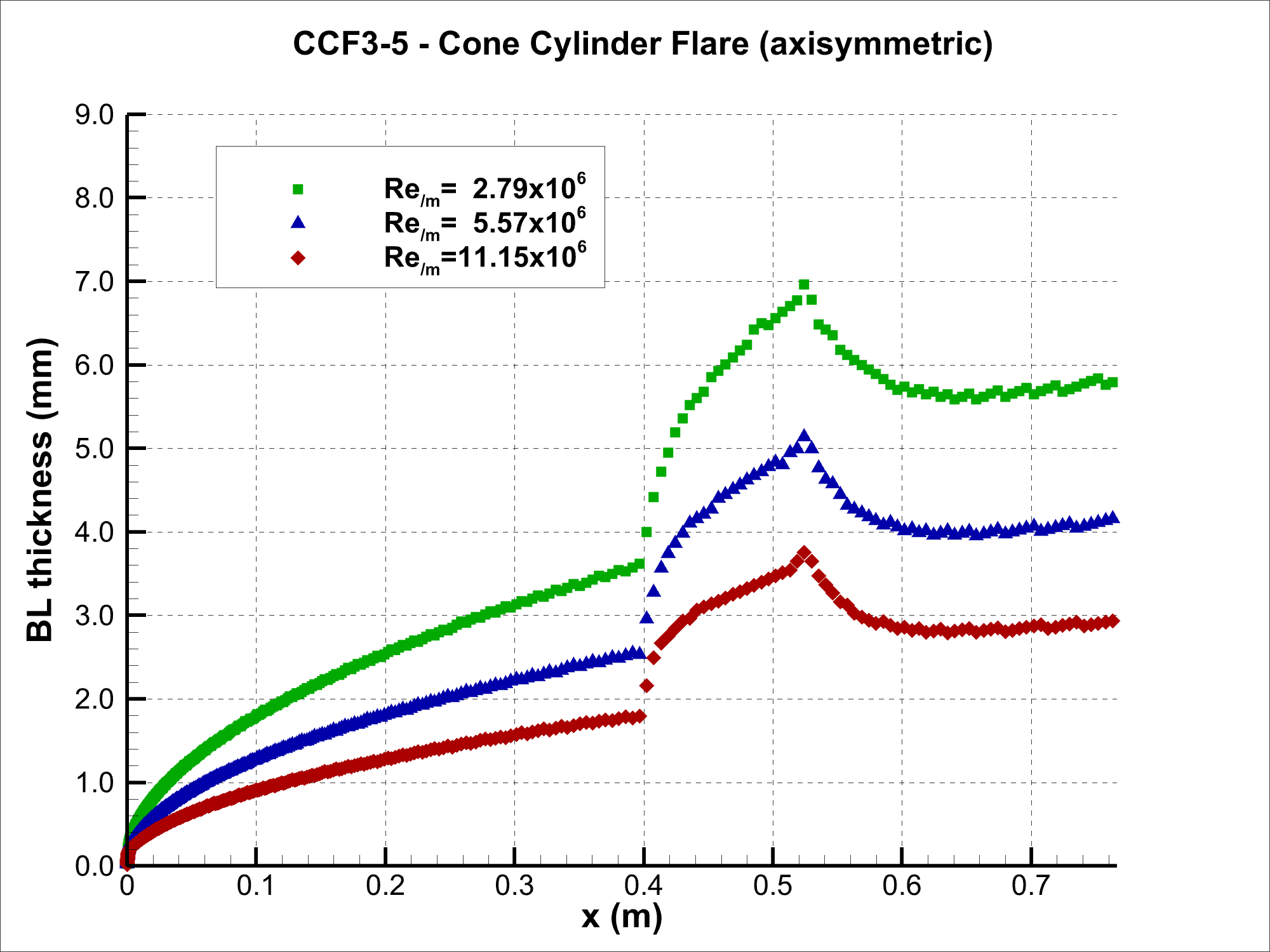}
	\includegraphics[width=0.46\linewidth]{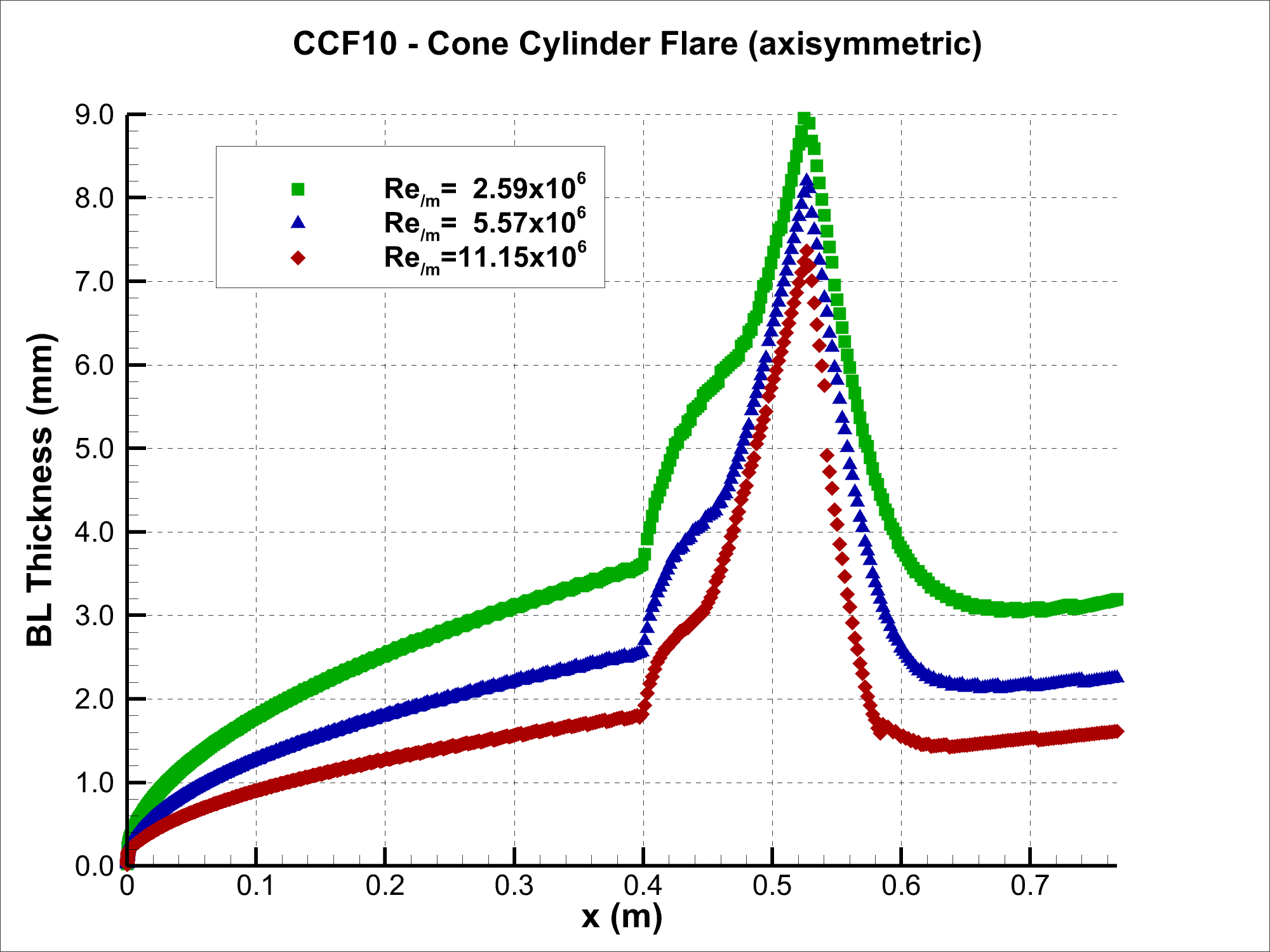}
	\caption{CCF3-5 / CCF10 - Boundary-layer thickness (edge of the shear-layer above the separation bubble for CCF10)}
	\label{fig:CCF3-5_10delta}
\end{figure}

%After careful checks, this boundary-layer edge detection method appears well suited and accurate for the present cases, which is mandatory for thorough stability analysis purposes.

The resulting boundary-layer thicknesses are presented in figure \ref{fig:CCF3-5_10delta}. Whatever the Reynolds number considered, the general trend of the boundary-layer thickness evolution is the same. 
After the classical increase of the boundary-layer thickness along the cone, a new abrupt and large thickness increase appears on the cylinder because of the influence of the flow expansion generated at cone-cylinder junction. 
Moving downstream to the flare, two different cases appear depending on the configuration:
\begin{itemize}
    \item CCF3-5: the adverse pressure gradient leads to a significant decrease of the boundary-layer thickness on the first half of the flare before a slight and progressive increase of the boundary-layer thickness further downstream ;
    \item CCF10: the separation bubble is a large zone which leads to a high increase of the detected boundary-layer thickness. Indeed, this limit includes the recirculation region and the mixing layer that develops above it (see figure \ref{fig:FlowSep2}).  A larger distance above the wall is concerned by viscous effects. The shear-layer stand-off distance is maximum above the cylinder-flare junction, it is of the order of 7 to 9 mm in the considered case. The increase of Reynolds number changes the length of the separation bubble (longer for high Reynolds numbers for these laminar solutions).\\
\end{itemize}

\begin{figure}
	\centering
	\includegraphics[width=0.43\linewidth]{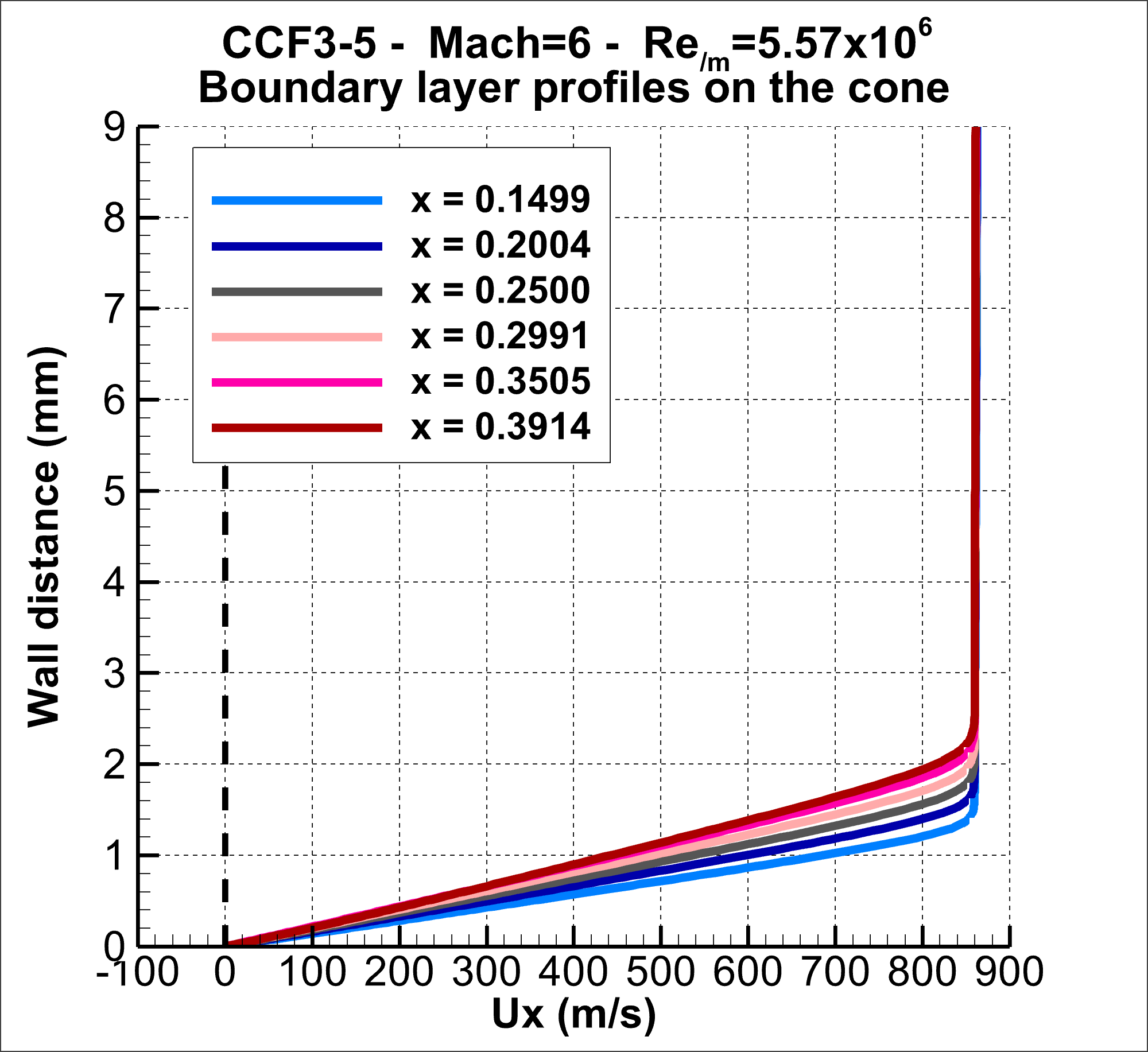}
	\includegraphics[width=0.43\linewidth]{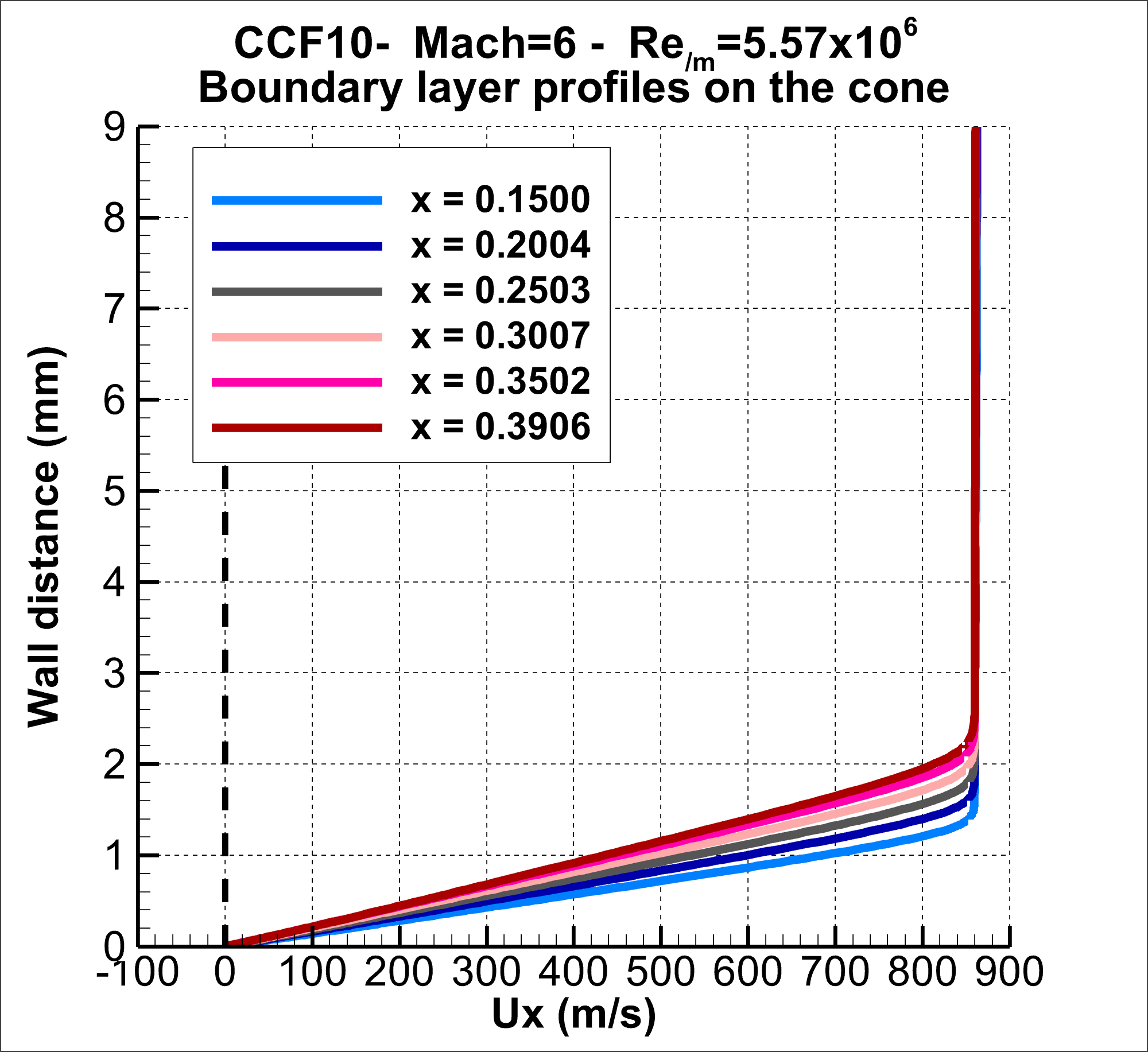}
	\caption{CCF3-5 / CCF10 - Boundary-layer profiles on the cone  at Re${/m}$ = 5.6x10$^{6}$}
	\label{fig:BLprofiles_Cone}
\end{figure}

\begin{figure}
	\centering
	\includegraphics[width=0.43\linewidth]{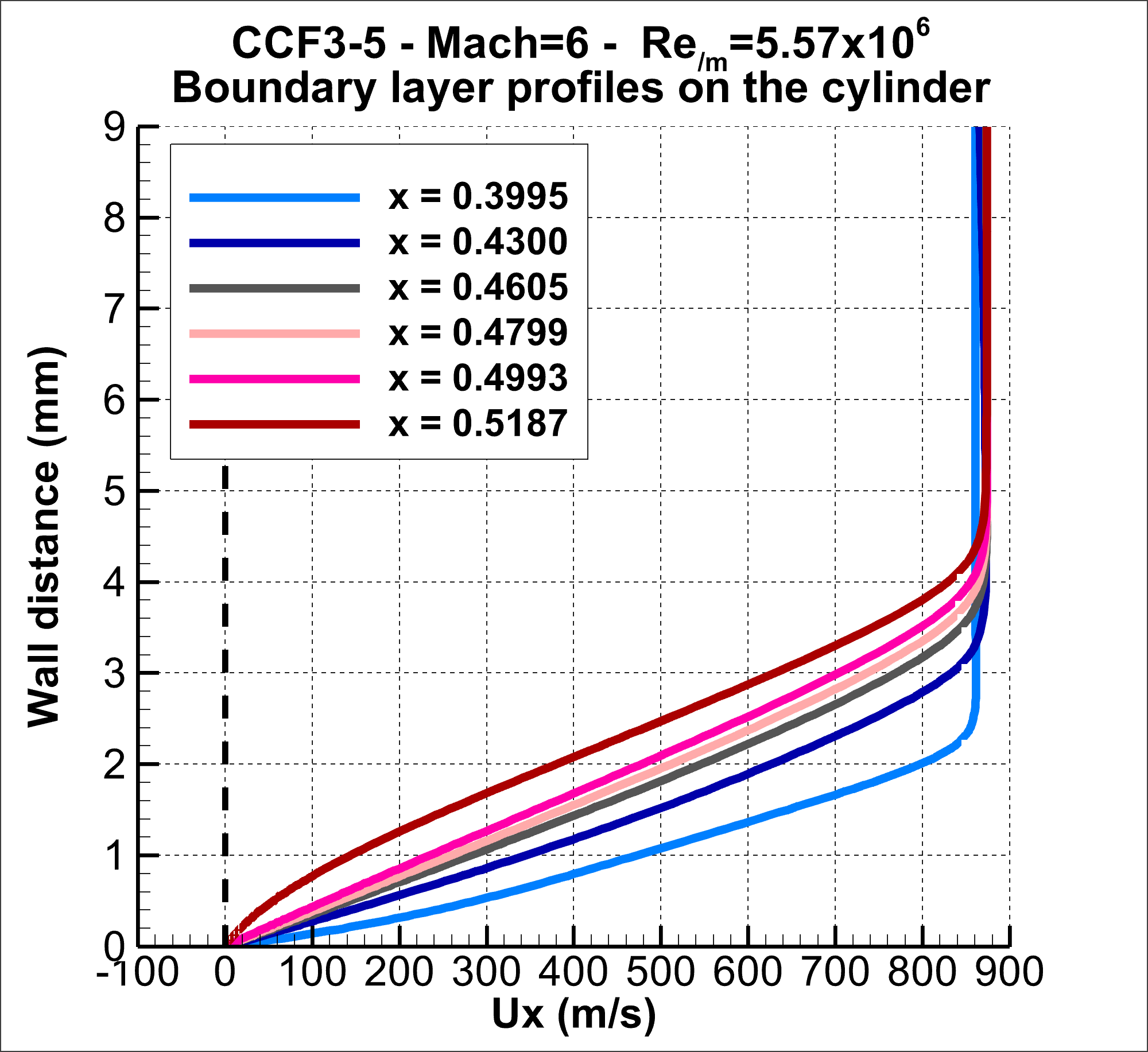}
	\includegraphics[width=0.43\linewidth]{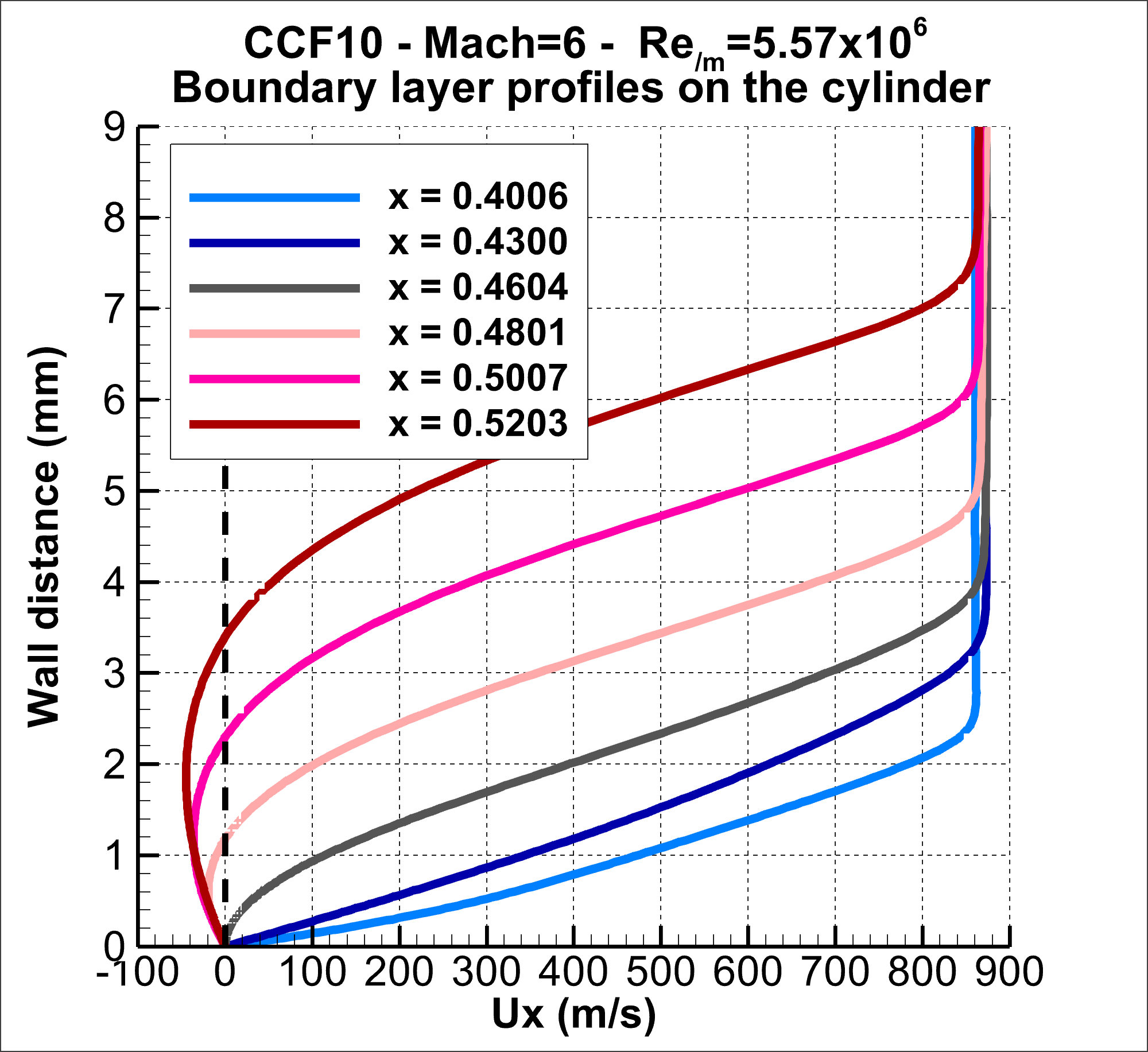}
	\caption{CCF3-5 / CCF10 - Boundary-layer profiles on the cylinder at Re${/m}$ = 5.6x10$^{6}$}
	\label{fig:BLprofiles_Cyl}
\end{figure}

\begin{figure}
	\centering
	\includegraphics[width=0.43\linewidth]{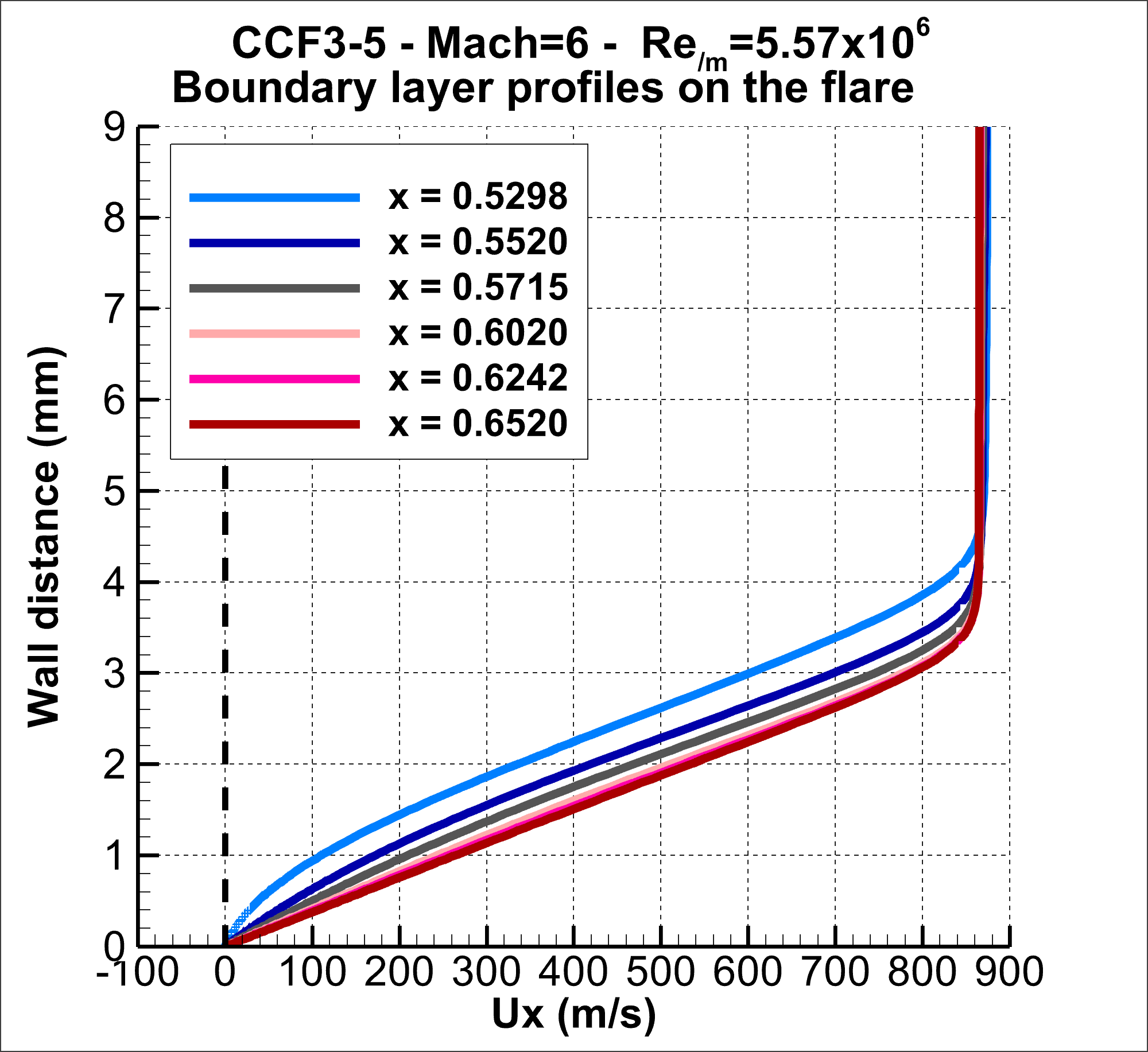}
	\includegraphics[width=0.43\linewidth]{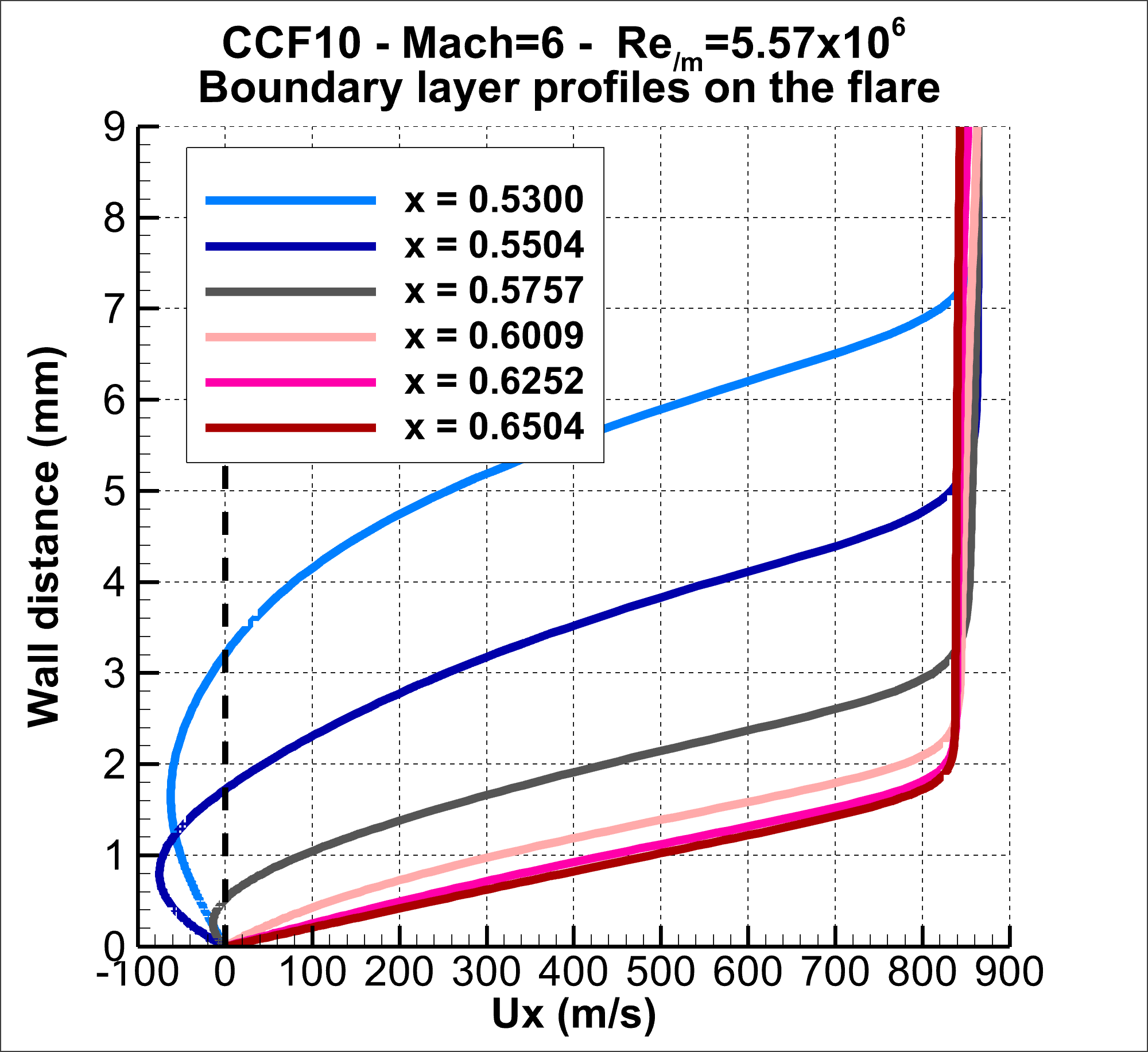}
	\caption{CCF3-5 / CCF10 - Boundary-layer profiles on the flare at Re${/m}$ = 5.6x10$^{6}$}
	\label{fig:BLprofiles_Flare}
\end{figure}

The boundary-layer profiles that develop on the cone, on the cylinder (accelerated flow) and on the flare (decelerated flow) are shown in figures \ref{fig:BLprofiles_Cone}, \ref{fig:BLprofiles_Cyl} and \ref{fig:BLprofiles_Flare} at  Re${/m}$=5.6x10$^{6}$ for both configurations with CCF3-5 on the left figures and CCF10 on the right ones. The profiles are strictly identical on the cone but, they are dramatically different on the cylinder and on the flare.\\

On CCF10, negative velocity values are confirmed in the lower part of the boundary-layer upstream and downstream of the cylinder-flare junction. They are a good marker of the recirculation bubble extension. The unperturbed freestream flow is recovered at twice the height of the one detected for the attached case with very different velocity profiles for each case.
The flow separation has a big impact on the global aerodynamics of the flow and, as a consequence, it can be guessed that the impact on the boundary-layer stability will be significant.

%---------------
\subsection{About local stability analyzes: only the attached flow considered here}
%---------------
In the previous sections, the aerodynamic flows have been presented in detail to quantify the influence of the shape. In spite of this extensive study, the numerical stability analysis of the cone-cylinder-flare shape with flow separation is out of the scope of this paper because local stability analysis cannot handle such highly non-parallel flows relative to separation bubble.

Nevertheless experimental results will be presented on the separated case in the last part of this article, even though recent numerical studies have started to investigate the flow separated cases with more advanced stability tools (see \cite{paredJ061829}, \cite{li2022-3855}, \cite{benit2023bluntcone}, \cite{caillaud2023}).\\

%Here, the focus is on CCF3-5.

%===============
\section{Instability waves on CCF3-5}
%===============
\label{sec:instab_waves}

For a low disturbance freestream environment, there are at least four different instability mechanisms described by linear theory which can produce disturbance growth in a hypersonic boundary-layer: first-mode, second-mode, crossflow instabilities and Görtler vortices \cite{stetson1990hypbl}.

\begin{figure}
	\centering
	\includegraphics[width=0.98\linewidth]{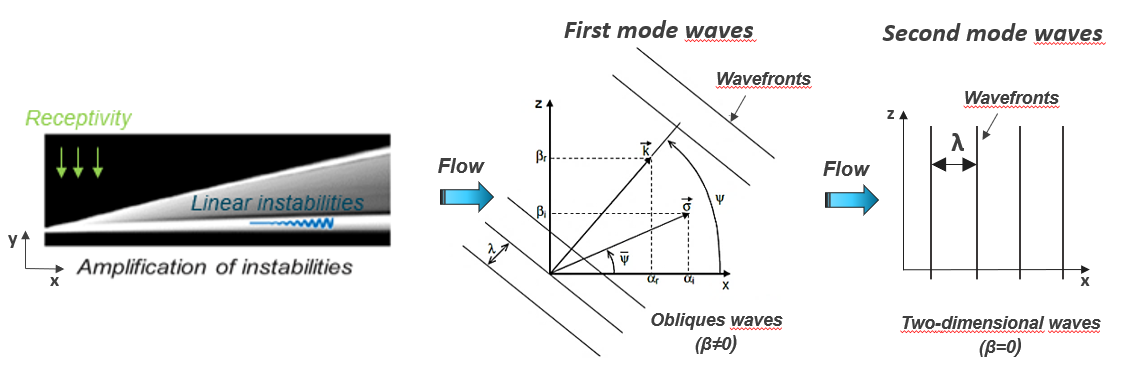}
	\caption{Linear instabilities - 1st and 2nd modes}
	\label{fig:Linear1st2nd}
\end{figure}

%----------
\subsection{Main instability mechanisms}
%----------
%----------
\subsubsection{First-mode instabilities}
%----------
The first-mode wave is an inflectional instability which can be thought as the compressible counterpart of Tollmien-Schlichting waves in incompressible flow \cite{thome2019BL}. It occurs for high subsonic and moderate supersonic flows and it is most amplified when the wavefronts are oblique to the stream direction (see figure \ref{fig:Linear1st2nd}). The wave angle $\psi$ of the most unstable disturbance increases rapidly with Mach number and is in the range from 55°-60° above a Mach number of 1.5 (see left illustration in figure \ref{fig:CCF3-5edgeMach}). First-mode instabilities are less amplified but on a larger distance than second-mode waves, so they can reach comparable amplifications, as seen later in the paper. The estimated first-mode instability frequency is calculated as:

\begin{equation}
  f_{\,1st\,mode} \simeq \frac{V_e}{10\delta}
  \label{Eq.1stMode}
\end{equation}

From a physical point of view, the first-mode fluctuations lie near the boundary-layer edge close to the generalized inflection point. As a consequence, the associated pressure fluctuations are generally difficult to detect with wall pressure sensors.

%----------
\subsubsection{Second-mode instabilities}
\begin{figure}
	\centering
    \includegraphics[width=0.51\linewidth]{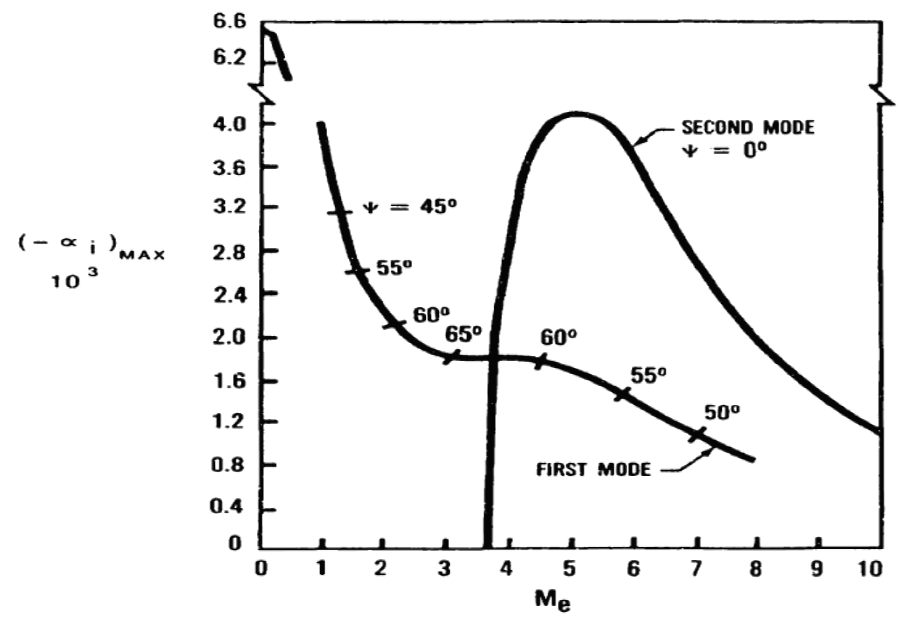}
    \includegraphics[width=0.46\linewidth]{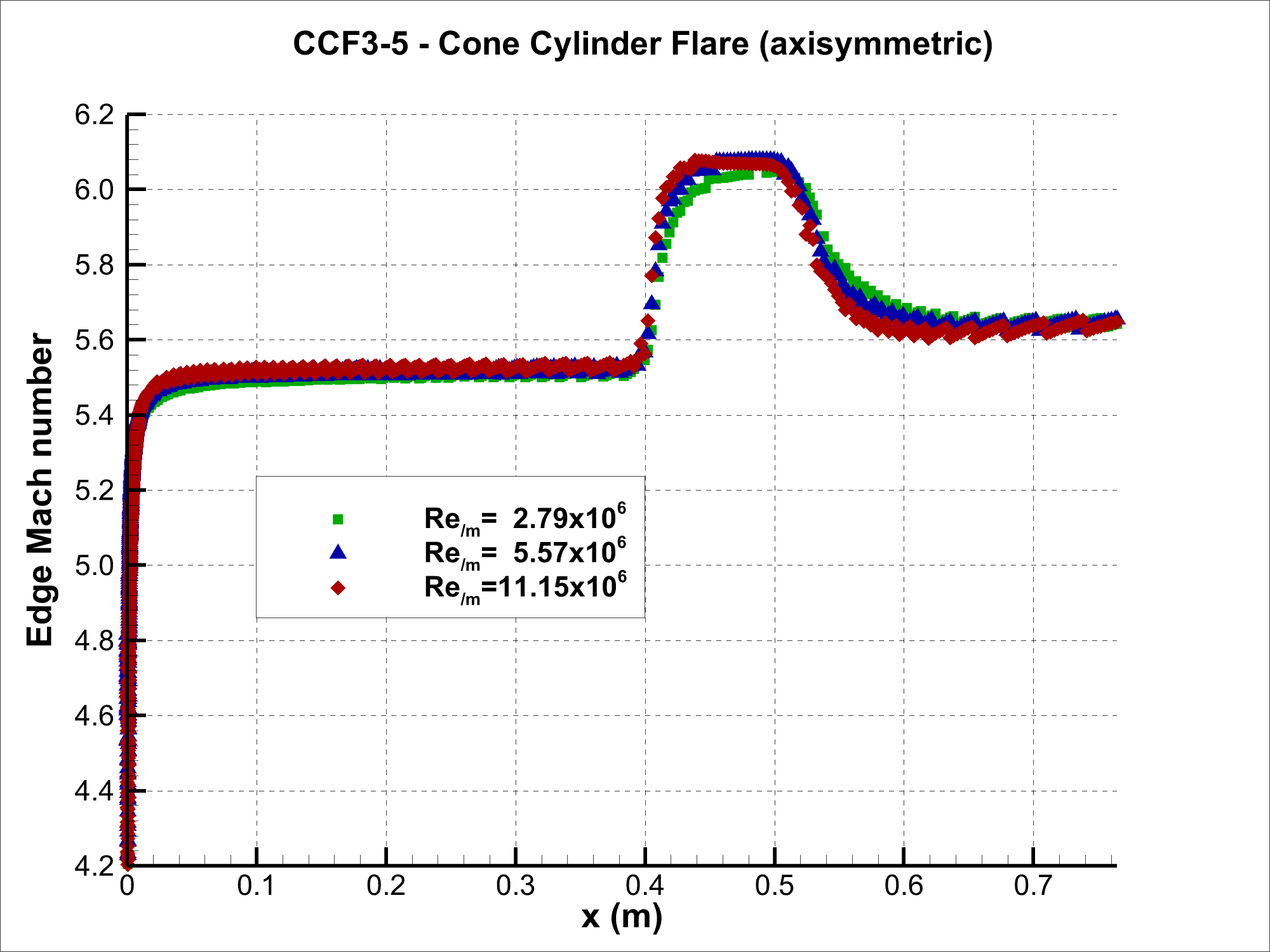}
	\caption{First-mode and second-mode spatial amplification rates (from \cite{stetson1990hypbl}) - Edge Mach number on CCF3-5 (illustration on the right)}
	\label{fig:CCF3-5edgeMach}
\end{figure}

%-----------
In hypersonic flows above an edge Mach number of 4.5, the second-mode waves are generally the most amplified on planar and conical geometries at 0 degree angle of attack.
These inviscid instabilities are most unstable as two-dimensional disturbances ($\psi$=0 °, see figure \ref{fig:CCF3-5edgeMach}). They are acoustic instabilities propagating inside the laminar boundary-layer, which behaves as an acoustic waveguide \cite{fedorov2011stab}. Knowing that the wavelength of the second-mode instability is about twice the boundary-layer thickness \cite{kendall1975WTexp}, the estimated second-mode frequency is calculated as:

\begin{equation}
  f_{\,2nd\,mode} \simeq \frac{V_e}{2\delta}
  \label{Eq.2ndMode}
\end{equation}

The second mode lies below the sonic line near the wall. Consequently, the associated pressure fluctuations are generated in the lower part of the boundary-layer which is very well suited for pressure measurements at the wall.

%----------
\subsubsection{Cross-flow instabilities}
%----------
These disturbances result from an inflectional instability of the cross-flow velocity profile that is observed for configurations at angle of attack or for 3D shapes \cite{moyes2017jfm}. They have rapid growth rates and may be the dominant disturbances in three-dimensional flows. They can have both traveling or stationary forms.

%----------
\subsubsection{Görtler vortices}
%----------
Görtler vortices result from centrifugal instabilities due to the concave streamline curvature (concave walls or other specific shapes) which produces counter-rotating streamwise vortices \cite{saric94gortler}. Görtler vortices can dominate the transition process under some conditions.\\

Local stability theory provides a valuable tool to study parametric effects and works well to describe the features of these mechanisms.

\subsection{Dominant instabilities on CCF3-5}

The edge Mach number detected on CCF3-5 with the total enthalpy criterion is shown in the right picture of figure \ref{fig:CCF3-5edgeMach}.
The value of $M_{e} \sim 5.5$ on the cone is in good agreement with the Taylor-Maccoll estimation of $M_{e} \sim 5.6$ for a 5-degree half-angle nearly-sharp cone at a freestream Mach number of 6.0 at moderate Reynolds numbers.
Downstream, the edge Mach number increases to around $M_{e} \sim 6.1$ on the cylinder due to the flow expansion and finally decreases to around $M_{e} \sim 5.65$ on the flare after the recompression at the cylinder flare junction. 

It is known that the second-mode instabilities grow more rapidly at high Mach number values and on cold walls. Here, the high edge Mach number and the limited wall temperature ($T_{wall} \sim 300K$) suggest that the second-mode waves will be the dominant instability as stated by Mack's theory \cite{mack1984BLStabl}.

Other instabilities such as G\"{o}rtler vortices could also be present in the concave part of the geometry but, for small flare angle on CCF3-5, these instabilities should not be predominant. The crossflow instabilities will not occur on this axisymmetric flow.\\

So, here, the dominant instability is the second-mode of Mack. The slowly growing first-mode will also be present in the boundary-layer but, at lower amplitude levels . This will be confirmed in the following sections.

%Remark: for larger flare angles with a separation bubble, G\"{o}rtler instabilities, streaks or even other instability mechanisms should be considered for the study of the transition process. 

%---------------
\section{Stability analyses: numerical study on CCF3-5}
%---------------
\label{sec.stab_analysis}
The stability analyses are realized thanks to two stability codes, the STABL software suite from the University of Minnesota \cite{johnson2005stabl} and the Mamout code from ONERA developed by Dr. Brazier.

%==========
\subsection{Stability analysis solvers}
%==========
%--------------------
\subsubsection{STABL}
%--------------------
The main part of the stability analyses are performed using the PSE-Chem solver, which is a part of the STABL software suite. PSE-Chem solves the reacting, two-dimensional, axisymmetric, linear parabolized stability equations (PSE) to predict the amplification of disturbances as they interact with the boundary-layer.
The PSE-Chem solver includes finite-rate chemistry and translational-vibrational energy exchange. Both linear stability theory (LST) and PSE analyses were performed using the PSE-Chem code. 

For the LST analysis, a parallel flow assumption is made by neglecting derivatives of mean flow quantities in the direction of the computational coordinate along  the body. The PSE is expected to be more accurate as it does not assume fully parallel flow. Spatial amplification rates are found for given disturbance frequencies and surface locations. Linear stability theory is used to provide the initial wavenumber for the PSE marching procedure \cite{robarge2005lambl}.

Generally, transition occurs when the N-factor (see detail in section \ref{sec.2ndmode_Nfactor}) reaches a prescribed value of N$_T$, estimated empirically and depending mainly on the free-stream turbulence level.

%------------------------
\subsubsection{Mamout}
%------------------------
Stability analyses have also been carried out on the same configuration with the ONERA in-house code Mamout. This code solves the local linear stability equations for an incompressible fluid, an ideal gas or a chemical equilibrium mixture, with the parallel flow assumption. The present computations use the perfect gas model. The resulting one-dimensional differential eigenvalue problem can be solved thanks to several numerical schemes, among which are the Chebyshev polynomial collocation method and high-order compact schemes. The computational grid can be split in multiple sub-domains. Outside of the boundary-layer, the fluctuation is matched to the analytical solution obtained for a uniform base flow.

\begin{figure}
	\centering
	\includegraphics[width=0.48\linewidth]{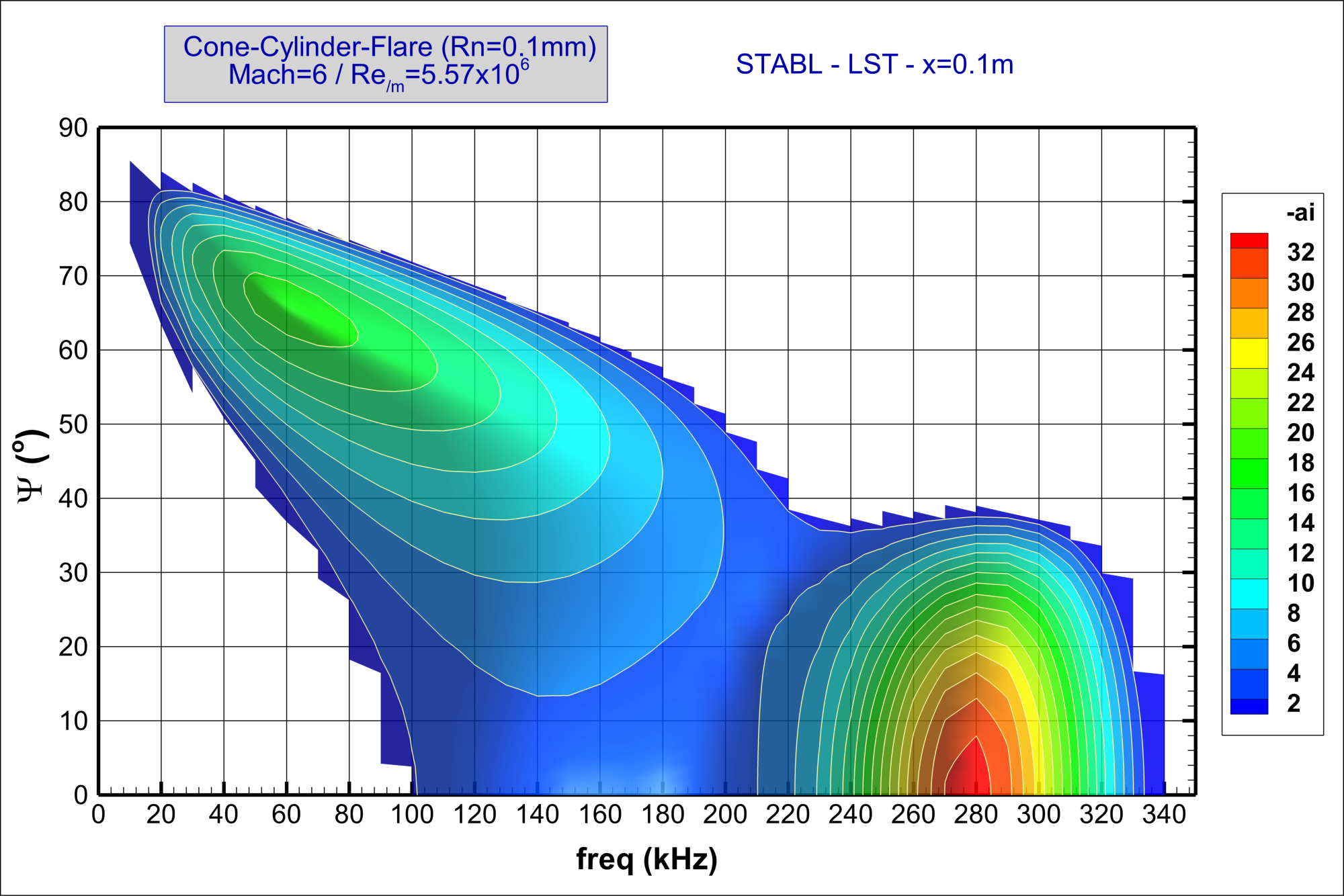}
	\includegraphics[width=0.48\linewidth]{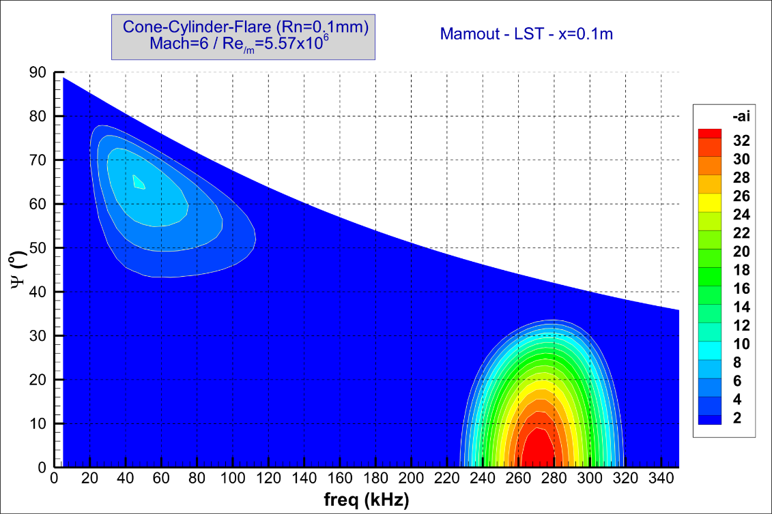}
	\caption{STABL / Mamout (CCF3-5) - Amplification rates and wave propagation angles on the cone at Re${/m}$ =5.6x10$^{6}$}
	\label{fig:max_ai_LST_STABL_Mamout_x0_1m}
\end{figure}

%------------------------------------------
\subsection{Stability analysis results}
%------------------------------------------
The same DPLR2D laminar flow solutions are used as input for STABL and Mamout stability analyses.

%---------------
\subsubsection{Stability analysis of first-mode and second-mode waves on CCF3-5}
\begin{figure}
	\centering
	\includegraphics[width=0.325\linewidth]{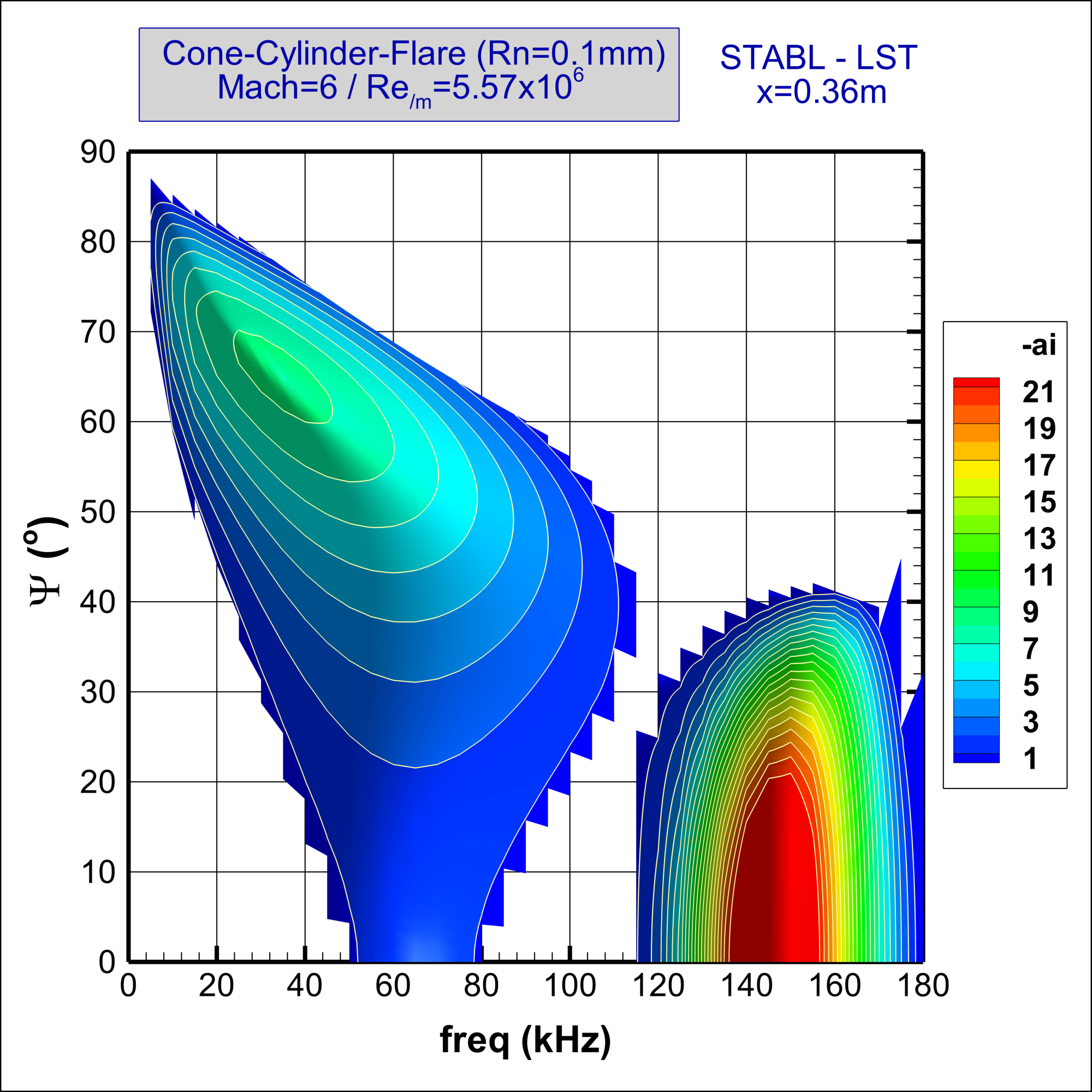}
	\includegraphics[width=0.325\linewidth]{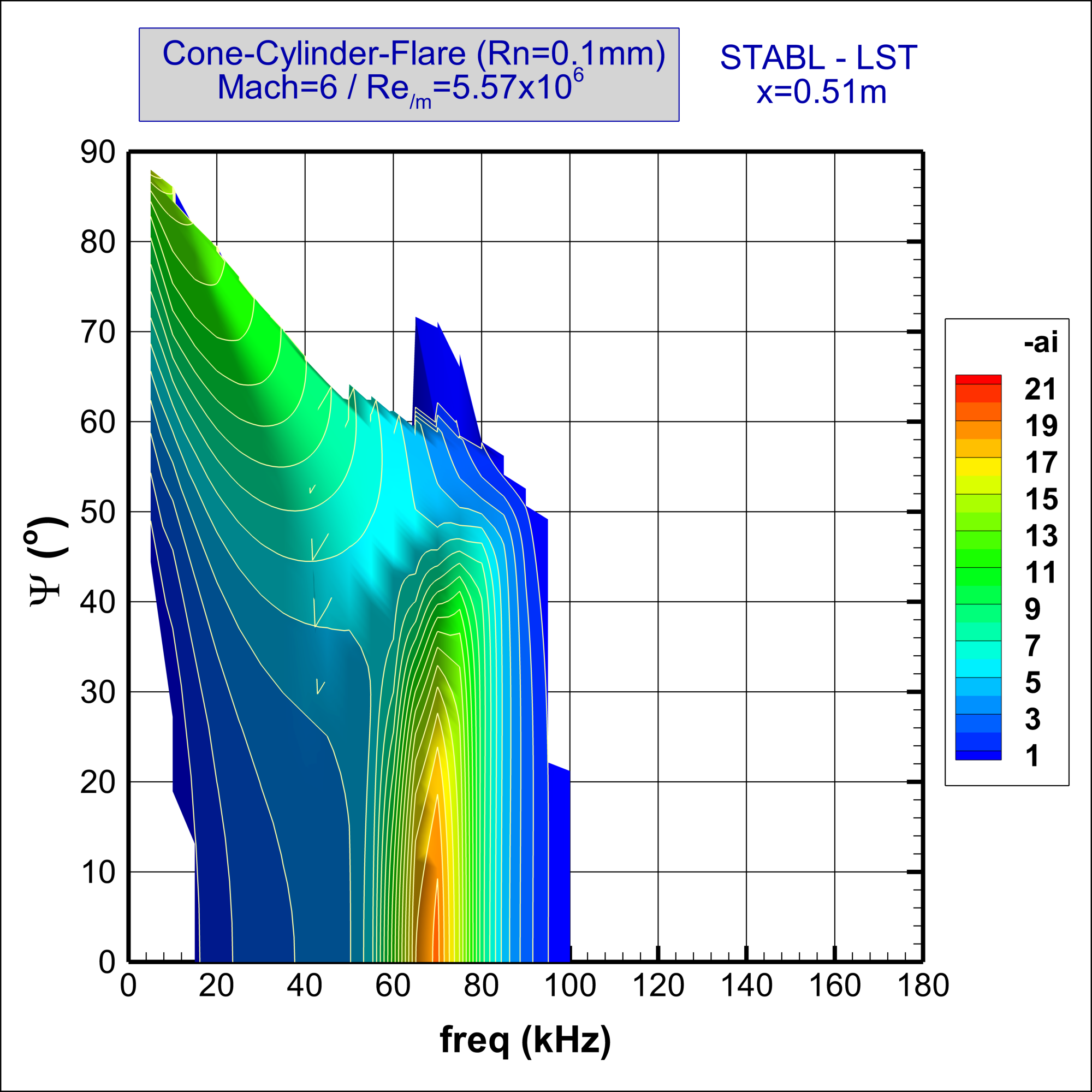}
	\includegraphics[width=0.325\linewidth]{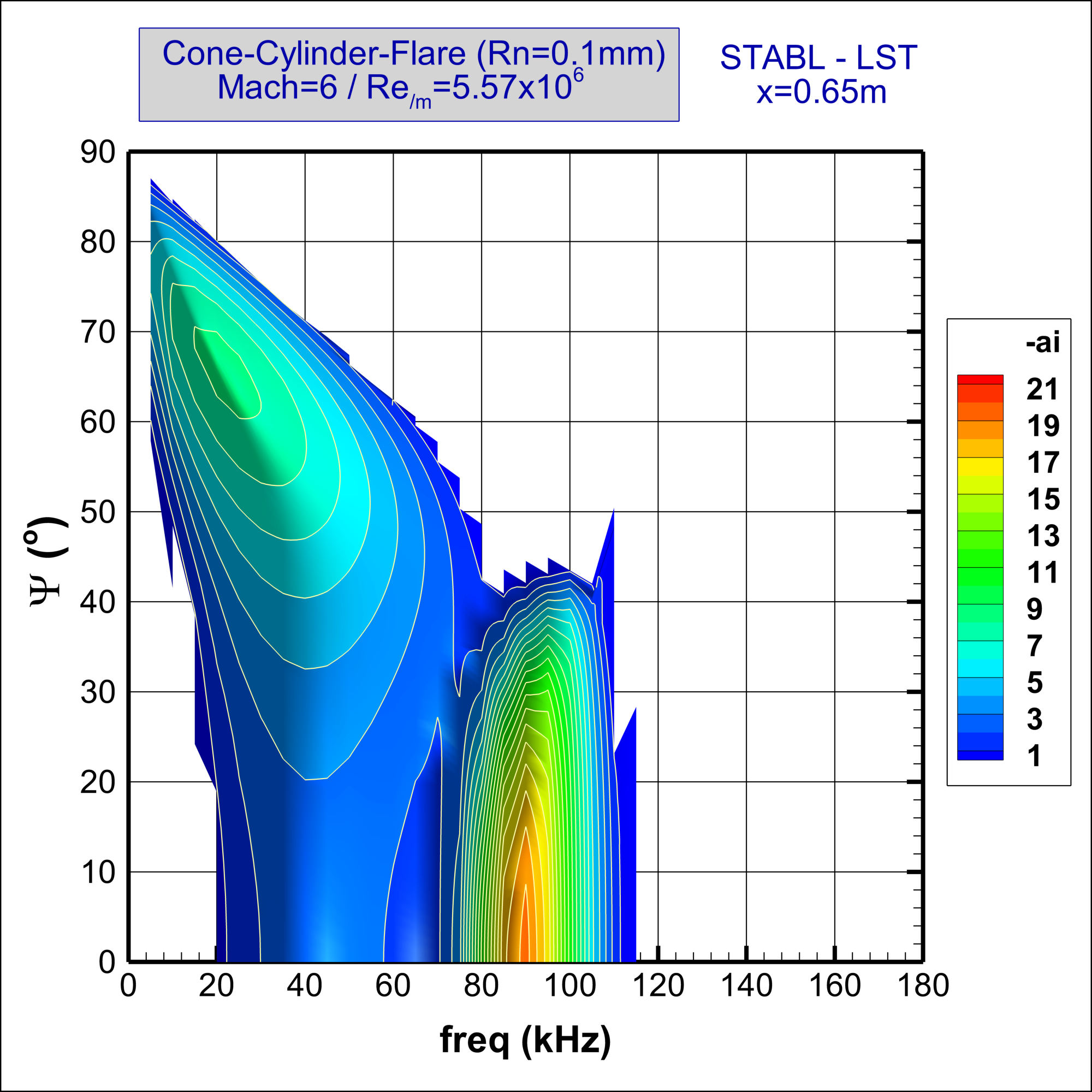}
	\caption{STABL (CCF3-5) - Amplification rates and waves propagation angles on the cone, on the cylinder and on the flare at Re${/m}$ =5.6x10$^{6}$}
	\label{fig:ai_psi_LST_STABL_3xloc}
\end{figure}

%----------------
The amplification rates and wave angles were extracted at four different locations, two on the cone, one on the cylinder and one on the flare (see figures \ref{fig:max_ai_LST_STABL_Mamout_x0_1m} and \ref{fig:ai_psi_LST_STABL_3xloc}).

\begin{figure}
	\centering
	\includegraphics[width=1.0\linewidth]{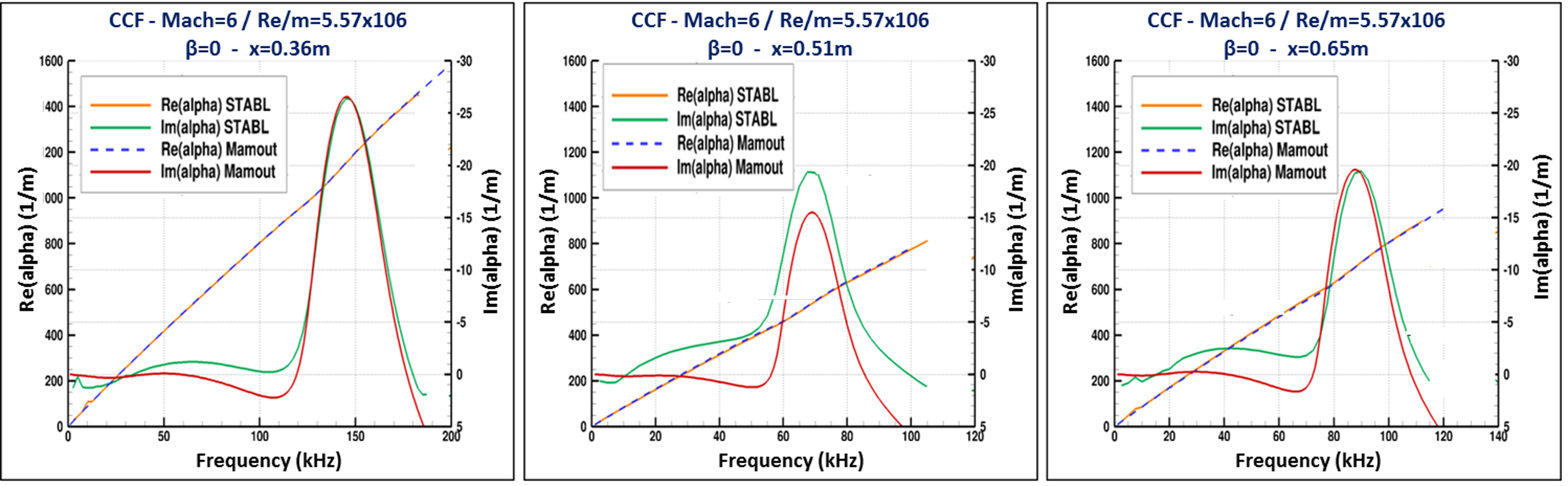}
	\caption{STABL / Mamout (CCF3-5) - Wavenumber and maximum amplification in LST mode on the cone, on the cylinder and on the flare at Re${/m}$ =5.6x10$^{6}$}
	\label{fig:max_ai_LST}
\end{figure}

As detailed by \cite{arnal1993BLTrans}, these stability diagrams indicate that first-mode and second-mode are simultaneously amplified in the boundary-layer. The amplification rates confirm the dominance of the higher frequency second-mode for two-dimensional waves ($\psi=0^{o}$) whatever the considered location on the configuration. The first-mode instabilities are more amplified as oblique waves with sensible levels obtained for wave propagation angles between 50$^{o}$ and 80$^{o}$.

The extracts from STABL and Mamout in figure \ref{fig:max_ai_LST_STABL_Mamout_x0_1m} indicate the amplification rates of the waves on the cone at x=0.1m. The first-mode is visible for oblique waves with a maximum amplification detected for frequencies around 40 kHz and wave angles between $65^{o}$ and $70^{o}$. The second-mode is, as expected, clearly the dominant mode with a maximum amplification rate for the two-dimensional waves ($\psi \sim 0^{o}$) at frequencies around 260-280 kHz.

Three other locations are represented in figure \ref{fig:ai_psi_LST_STABL_3xloc} at the end of the cone, on the cylinder and on the flare. The most amplified frequencies are very different on each section due primarily to changes in the boundary-layer thickness. The frequencies of the amplified waves are narrower on the last two sections of the configuration but the second-mode is still the dominant instability.

A quantitative comparison between STABL and Mamout for the streamwise wavenumber and the maximum amplification rate prediction is provided in figure \ref{fig:max_ai_LST}. The predicted wavenumbers (real part of the streamwise wavenumber) are nearly exactly the same with both codes. For the amplification rates (imaginary part), the agreement between the two codes is very similar on the cone and on the flare. On the cylinder, the most amplified frequencies correspond very well but the waves are more amplified with STABL (about 15 percent higher). It seems that the damping influence of the pressure gradient is felt differently by the two codes. A slight difference can also be observed in the low frequency content corresponding to the small amplification levels of the two-dimensional first-mode waves.

As stated before, the second-mode is the dominant instability on CCF3-5.

%---------------
\subsubsection{Second-mode instabilities: amplification rates all along the geometry}
%----------------
The second-mode disturbances are highly "tuned" to the boundary-layer thickness resulting in considerable selectivity in the disturbance frequencies which are most amplified \cite{stetson1990hypbl}. Seeing the boundary-layer thickness evolution shown previously in figure \ref{fig:CCF3-5_10delta}, it can be stated that the thin boundary-layer on the cone will generate higher frequency disturbances than the thicker boundary-layer on the cylinder and on the flare. The question here is the following: what will be the impact of the pressure gradients (expansion and recompression) on the boundary-layer stability?

\begin{figure}
	\centering
%	\begin{subfigure}[b]{0.49\linewidth}
	\includegraphics[width=0.48\linewidth]{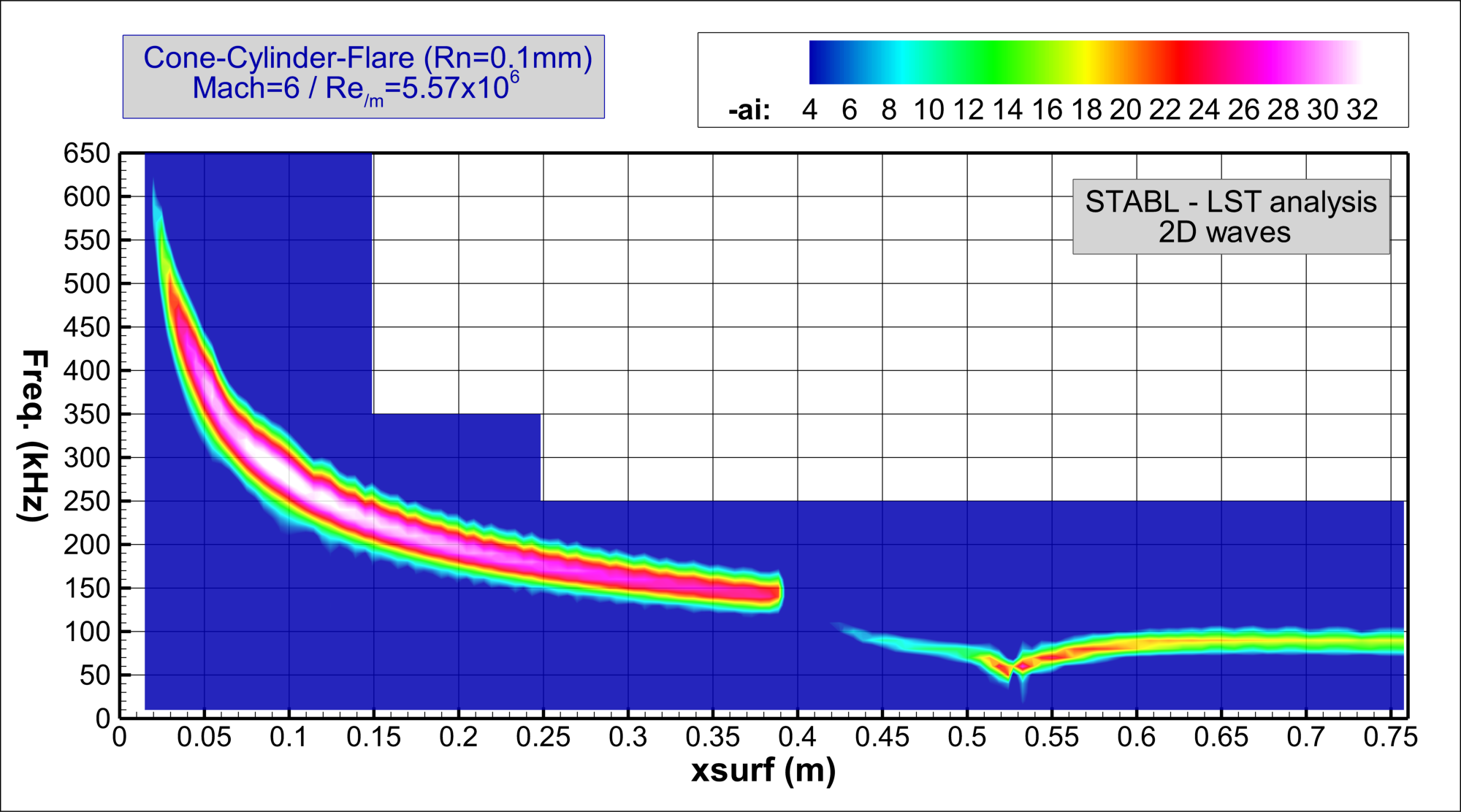}
%	\caption{STABL at Re${/m}$ = 5.6x10$^{6}$}
%	\end{subfigure}
%	\begin{subfigure}[b]{0.49\linewidth}
	\includegraphics[width=0.48\linewidth]{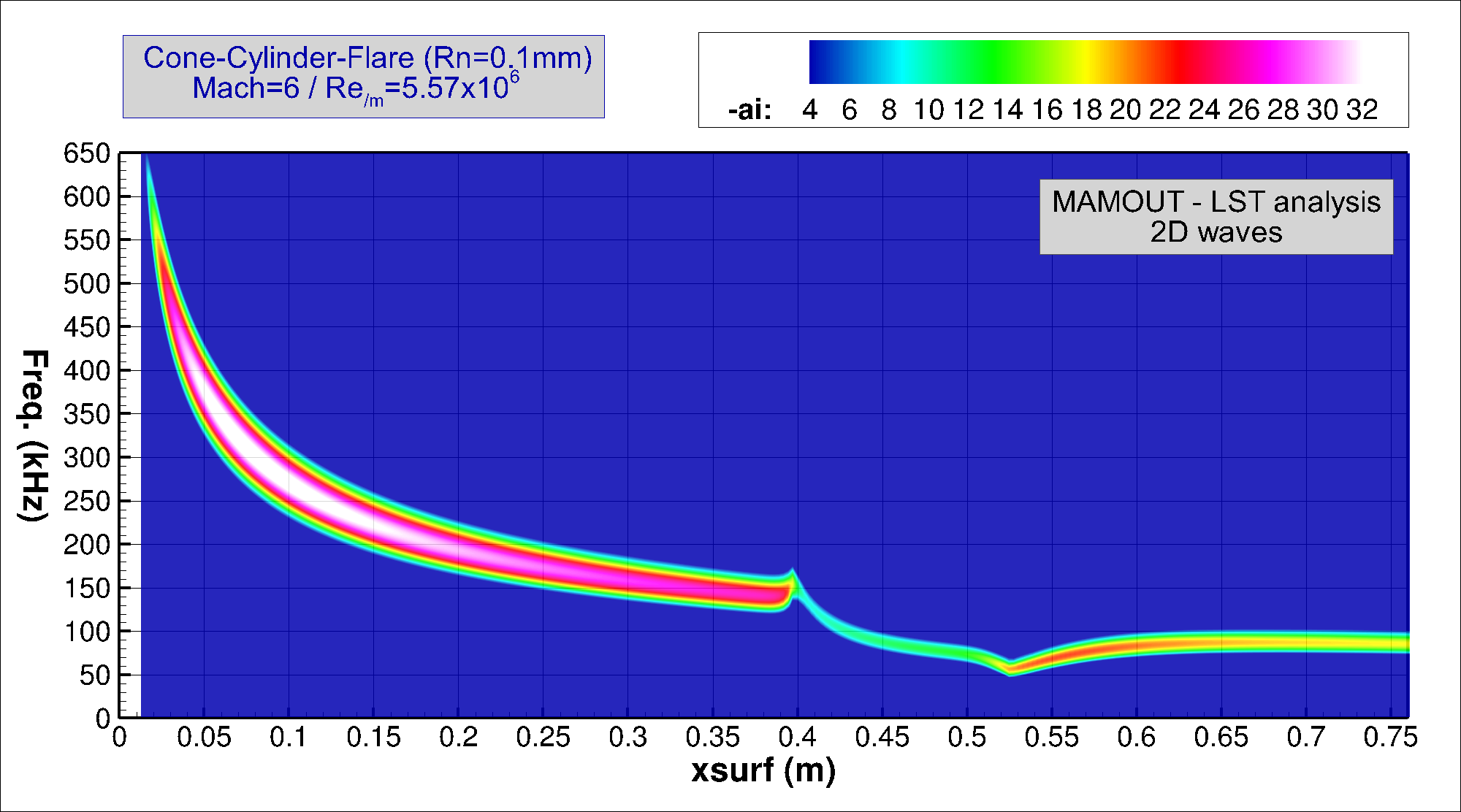}
%	\caption{Mamout at Re${/m}$ = 5.6x10$^{6}$}
%	\end{subfigure}
	\caption{STABL / Mamout (CCF3-5) - Amplification rates from LST (Re${/m}$ = 5.6x10$^{6}$)}
	\label{fig:amplif1}
\end{figure}

\begin{figure}
	\centering
%	\begin{subfigure}[b]{0.49\linewidth}
	\includegraphics[width=0.48\linewidth]{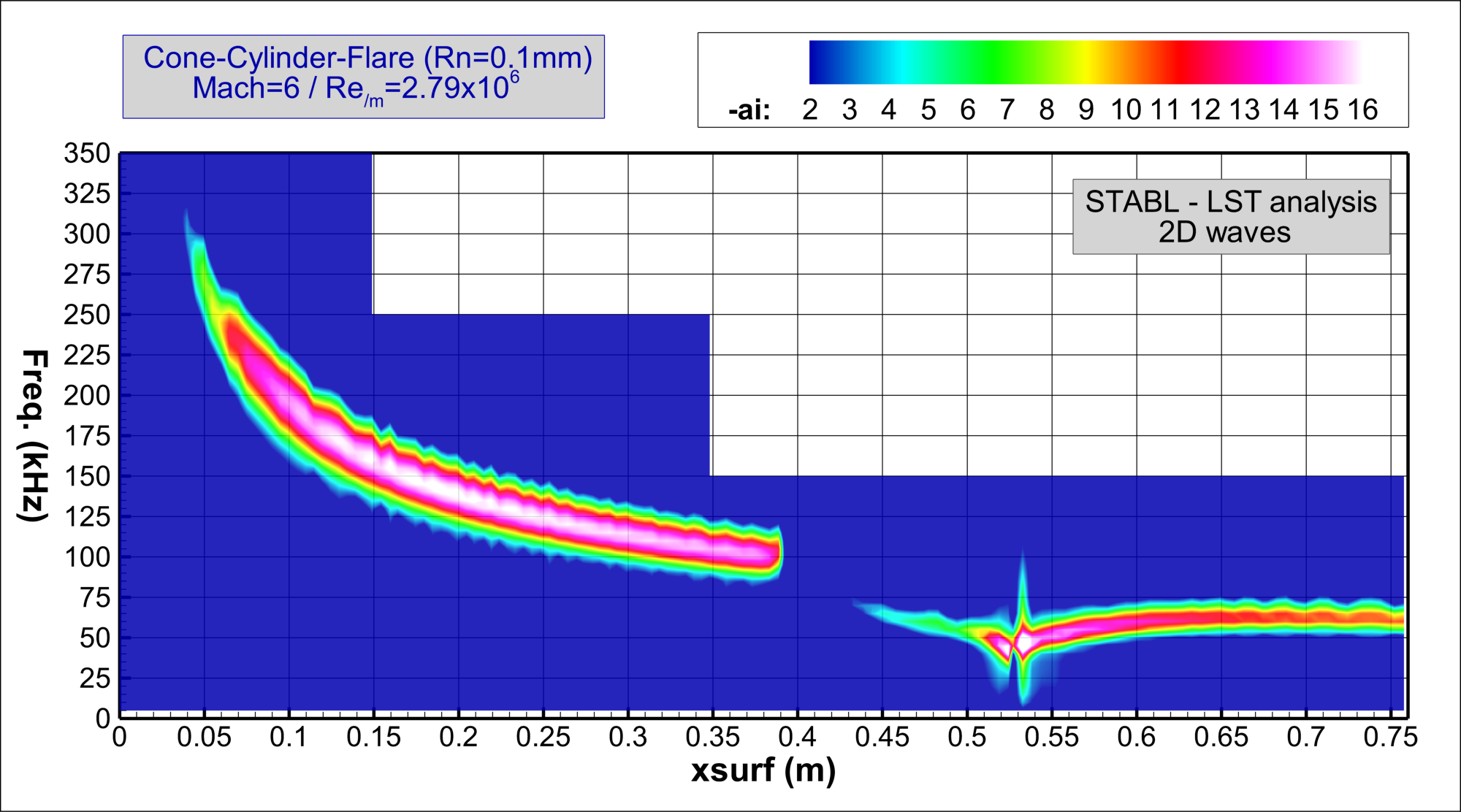}
%	\caption{STABL at Re${/m}$ = 2.8x10$^{6}$}
%	\end{subfigure}
%	\begin{subfigure}[b]{0.49\linewidth}
	\includegraphics[width=0.48\linewidth]{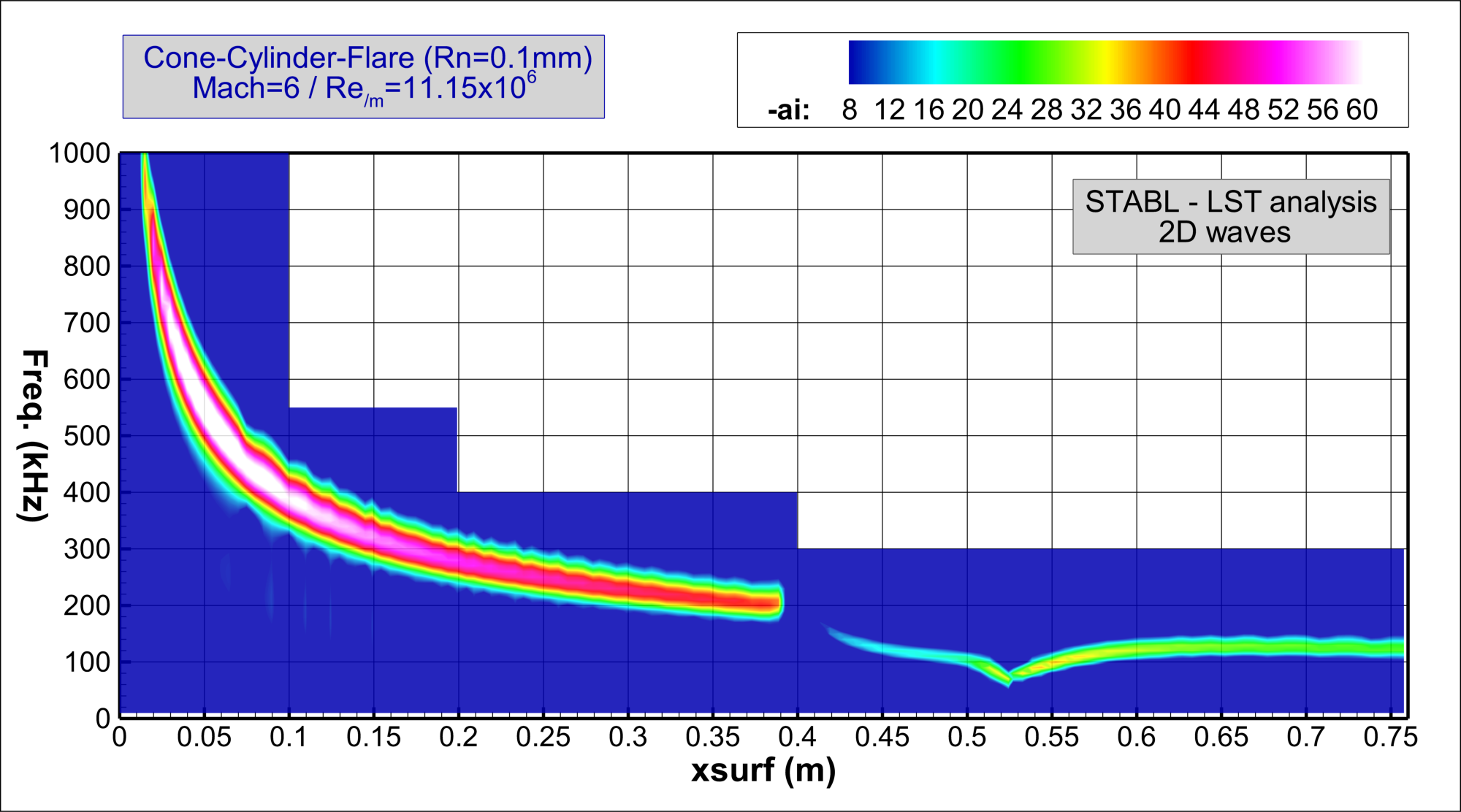}
%	\caption{STABL at Re${/m}$ = 11.2x10$^{6}$}
%	\end{subfigure}
	\caption{STABL (CCF3-5) - Amplification rates from LST (Re${/m}$ = 2.8x10$^{6}$ and 11.2x10$^{6}$)}
	\label{fig:amplif2}
\end{figure}

All the stability diagrams presented in figures \ref{fig:amplif1} and \ref{fig:amplif2} for different Reynolds numbers show the standard "thumb" curve on the cone. The higher frequencies at the beginning of the cone detune rapidly in the narrow amplified band but, as the disturbance waves proceed more downstream, become better tuned to the boundary-layer thickness and amplify at higher rates on longer periods of growth. As a consequence, the resulting amplified boundary-layer instabilities can reach high amplitude on the cone, potentially leading to the critical breakdown amplitude for moderate to high Reynolds numbers.

Concerning the cylinder part, it is known that a favorable pressure gradient has a stabilizing effect on the boundary-layer. This is definitively confirmed by the stability diagrams where no substantial amplification rates are visible on the main part of the cylinder. As shown in figure \ref{fig:CCF3-5_10delta}, the boundary-layer thickens very rapidly under the influence of the flow expansion and experiences a very rapid change all along the cylinder. The accelerated boundary-layer is very stable as confirmed by the amplification rate diagrams. The second-mode waves, initially amplified on the conical part, rapidly detune and lower frequency disturbances are only slightly amplified along the cylinder part. 

The phenomenon is totally different on the flare. It is interesting to remember here that adverse pressure gradients are commonly known as destabilizing for the boundary-layer. The boundary-layer thins rapidly just after the cylinder-flare junction but, once the pressure gradient is past, the boundary-layer keeps a relatively constant thickness and quite similar velocity profiles all along the flare (see figure \ref{fig:BLprofiles_Flare}). As a consequence, it can be guessed that, after an important destabilization at the cylinder-flare junction, a quite amplified frequency plateau could be observed on the flare. It is confirmed by figures \ref{fig:amplif1} and \ref{fig:amplif2} for all the Reynolds numbers considered here. The amplified waves stay in a narrow frequency range on the flare which can potentially generate high amplitude disturbances at the end of the flare.

All these general trends are confirmed for all the Reynolds numbers and also with the code-to-code comparison, STABL / Mamout, which points out a good agreement between these two numerical approaches (see figure \ref{fig:amplif1}). This verification gives an even larger confidence in the numerical stability predictions.

The Reynolds number effect on the amplification of the second-mode frequency waves is clearly highlighted by figure \ref{fig:amplif2}. The large increase of the boundary-layer thickness in the case of the lower Reynolds number leads to low frequency range for the second mode amplification: 100-300 kHz on the cone and a narrow band around 50-60 kHz on the flare. The higher Reynolds number considered leads logically to high frequency content: 200 to 1000 kHz on the cone and a long flat frequency range on the flare around 120-130 kHz with high amplification rate levels.

%---------------
\subsubsection{Second-mode instabilities: trapped acoustic waves}
%----------------
As explained by \cite{fedorov2011stab} and \cite{knisely2017soundrad}, the second-mode disturbances belong to the family of trapped acoustic waves. A region of supersonic mean flow relative to the disturbance phase velocity is present in the upper part of the boundary-layer while, in the lower part of the boundary-layer, the disturbances are supersonic relative to the mean flow.

As a consequence, this small region near the wall acts as an acoustic waveguide where the waves are trapped. Acoustic rays are reflected by the wall and cannot cross the sonic line. 
The second-mode waves developing in this near-wall region represent inviscid instabilities of acoustic nature which become the dominant instability in hypersonic flow. This supersonic disturbance region can be represented by plotting the relative Mach number defined as:

\begin{equation}
  M_{r} = \frac{\bar{u}-c_{r}}{\bar{a}}
  \label{Eq.Mr}
\end{equation}

\noindent where $\bar{u}$ is the mean flow velocity, $c_{r}$ is the disturbance propagation speed, and $\bar{a}$ is the mean flow speed of sound.

\begin{figure}
	\centering
%	\begin{subfigure}[b]{0.50\linewidth}
		\includegraphics[width=0.52\linewidth]{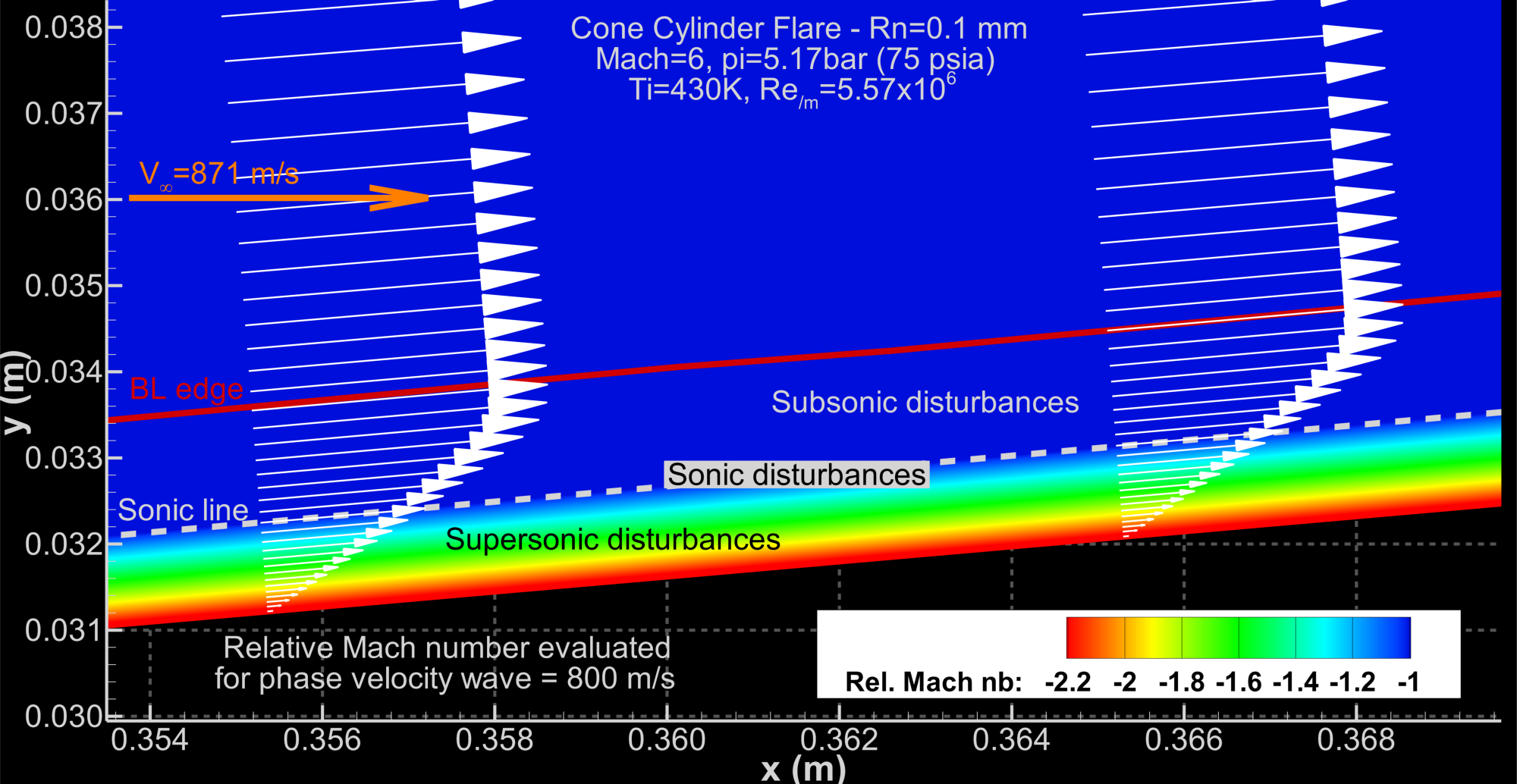}
%		\caption{Relative Mach number}
%		\label{fig:relatM1}
%	\end{subfigure}
%	\centering
%	\begin{subfigure}[b]{0.32\linewidth}
		\includegraphics[width=0.35\linewidth]{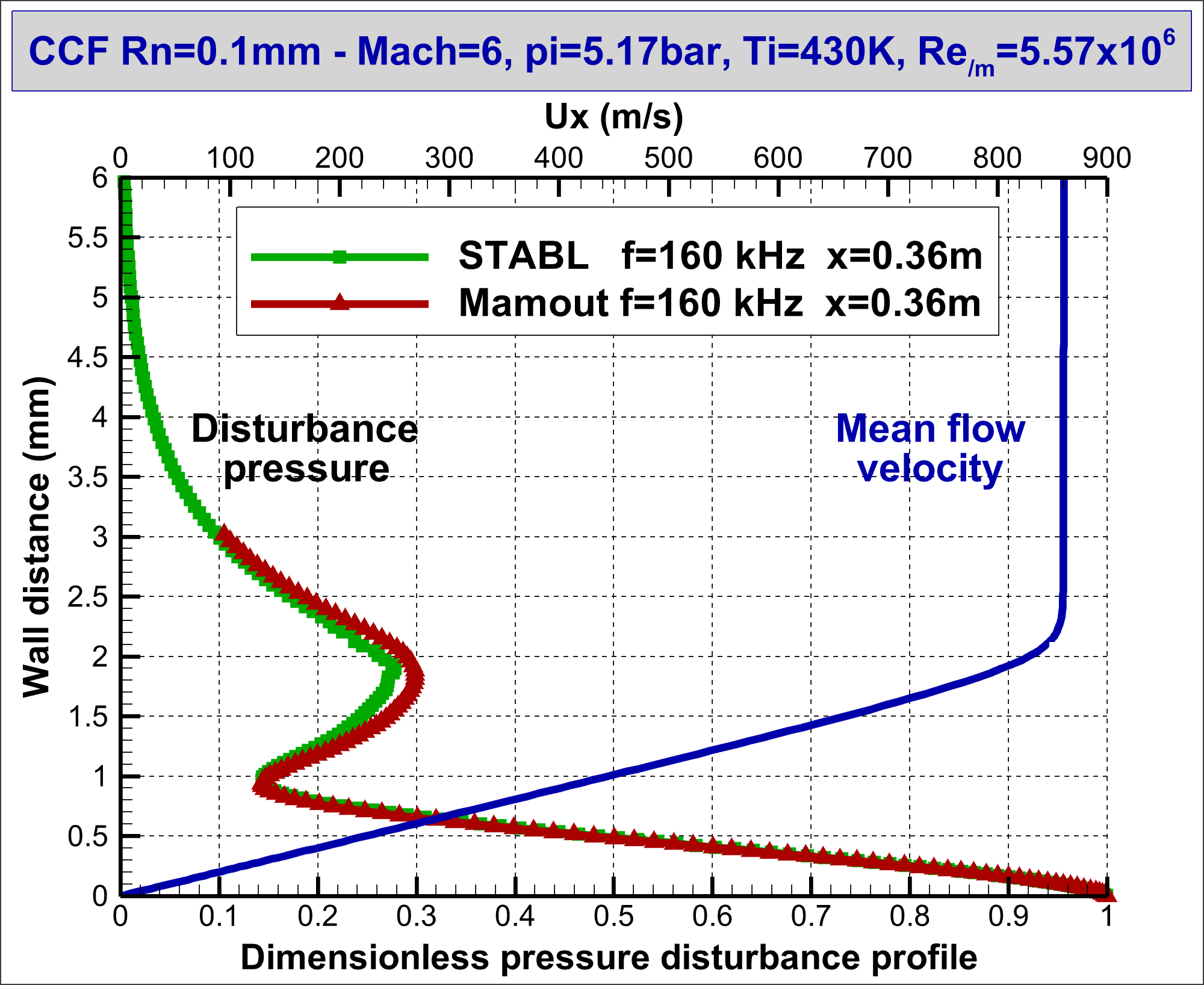}
%		\caption{Pressure disturbance profile}	
%		\label{fig:pdisturb1}
%	\end{subfigure}
	\caption{STABL (CCF3-5) - Relative Mach number and pressure disturbance profile at the end of the cone (Re${/m}$ = 5.6x10$^{6}$)}
           \label{fig:relatM.pdisturb1}
\end{figure}

\begin{figure}
	\centering
%	\begin{subfigure}[b]{0.50\linewidth}
		\includegraphics[width=0.52\linewidth]{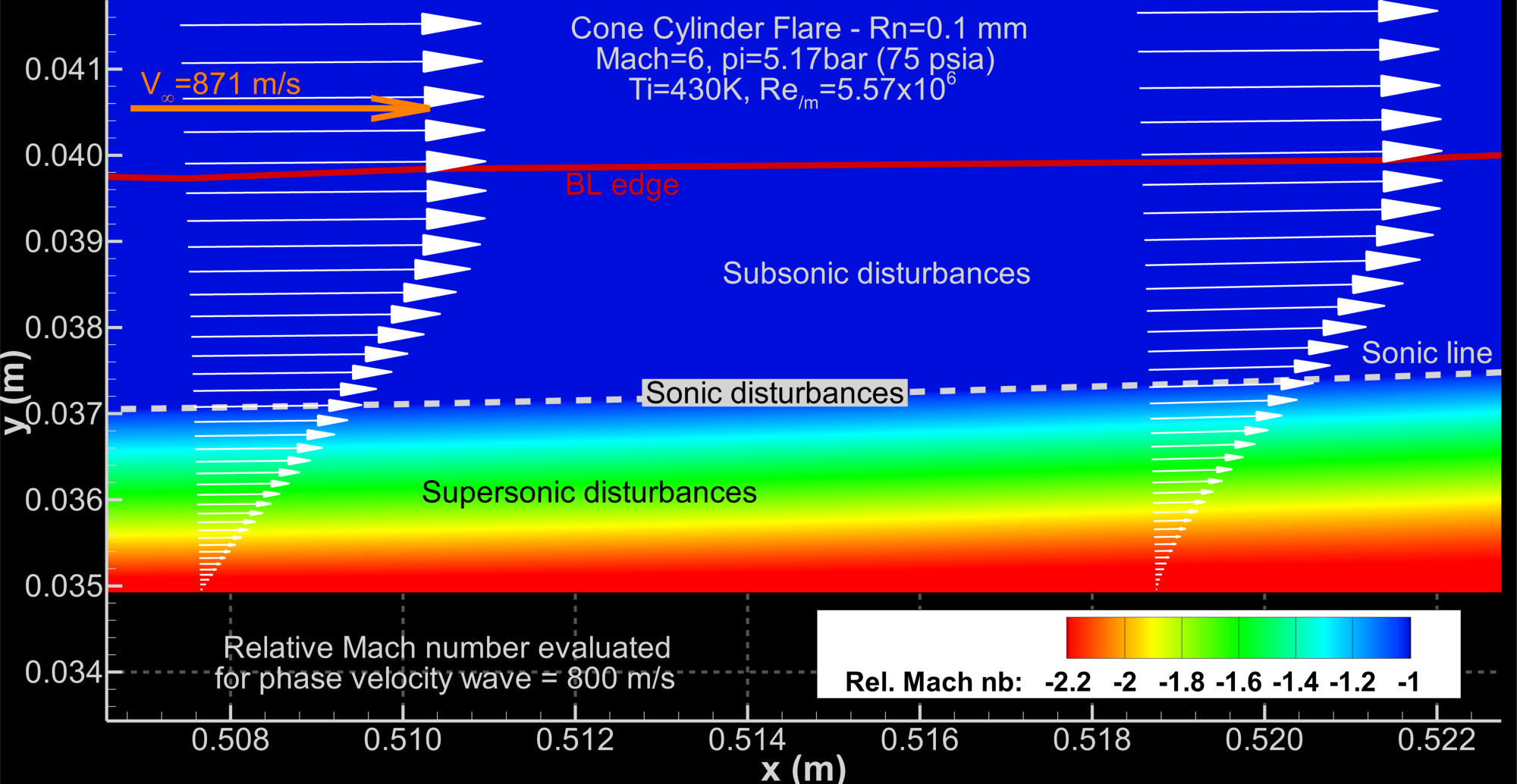}
%		\caption{Relative Mach number}
%		\label{fig:relatM2}
%	\end{subfigure}
%	\centering
%	\begin{subfigure}[b]{0.32\linewidth}
		\includegraphics[width=0.35\linewidth]{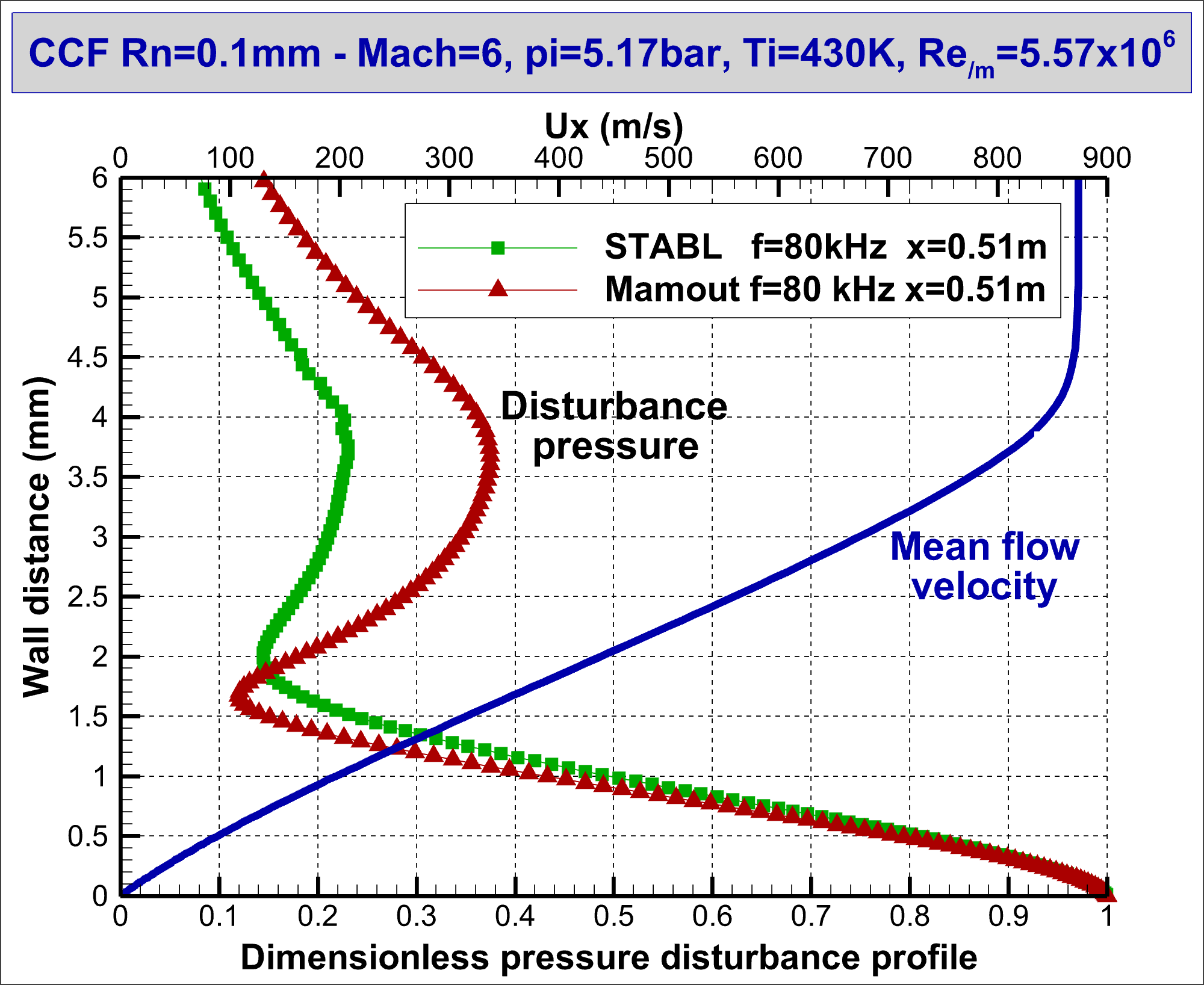}
%		\caption{Pressure disturbance profile}	
%		\label{fig:pdisturb2}
%	\end{subfigure}
	\caption{STABL (CCF3-5) - Relative Mach number and pressure disturbance profile at the end of the cylinder (Re${/m}$ = 5.6x10$^{6}$)}
	\label{fig:relatM.pdisturb2}
\end{figure}

\begin{figure}
	\centering
%	\begin{subfigure}[b]{0.50\linewidth}
		\includegraphics[width=0.52\linewidth]{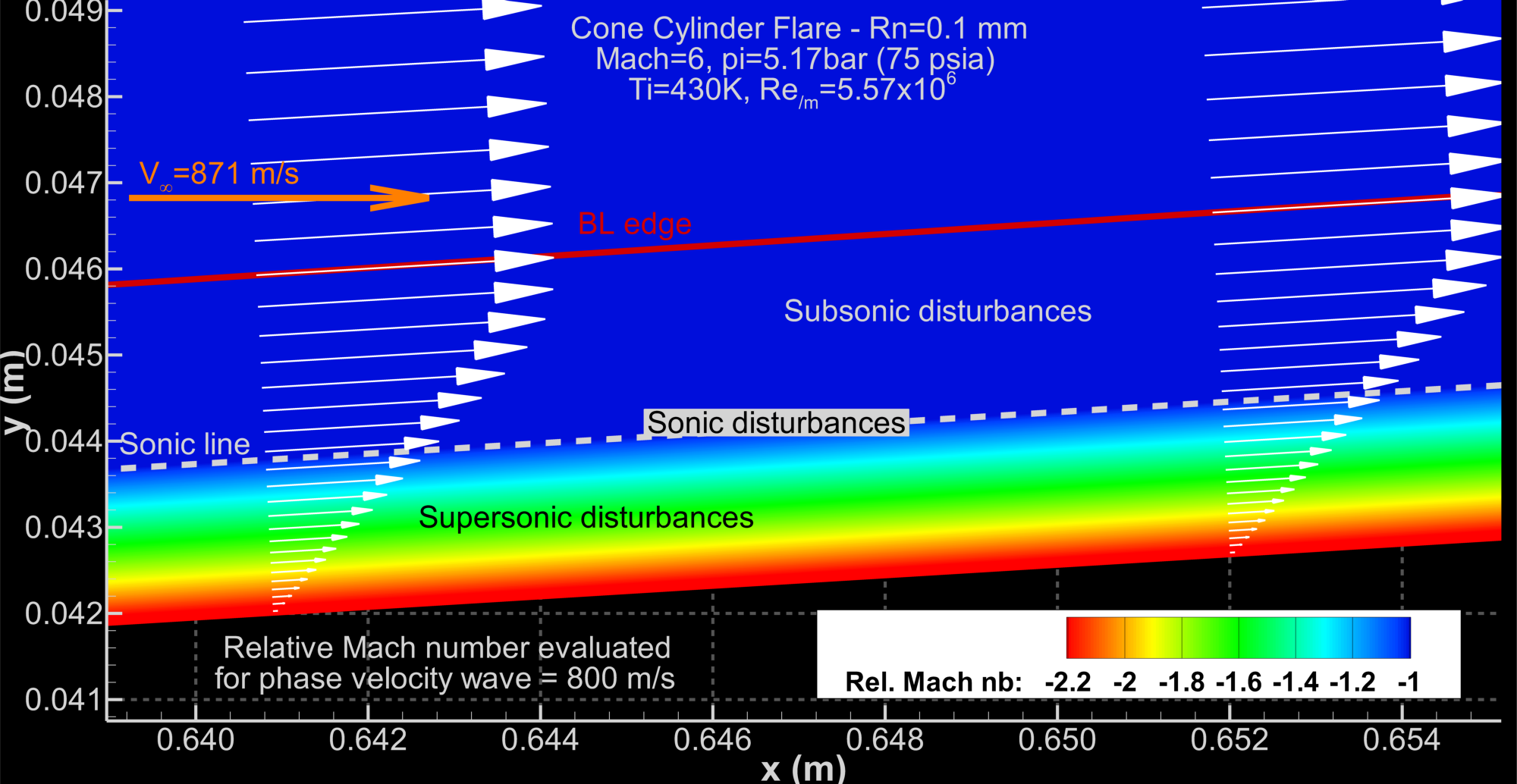}
%		\caption{Relative Mach number}
%		\label{fig:relatM3}
%	\end{subfigure}
%	\centering
%	\begin{subfigure}[b]{0.32\linewidth}
		\includegraphics[width=0.35\linewidth]{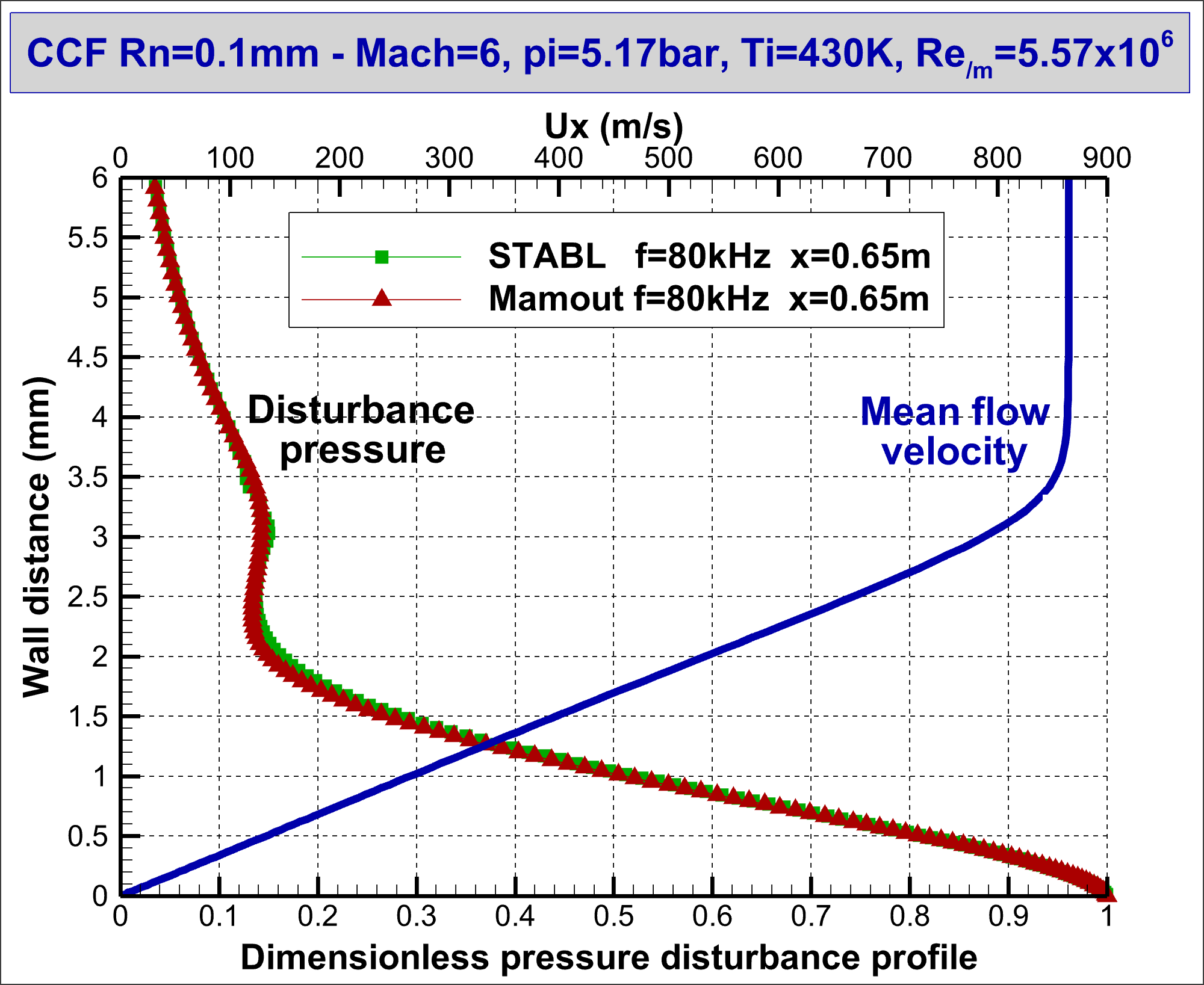}
%		\caption{Pressure disturbance profile}	
%		\label{fig:pdisturb3}
%	\end{subfigure}
	\caption{STABL (CCF3-5) - Relative Mach number and pressure disturbance profile in the middle of the flare (Re${/m}$ = 5.6x10$^{6}$)}
	\label{fig:relatM.pdisturb3}
\end{figure}

The region where the waves travel downstream supersonically relative to the mean flow is illustrated in figures \ref{fig:relatM.pdisturb1}, \ref{fig:relatM.pdisturb2} and \ref{fig:relatM.pdisturb3} respectively at the end of the cone, in the middle of the cylinder and at the end of the flare for a given phase velocity wave equal to 800 m/s ($\sim 0.92 V_{\infty}$). The supersonic region is very thin. The associated pressure disturbance has a maximum at the wall and decreases very rapidly in this supersonic region.
The disturbance profiles obtained with STABL and Mamout confirm that the main part of the second-mode lies below the sonic line (as also shown by \cite{chen2017modes}). %while the trace of the first-mode (2D waves only here) is visible near the boundary-layer edge close to the generalized inflection point. 

The high speed flow regime considered here leads logically to the dominance of the compressibility effects when compared to the inflectional effects. As a consequence, the second-mode is more unstable than the first-mode even if we consider the most-amplified oblique first-mode waves at $55-60^{o}$ (see figure \ref{fig:ai_psi_LST_STABL_3xloc}). The second-mode pressure disturbance peak is located at the wall.
This is a real asset to measure the amplitude of the second-mode waves by wall pressure measurements.

%---------------
\subsubsection{Second-mode instabilities: N-factors}
%----------------
\label{sec.2ndmode_Nfactor}

Knowing the evolution of an initial disturbance as it moves downstream from its starting point through the mean flow-field, the onset of transition can be estimated with the semi-empirical $e^N$ correlation method. 
Conveniently, this method is based on the linear amplification of a given instability wave by a factor of $e^{N}$ from the beginning of instability to the observed or predicted transition location \cite{redd96,schneider2008BAM6QT}.
In PSE-Chem, the N factor is defined as:

\begin{equation}
  N = \int_{s_{0}}^{s} [-\alpha_i + \frac{1}{2E}\frac{dE}{ds}] ds 
  \label{Eq.N}
\end{equation}

\noindent where not only the imaginary component of the streamwise wavenumber is taken into account (as in LST) but also the change in the kinetic energy of the shape function.

With this integrated growth of the linear instability waves, it is possible to correlate the onset of transition using N-factor values of about:
\begin{itemize}
    \item $N\sim3-5$ for conventional tunnel.
    \item $N>10$ for low noise environment. Some N-factor values at transition can be found in the literature : see \cite{schneider2015mechanisms} where computed N-factors for instability waves on a ﬂared cone show a delay of transition to N between 17 and 20 in quiet wind tunnel and also, \cite{li2015transit} for HIFiRE-1 in ascent where N factors at transition were about 13.
\end{itemize}

%\begin{table}
%\begin{center}
%\def~{\hphantom{0}}
%  \begin{tabular}{lccc}
%	   Environment &  Conventional &   Quiet   &   Flight  \\
%		  N-factor  &   4-5         &    8-10   &   12-14   \\
%	\end{tabular}
%	\caption{N-factor order of magnitude as a function of the envrionment}
%	\label{tab:coor} 
%	\end{center}
%\end{table}

\begin{figure}
	\includegraphics[width=0.48\linewidth]{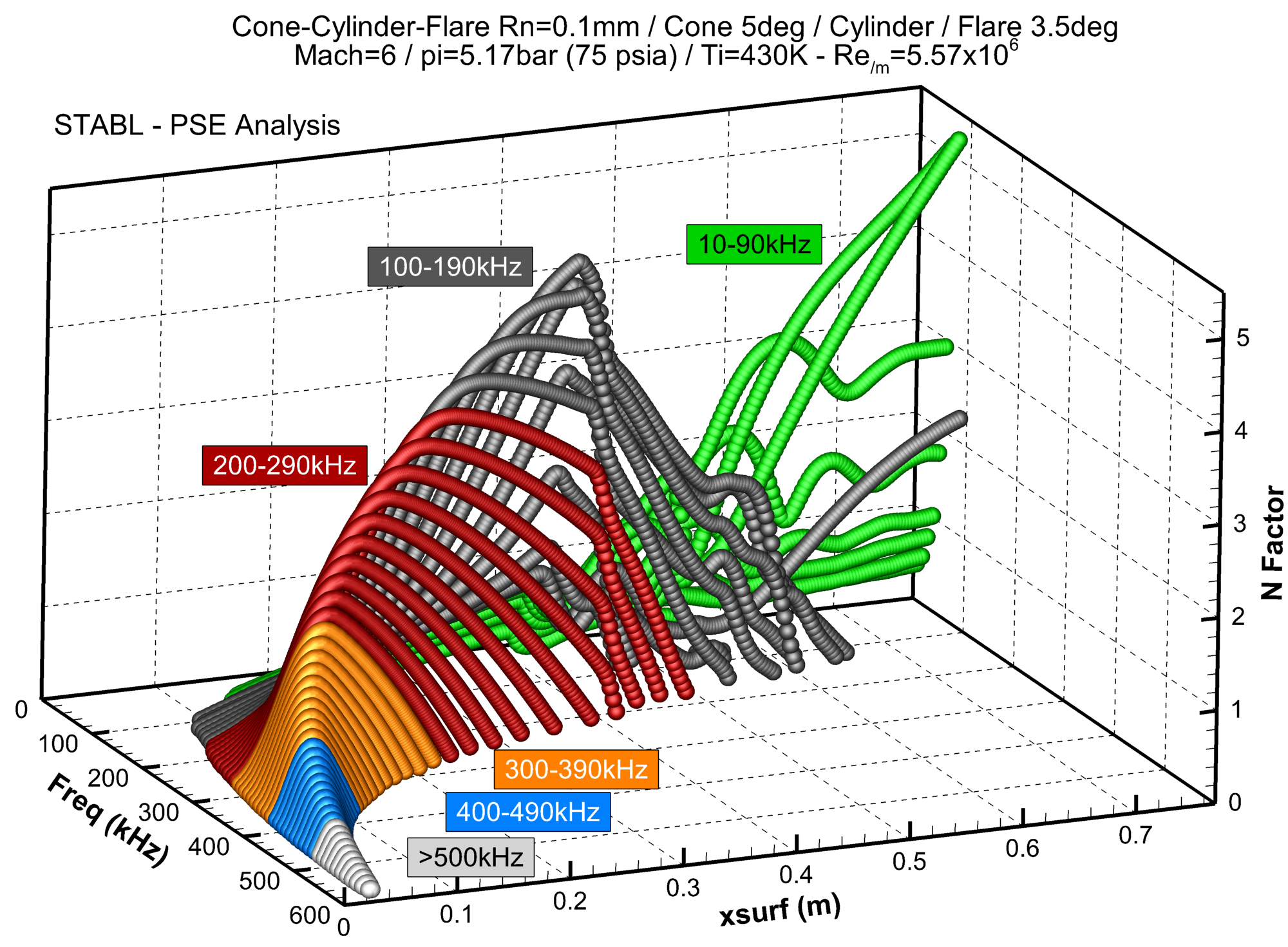}
	\includegraphics[width=0.48\linewidth]{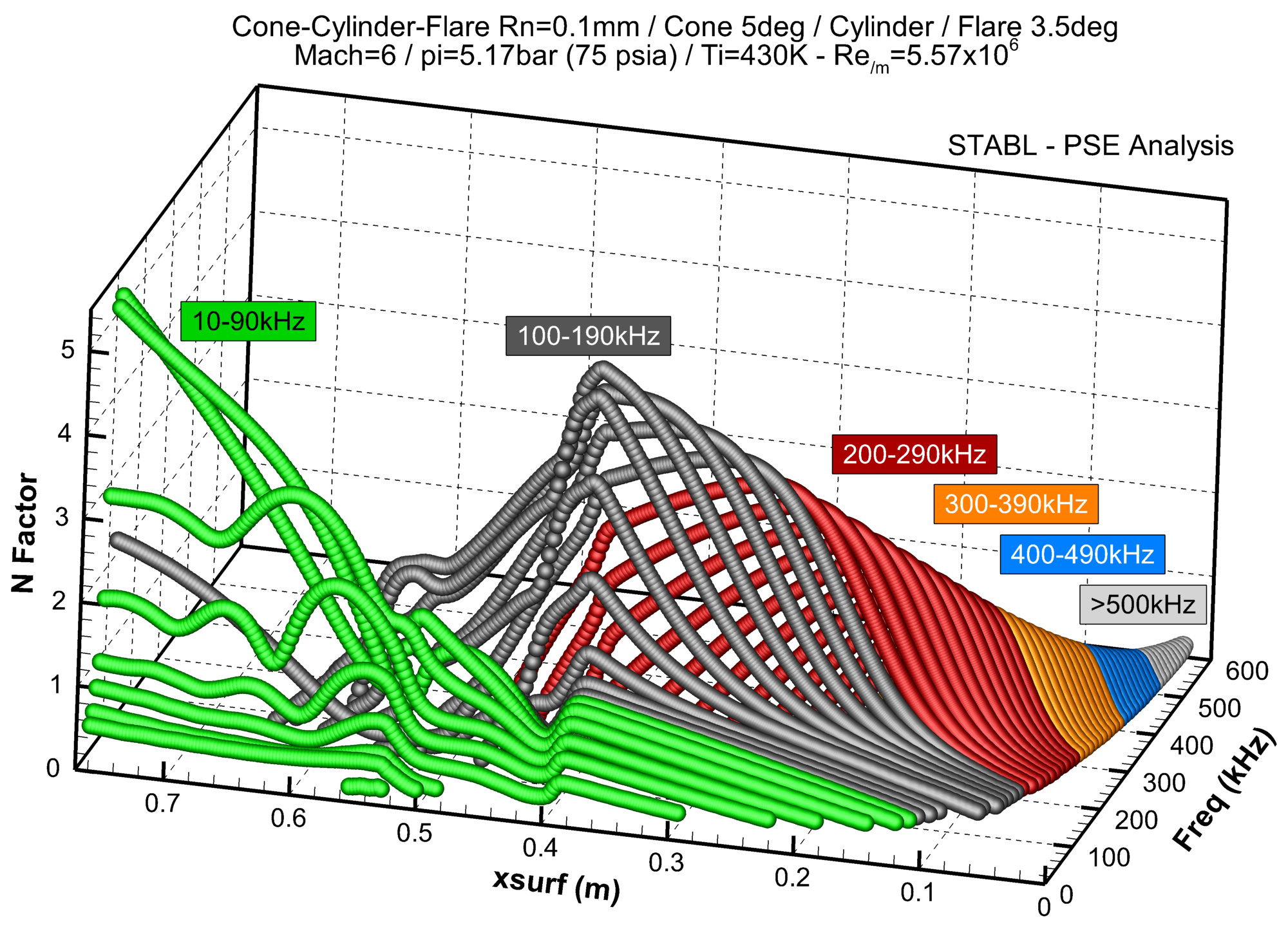}
	\caption{STABL (CCF3-5) - N factor evolution along the object at Re${/m}$ = 5.6x10$^{6}$}
	\label{fig:Nfactor_3D}
\end{figure}

\begin{figure}
	\centering
	\includegraphics[width=\linewidth]{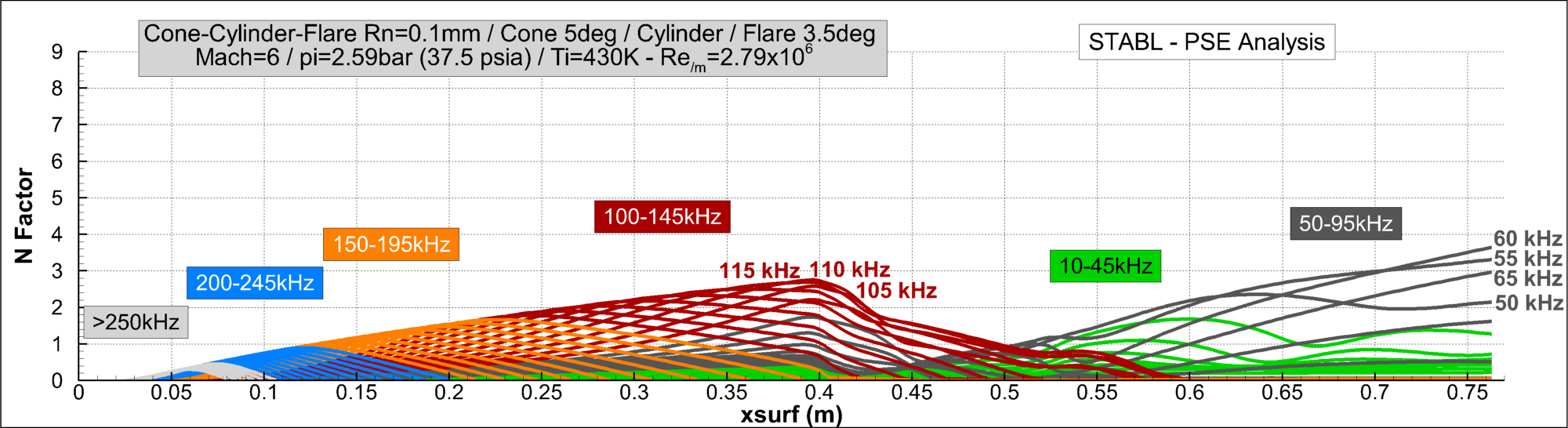}
%	\caption{N factors by frequency from STABL-PSE analysis at Re${/m}$ = 2.8x10$^{6}$}
%	\label{fig:NfactorsRe1}
%\end{figure}
%
%\begin{figure}
%	\centering
	\includegraphics[width=\linewidth]{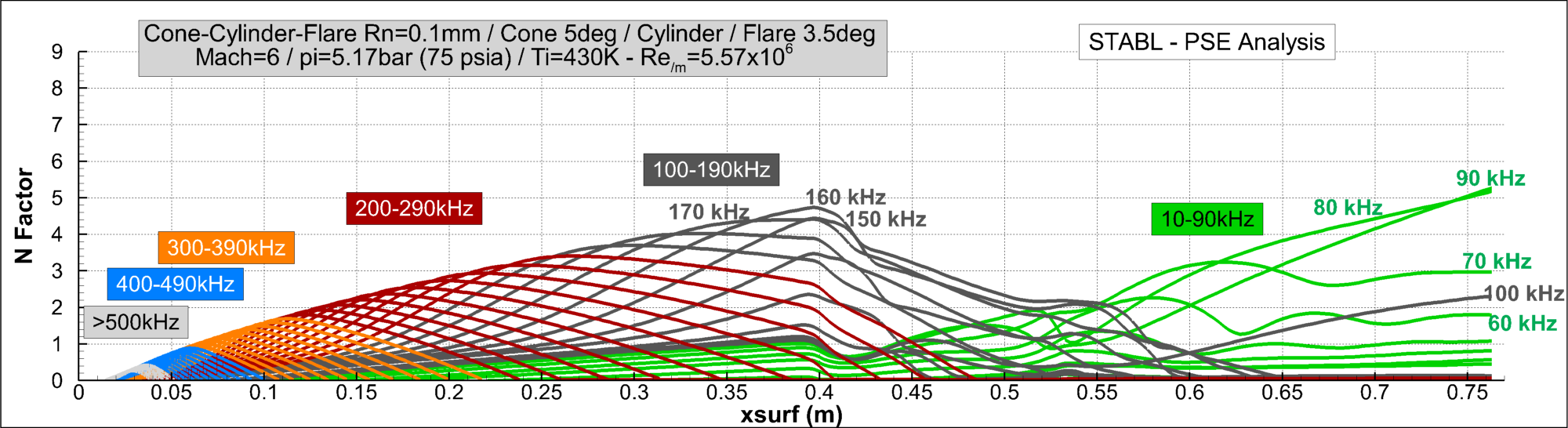}
%	\caption{N factors by frequency from STABL-PSE analysis at Re${/m}$ = 5.6x10$^{6}$}
%	\label{fig:NfactorsRe2}
%\end{figure}
%
%\begin{figure}
%	\centering
	\includegraphics[width=\linewidth]{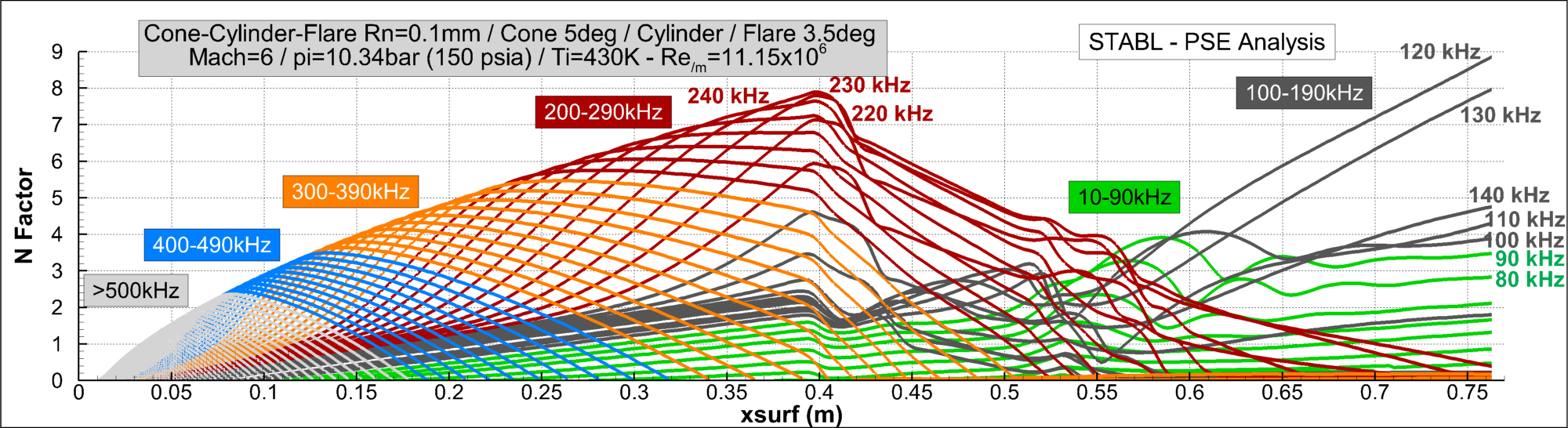}
	\caption{STABL (CCF3-5) - N factors by frequency from PSE analysis at Re${/m}$ = 2.8x10$^{6}$, 5.6x10$^{6}$ and 11.2x10$^{6}$}
	\label{fig:NfactorsRe1-2-3}
\end{figure}

Figure \ref{fig:Nfactor_3D} shows a tridimensional N factor distribution along the cone-cylinder-flare configuration (CCF3-5) at Re${/m}$ =5.6x10$^{6}$ obtained in PSE mode for the two-dimensional disturbances only ($\psi$= 0°).
The wave amplitude increase on the cone is classical and well correlated with the boundary-layer thickness. The maximum amplitude of the waves increases when the frequency decreases ; the greatest N factor on the cone is obtained for 160 kHz (see also figure \ref{fig:NfactorsRe1-2-3}).
The disturbances are damped by the flow expansion occurring at the cone cylinder junction. The amplitude of low frequency disturbances starts to increase on the second part of the cylinder and some of them are farther amplified on the flare, leading to high N factor levels at the end of the configuration for a very narrow frequency range (80-90 kHz) for Re${/m}$ =5.6x10$^{6}$.

For this intermediate Reynolds number, N factors slightly lower than 5 are reached at the end of the cone and slightly greater than 5 at the end of the model. Such N factors values can lead to turbulent breakdown in a noisy environment.

The integrated N factors for the other two Reynolds numbers, Re${/m}$ = 2.8x10$^{6}$ and Re${/m}$ = 11.2x10$^{6}$, are shown in figure \ref{fig:NfactorsRe1-2-3}. The general trend is the same as the previous one with a remarkable N factor increase along the cone and along the flare but, for different frequencies. This frequency change can be related to the very different boundary-layer thickness for these three different Reynolds numbers.

At Re${/m}$ = 2.8x10$^{6}$, the maximum wave amplitude is limited with N factor values just below 3 at the end of the cone and around 3.6 at the end of the flare. From an experimental point of view, these levels are probably too low to be detected in quiet conditions but measurements can be considered in conventional conditions for this case.

At Re${/m}$ = 11.2x10$^{6}$, high N factor values are reached: around 8 at the end of the cone and almost 9 at the end of the flare. This case will be turbulent in conventional conditions but represents an ideal candidate for quiet experiments. Indeed, in quiet tunnel conditions, without noise radiated from the boundary-layers on the nozzle walls, this case may generate critical wave amplitudes, very near to transition breakdown or possibly turbulent at the end of the configuration.

%---------------
\subsubsection{A look at the first-mode N-factors}
%----------------
As shown previously in figures \ref{fig:max_ai_LST_STABL_Mamout_x0_1m} and \ref{fig:ai_psi_LST_STABL_3xloc}, whatever the location on the object, the first-mode instability has lower amplification rates than the second-mode but, is at the same time, known to have longer zones of amplification than the second-mode. As a consequence, the first-mode waves can potentially reach significant amplitudes on long enough configurations, on the conical part for example. 

\begin{figure}
	\centering
	\includegraphics[width=0.46\linewidth]{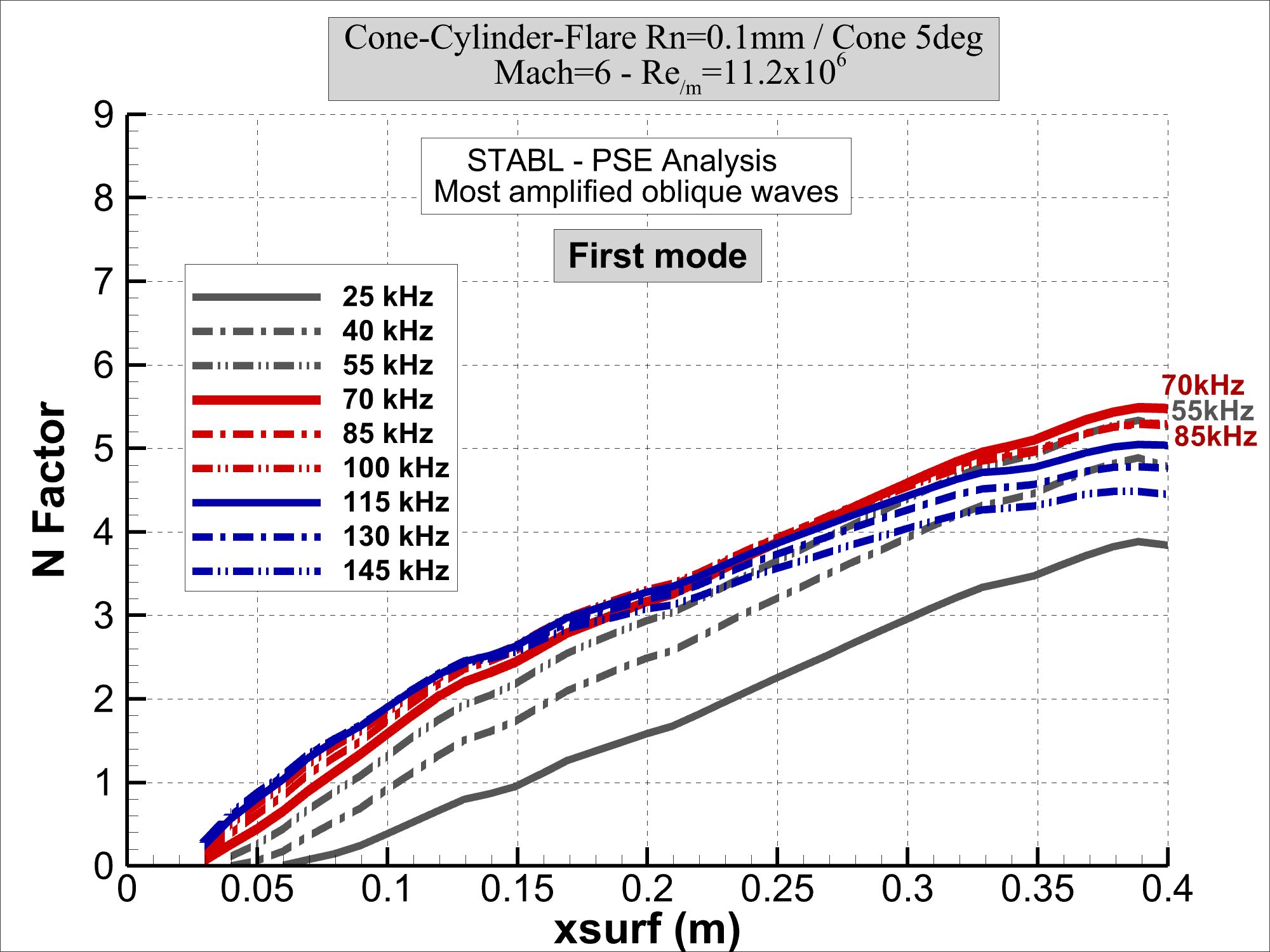}
	\includegraphics[width=0.46\linewidth]{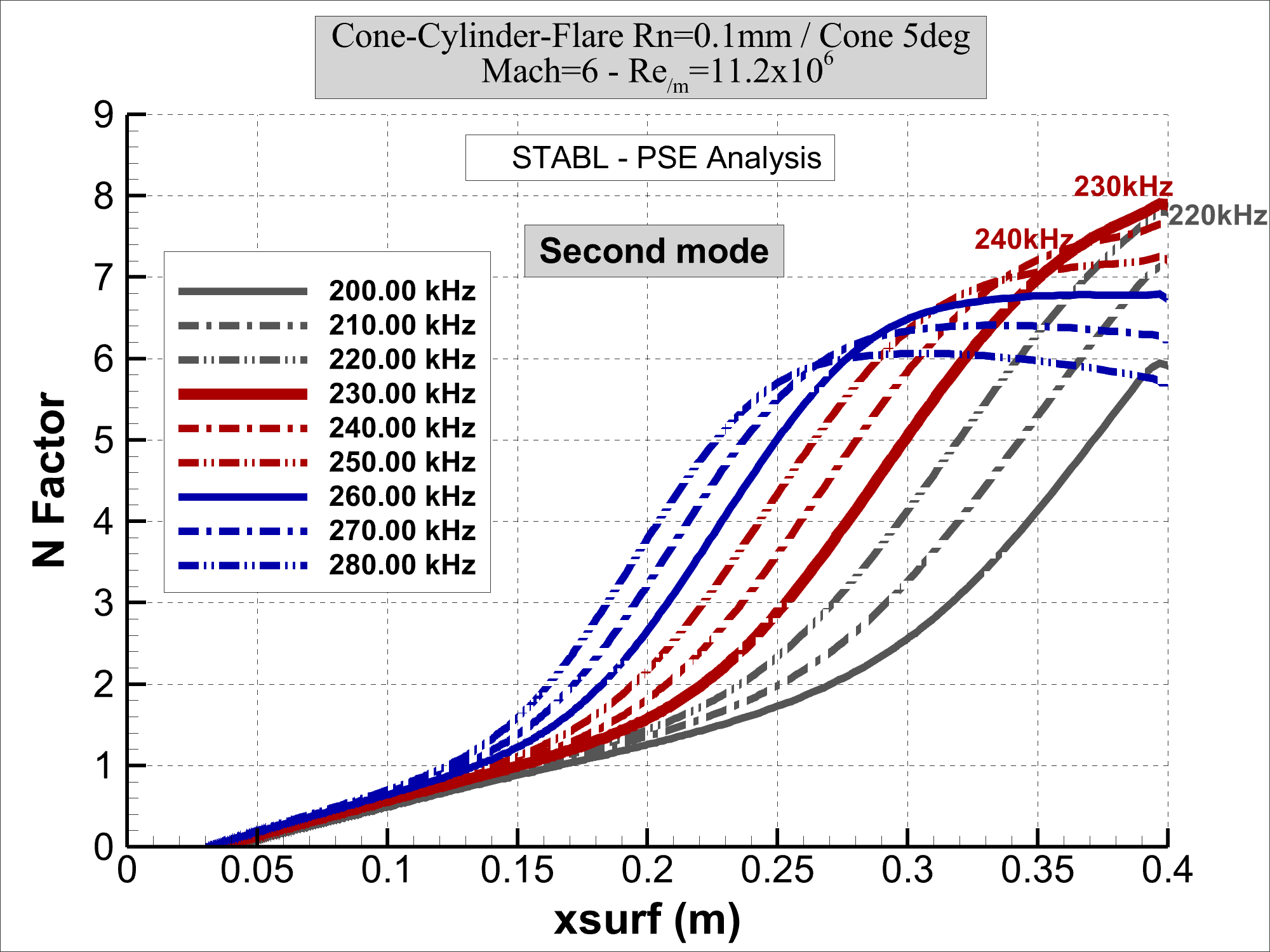}
	\caption{STABL (CCF3-5) - 1st mode / 2nd mode - N-factor evolution on the cone}
	\label{fig:CCF_1st_vs_2nd_mode_on_cone}
\end{figure}

This is evaluated on the cone in figure \ref{fig:CCF_1st_vs_2nd_mode_on_cone} where N-factors are plotted for the highest Reynolds number considered in this study (Re${/m}$ = 11.2x10$^{6}$).
The most amplified first-mode waves are at frequencies around 55-70kHz, they reach N-factor levels around 5.5 at the end of the cone.
Second-mode N-factor levels are near 8 just upstream of the cone-cylinder junction for frequencies in the range 220-240 kHz.

So, first-mode N-factors are significant at the end of the cone but lower than the second-mode N-factors. The second-mode is the dominant instability at the end of the cone. For attached boundary-layers, as this is the case here, this statement can be generalized to the full configuration. 
This confirms our strategy to focus on the second-mode for this hypersonic study on CCF3-5.\\

Remark: for flow separated cases as CCF10, the first-mode should also be looked at carefully. Indeed, this mode seems to be less sensitive to flow expansion than the second mode and possible non-linear interactions could occur within the bubble region (see comments on this point in \cite{lugrin2022TransSBLI}).\\\\

In parallel of the previous numerical studies, experiments have been realized by Dr. Benitez (2021) in the BAM6QT wind tunnel. Both geometries, CCF3-5 and CCF10, were tested under both conventional noise and quiet flow with high frequency pressure fluctuation sensors. 
After a brief presentation of the wind tunnel and models, experimental results will be presented as well as comparisons with numerical predictions.

%==================
\section{Wind tunnel: Boeing / AFOSR Mach-6 Quiet Tunnel}
\begin{figure*}
	\centerline{\includegraphics[width=1.0\linewidth]{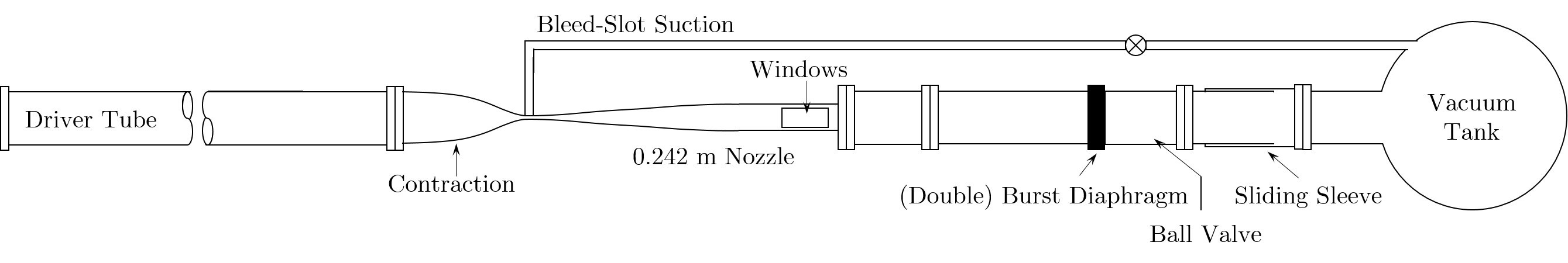}}
	\caption{BAM6QT schematic}
	\label{fig:bam6qt_diagram}
\end{figure*}

%==================
\label{sec:BAM6QT}
The Boeing/AFOSR Mach 6 Quiet Tunnel (BAM6QT) is located at Purdue University \cite{schneider2008BAM6QT}. The BAM6QT is a Ludwieg tube that is capable of running with conventional noise or quiet flow with run times of up to 6 seconds. The test section is located in the downstream end of the diverging section of the nozzle and includes optical access with a variety of windows. A diagram of the tunnel can be found in figure \ref{fig:bam6qt_diagram}.

The BAM6QT consists of a 37.3 meter driver tube connected to a nozzle with a converging-diverging design, which exhausts into a 113 cubic meter vacuum tank. To avoid condensation, the air is heated and dried. The tunnel is equipped with a diaphragm system located downstream of the test section, consisting of two thin aluminum sheets separated by an air gap; the air in this gap is evacuated to start a run. During the time between reflections of the expansion wave in the driver tube, steady flow is achieved for approximately 200 ms.  %The tunnel is started for up to 6 seconds per run. 
By obtaining data during different stages of this reflection cycle, a range of unit Reynolds numbers can be measured in a single run.

To obtain quiet flow, a combination of several features are implemented to reduce disturbances and keep the boundary layer on the nozzle laminar, including a polished nozzle, particle filter, and bleed slots at the throat. Together, these features allow the tunnel to operate with very low freestream noise levels (less than 0.02\%, close to free-flight levels). However, at high enough unit Reynolds numbers the flow will still be noisy. Since a turbulent boundary layer along the diverging portion of the nozzle is thicker than a laminar one, the effective diameter of that part of the nozzle is smaller.  This results in Mach 5.8 flow, as opposed to the Mach 6.0 achieved during a quiet run.

%===============================================
\section{The wind-tunnel models: CCF3-5 and CCF10}
%===============================================
The cone-cylinder-flare models have been produced in the Aerospace Sciences Laboratory of Purdue University. The models are divided into three sections: a stainless steel nose tip, an aluminum mid-body containing the initial 5 degrees cone and the first half of the cylinder, and a PEEK base including the latter half of the cylinder and the conical flare (see figure \ref{fig:CCF3-5_Model} and \ref{fig:CCF10_Model}). The nose tip was made to be as sharp as could be produced with a nose radius equal to 0.1mm based on microscope visual inspection.

%--------------------------------------------
\subsection{The CCF3-5 cone-cylinder-flare model}
%--------------------------------------------
On CCF3-5, 11 sensor holes are located along the centerline of the model, with 3 additional holes located azimuthally at the end of the first cone. 
All 14 sensor holes were utilized, with all but one being occupied by PCB 132B38 pressure sensors (see detail in section \ref{sec.instrum}).

Infrared thermography has been used to provide the heat fluxes on the model with a particular focus on the compression corner region. 

\begin{figure}
	\centering
    \includegraphics[width=0.80\linewidth]{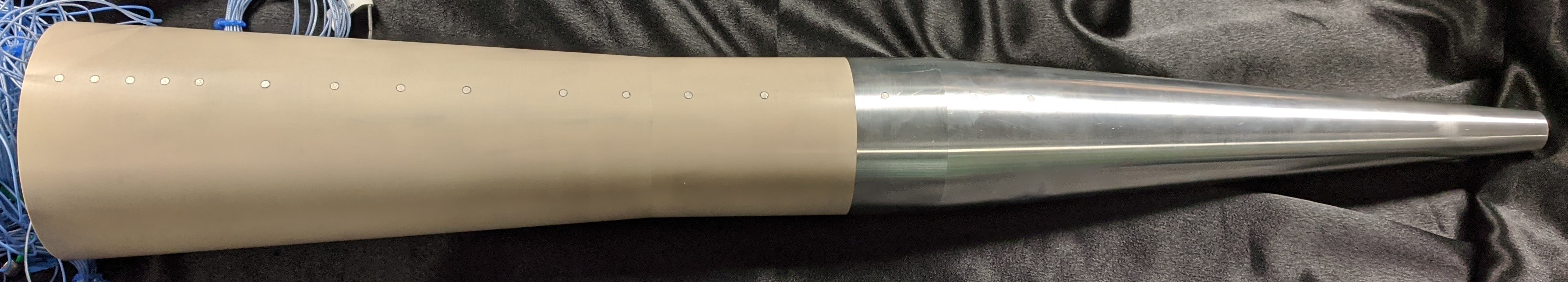}
	\caption{CC3-5 Purdue model with PCB sensors for BAM6QT experiments}
	\label{fig:CCF3-5_Model}
\end{figure}

Figure \ref{fig:CCF3-5_Model} displays the installed model with the locations of each sensor. PCB132B38 sensors are used for the measurement of the high-frequency pressure fluctuations relative to the development of boundary-layer instabilities. 

The four PCB sensors located 90$^\circ$ apart azimuthally at the same downstream location along the 5-degree cone help zero the angle of attack and slideslip of the model. Indeed, when at 0.0$^\circ$, the second-mode peak frequencies from all four sensors should theoretically be the same. With this method, an angle of attack very close to zero is realized.
%In practice, the model was adjusted until the four peaks were within 4 % of the mean peak frequency.

%--------------------------------------------
\subsection{The CCF10 cone-cylinder-flare model}
\begin{figure}
	\centering
	\includegraphics[width=0.68\linewidth]{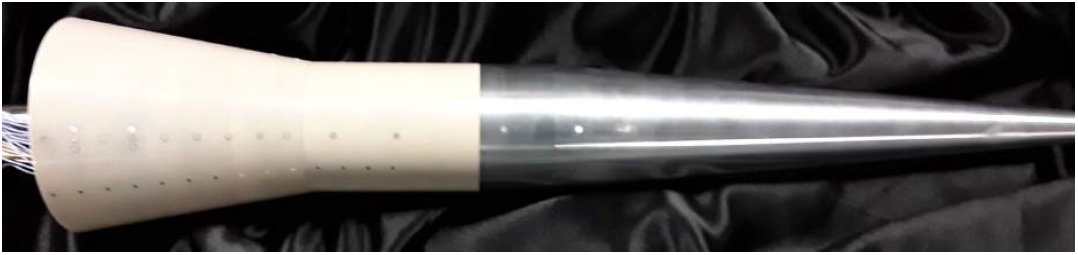}
	\caption{CC10 Purdue model with Kulite and PCB sensors for BAM6QT experiments}
	\label{fig:CCF10_Model}
\end{figure}

%--------------------------------------------
The first two sections of the model are the same on CCF10 as on CCF3-5. So this is the same PCB distribution on both models through the first half of the cylinder. On CCF10, the aft-part of the cylinder and the flare are made of PEEK in order to image it with infra-red thermography. The centerline is equipped with 2 PCB on the cylinder part and 9 PCB on the flare as visible in figure \ref{fig:CCF10_Model}.

%===============================================
\section{Surface pressure-fluctuation measurements}
%===============================================
\label{sec.instrum}
%--------------------------------------------
%\subsection{Surface pressure-fluctuation sensors}
%--------------------------------------------
Surface pressure-fluctuation measurements were acquired with PCB132B38 sensors as well as Kulite XCE-062-15A and 5A sensors. These sensors are commonly used for instability measurements. Here, a short description of both sensors is given for comparison, even if, only the PCB measurements will be presented in the following sections (see \cite{benit2021dissert} dissertation for extensive detail on the Kulite measurements).
%These sensors are commonly used for instability measurements, see \cite{benit2021dissert} dissertation for extensive detail on PCB measurements.

\subsection{PCB sensors}
The PCB sensors, manufactured by PCB Piezotronics, are high-pass filtered above 11 kHz and have a high-frequency response limit of 1 MHz, albeit resonance of the sensor has been observed for frequencies as low as 300kHz. Their nominal resolution of 6.89 Pa coupled with a rise time of less than 3 $\mu$s provides a voltage which can be scaled using the factory calibration to obtain pressure measurements.  PCB132B38 sensors have a surface diameter of 3.2 mm.

\subsection{Kulite sensors}
Kulite XCE-062-15A and 5A sensors were used to acquire lower-frequency disturbances and static pressures.  These transducers are smaller than PCBs, with a 1.7 mm, but have a slower response time.  This lower response time, coupled with a large resonance peak, restrict their useful output to below 270 kHz. The sensors were calibrated by manually changing the test section pressure and recording voltages across a known range.\\

For both sensors, PCB and Kulite, power spectral densities (PSDs) of the pressure signals are used to determine at what frequencies instabilities exist along the surface, if any are present.

%==================
\section{BAM6QT - Conventional runs on CCF3-5}
%==================
Surface pressure power spectral densities (PSDs) were obtained along each of the three sections of the model at an extensive range of freestream unit Reynolds number (Re${/m}$ between 2 and 10 million) in conventional noise. 

%-----------------
\subsection{Pressure fluctuation measurements (conventional noise)}
%-----------------
%-----------------
\subsubsection{Conical part}
\begin{figure}
	\centering
	\includegraphics[width=0.46\linewidth]{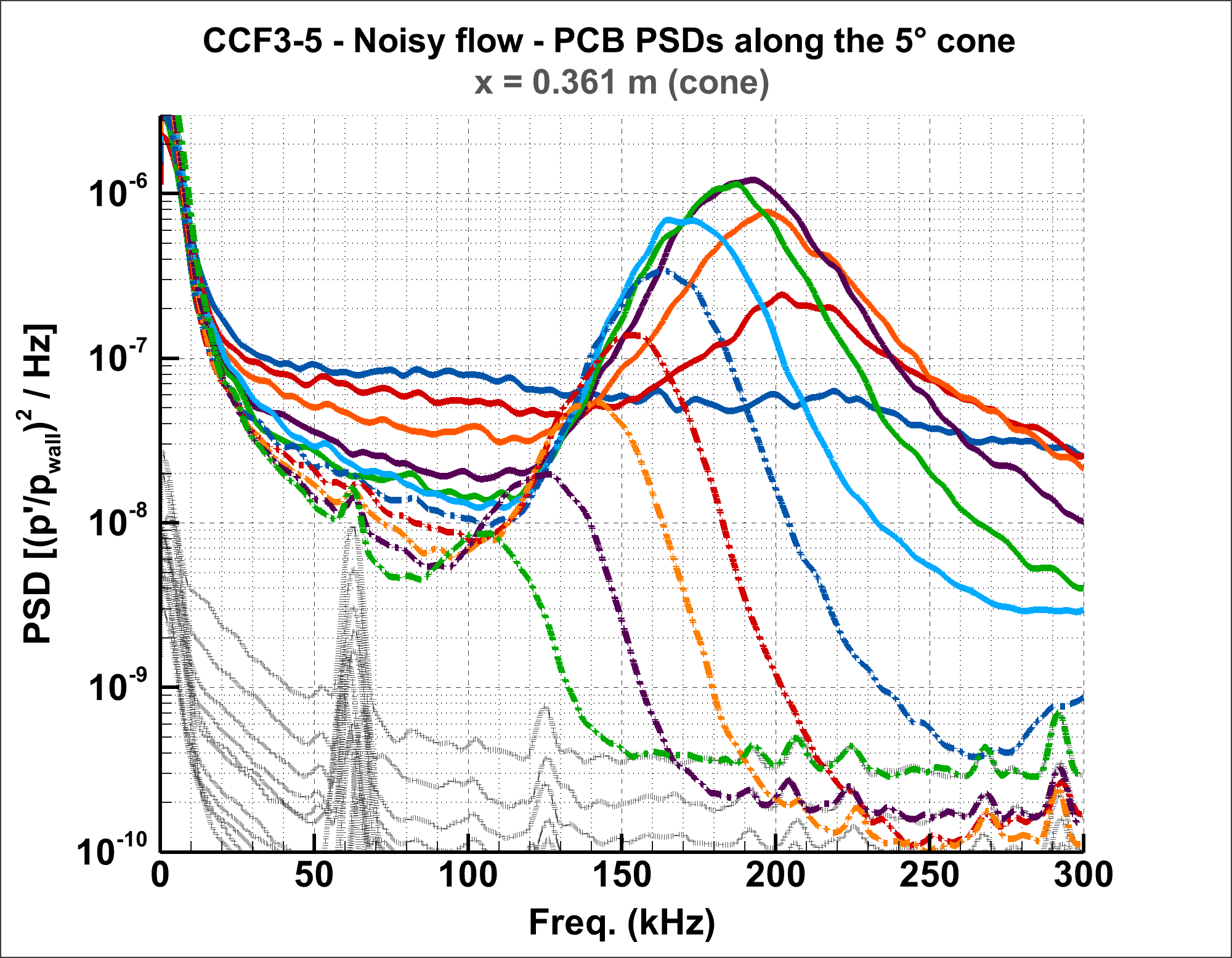} % Fig 4.15 Liz Thesis
	\includegraphics[width=0.46\linewidth]{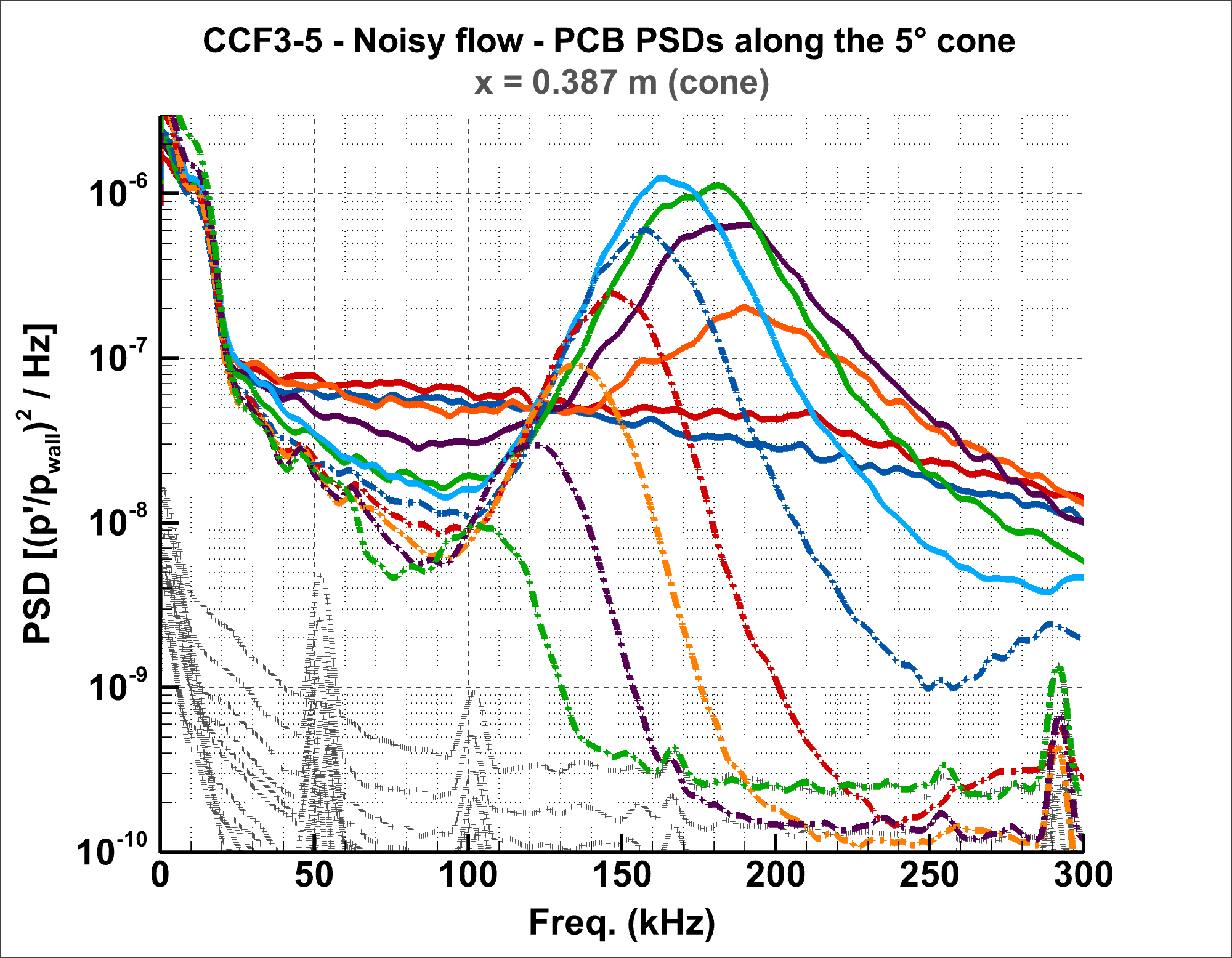} % Fig 4.15 Liz Thesis
	\caption{CCF3-5 / BAM6QT Conventional noise - PCB PSDs along the 5 degrees cone (same Reynolds number legend as figure \ref{fig:CCF10_Noisy_Cone_PCB_HighFreq} with additional pre-run noise in grey)}
	\label{fig:CCF10_Noisy_Cone_PCB}
\end{figure}

%-----------------
The PSDs obtained along the 5 degrees cone in conventional noise are presented in figure \ref{fig:CCF10_Noisy_Cone_PCB} at two different locations near the end of the cone (x=0.361m and x=0.387m). The spectra are normalized by wall pressure, resulting in noise floors that vary with unit Reynolds number.
These experimental wall surface pressure fluctuations show a clear identification of second Mack mode waves with peaks visible from 100 to 200 kHz. 
The peak shifts towards high frequency and larger amplitude with increasing freestream unit Reynolds number as the boundary-layer becomes thinner.

\begin{figure}
	\centering
	\includegraphics[width=0.46\linewidth]{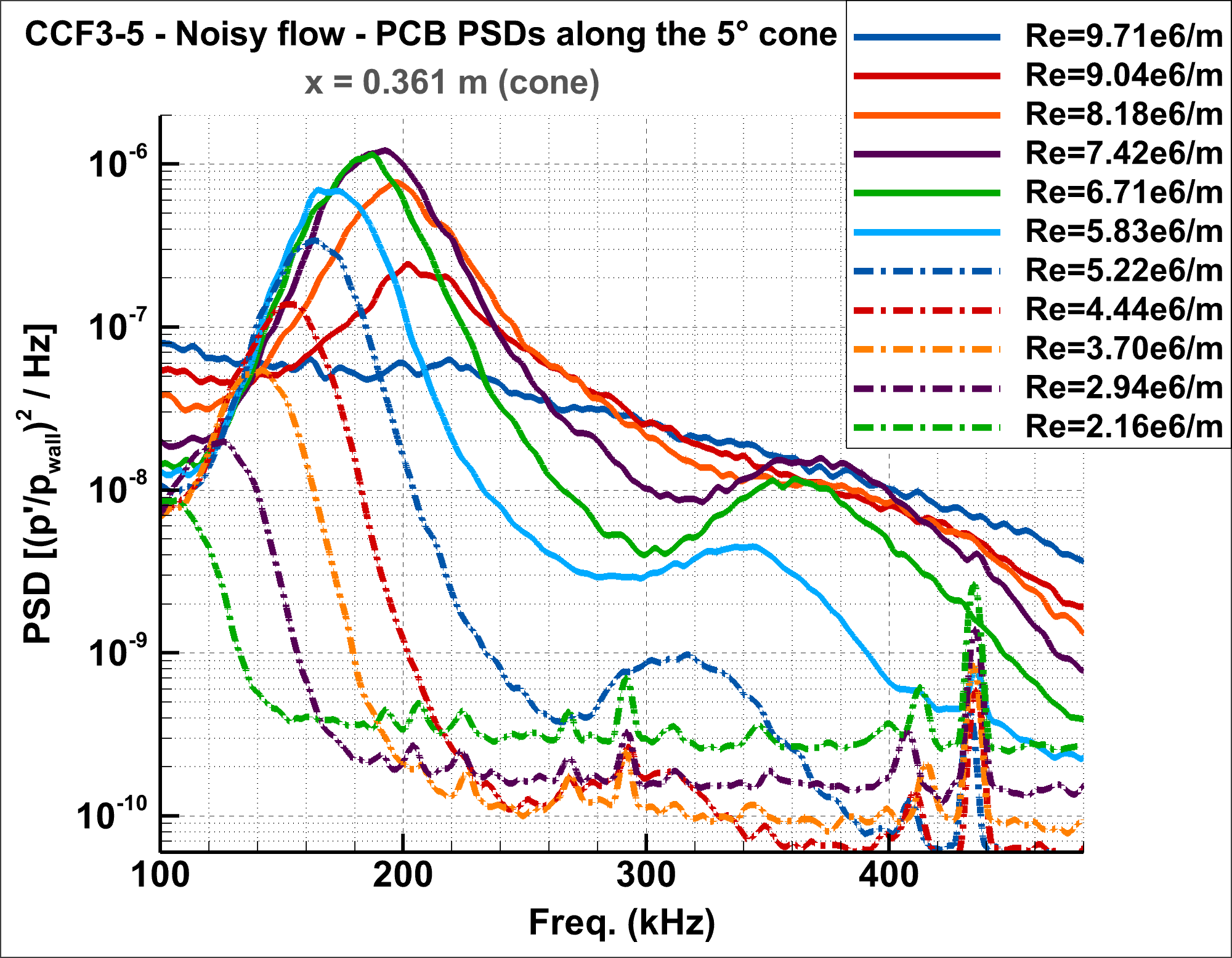}
	\includegraphics[width=0.46\linewidth]{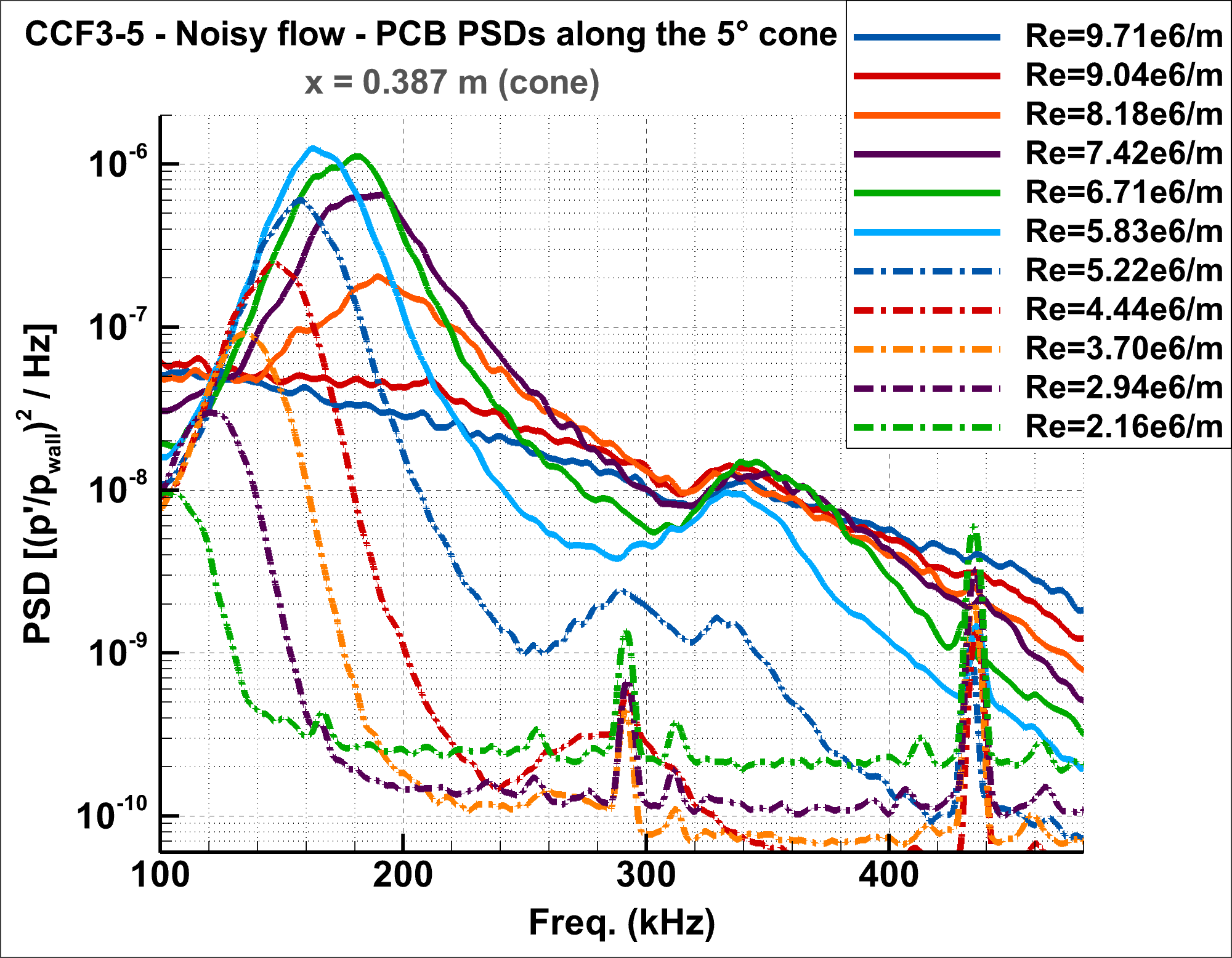}
	\caption{CCF3-5 / BAM6QT Conventional noise - PCB PSDs along the 5 degrees cone (extended frequency range: non-linear interaction broadening and harmonics)} 
	\label{fig:CCF10_Noisy_Cone_PCB_HighFreq}
\end{figure}

For unit Reynolds numbers around 5 million, harmonics and a progressive broadening of the spectra are observed. This is a sign of non-linear saturation as confirmed on figure \ref{fig:CCF10_Noisy_Cone_PCB_HighFreq} on an extended frequency range. The presence of non-linearity is confirmed by the spectra where most Reynolds number clearly display a harmonic at twice the peak frequency of the second mode.
Finally, the boundary-layer becomes turbulent at the end of the cone for unit Reynolds number around 9 million. 
This is characterized by the broadband spectra with no peaks, indicating a turbulent distribution of energy on all the frequencies. 
It is also interesting to look at the very rapid evolution of the spectra from the first location x=0.361m to the second one 26 mm apart ; the spectrum at Re${/m}$ = 9x10$^{6}$ has become turbulent or intermittently turbulent at the second location while still in the non-linear interaction process at the first location. 
The sharp peaks that are present in the pre-run spectra are due to electronic noise, rather than an aerodynamic effect.

This 5 degree cone is very efficient to study the second-mode waves evolution through an extended Reynolds number range, from fully laminar to turbulent.

%-----------------
\subsubsection{Cylinder part}
%-----------------

\begin{figure}
	\centering
	\includegraphics[width=0.46\linewidth]{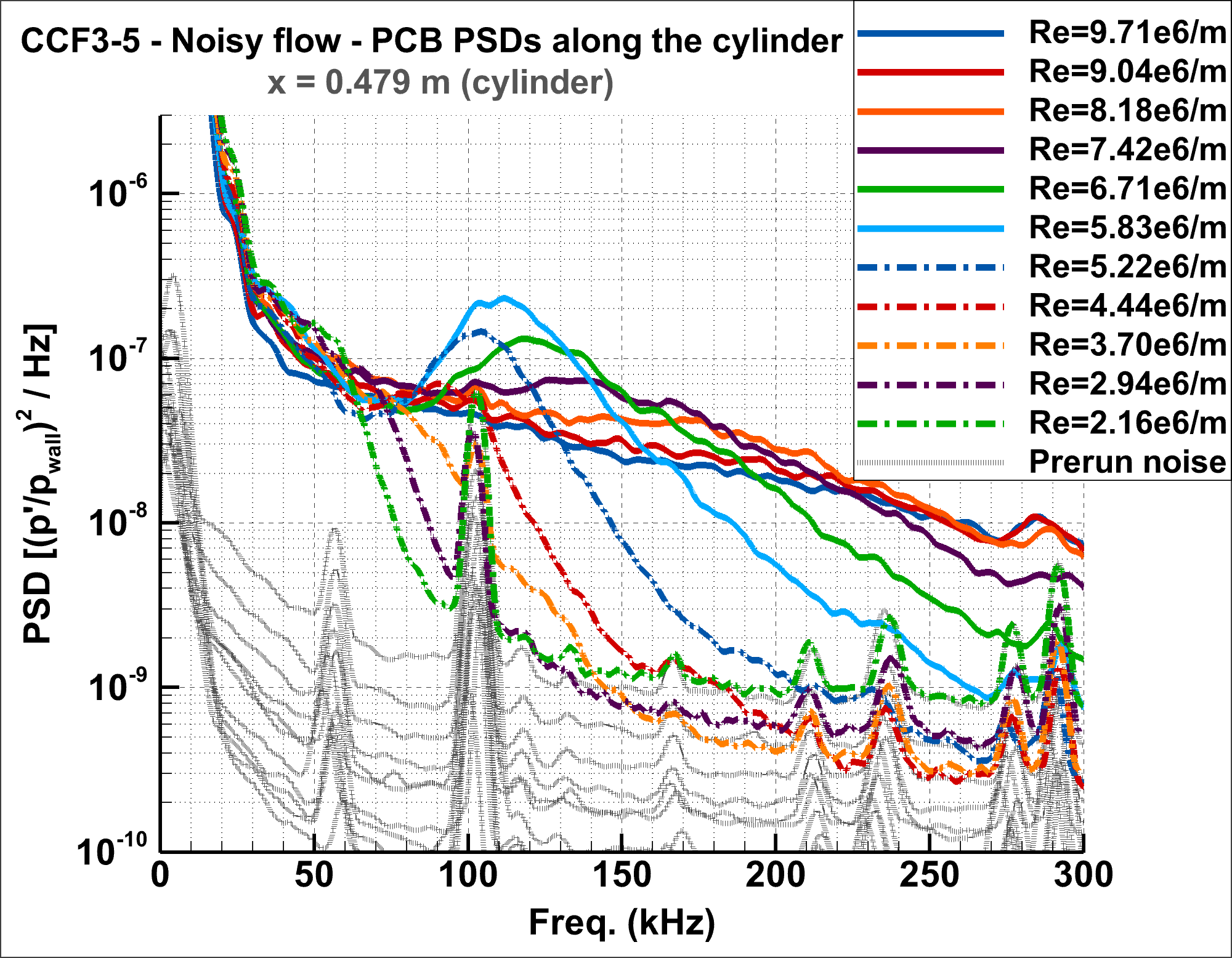} % Fig 4.15 Liz Thesis
	\includegraphics[width=0.46\linewidth]{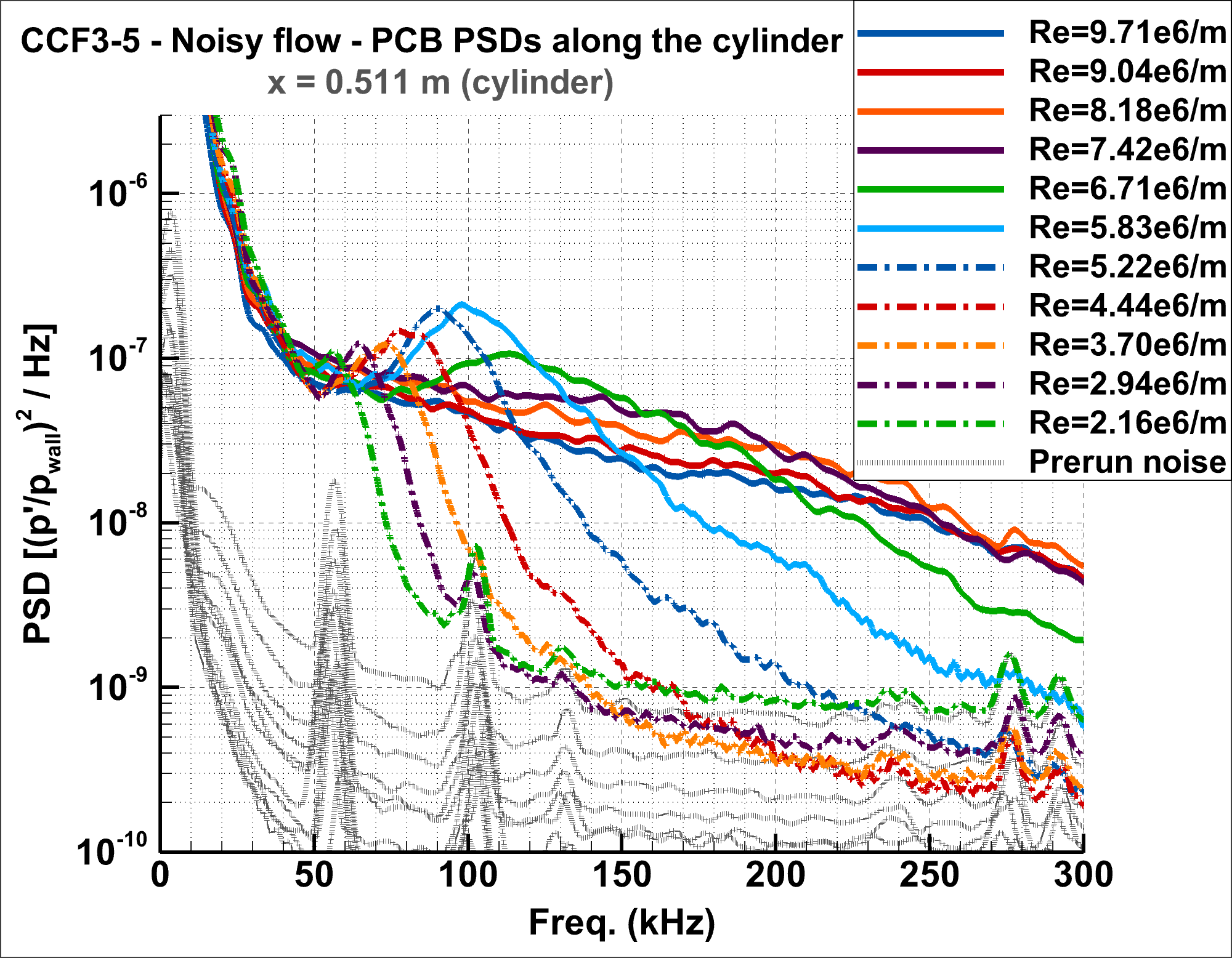} % Fig 4.15 Liz Thesis
	\caption{CCF3-5 / BAM6QT Conventional noise - PCB PSDs along the cylinder} 
	\label{fig:CCF10_Noisy_Cyl_PCB}
\end{figure}

Along the cylinder, going through the flow expansion strongly decreases the second-mode amplitude of the waves amplified on the conical part (see figure \ref{fig:CCF10_Noisy_Cyl_PCB}). Then, along the cylinder, the second-mode is progressively damped and decreases in peak frequency with the thickening of the boundary layer.
This increased boundary layer (about two times larger than on the cone) can potentially maintain or amplify second-mode instabilities coming from the cone, but this concerns only frequencies about twice as low as the most amplified ones on the conical part.
These peaks, like the second-mode, increase in peak frequency with increasing Reynolds number.

These experimental data confirm that turbulence in a cone-cylinder shape (considered in the same conditions as the present study) will start at the rear part of the cone before becoming turbulent on the cylinder. 

The flow expansion acts as an efficient damping mechanism for second-mode instabilities.

\begin{figure}
	\centering
	\includegraphics[width=0.46\linewidth]{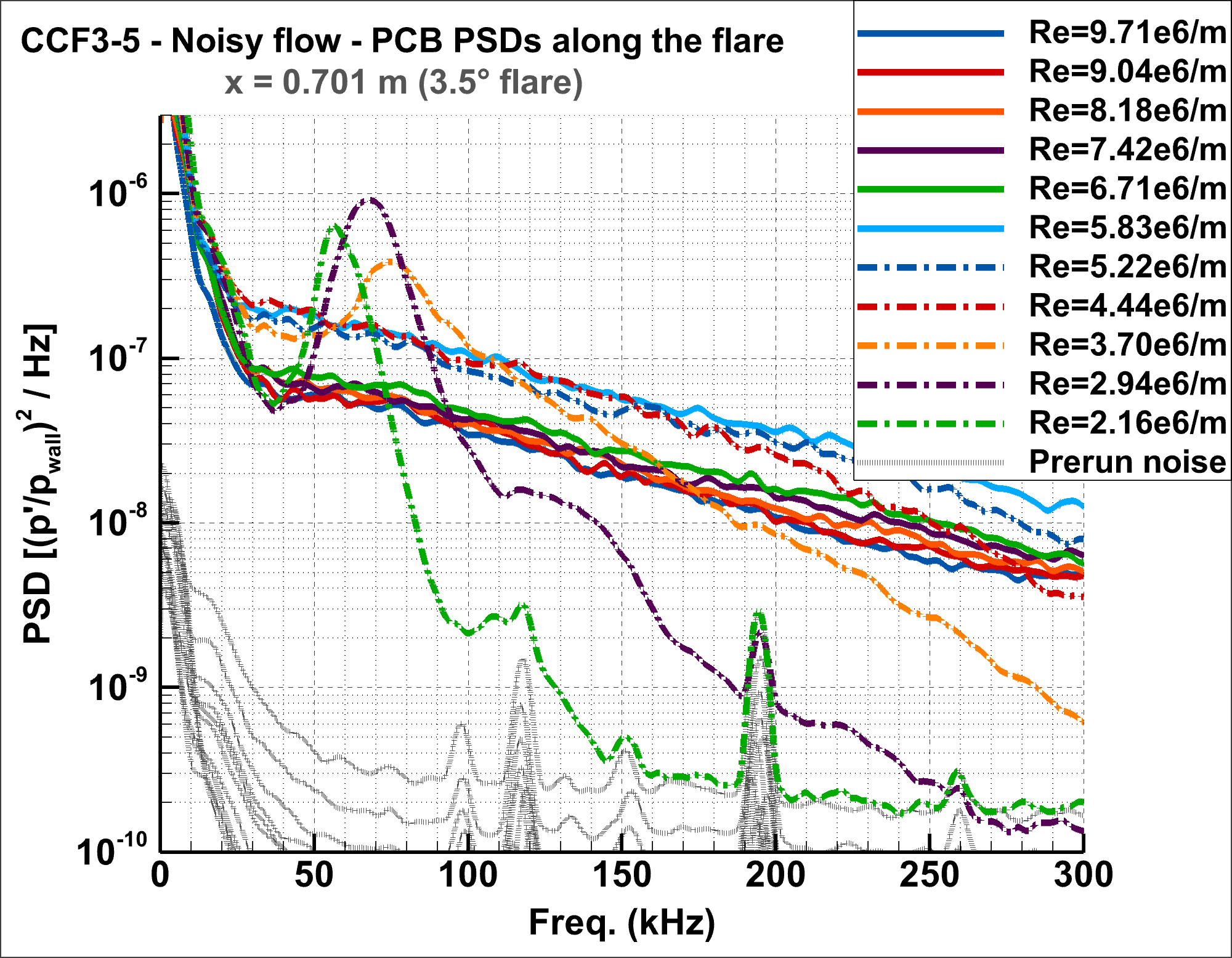} % Fig 4.16 Liz Thesis
	\includegraphics[width=0.46\linewidth]{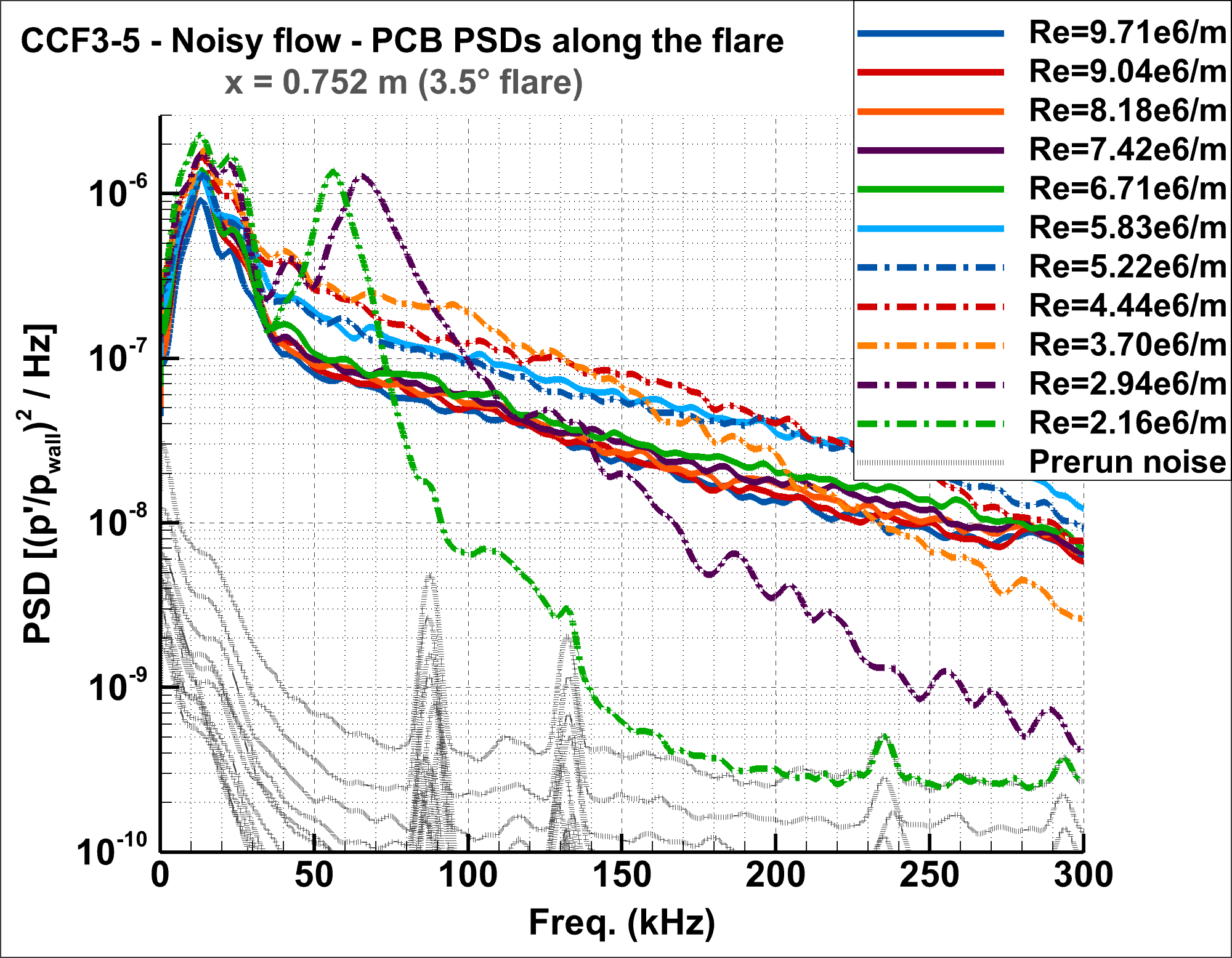} % Fig 4.16 Liz Thesis
	\caption{CCF3-5 / BAM6QT Conventional noise - PCB PSDs along the 3.5° flare} 
	\label{fig:CCF10_Noisy_Flare_PCB}
\end{figure}

%-----------------
\subsubsection{3.5° flare part}
%-----------------
Along the flare, for the lowest Reynolds number, the peak frequencies and amplitudes increase with unit Reynolds number as shown in figure \ref{fig:CCF10_Noisy_Flare_PCB}. 
The spectra are narrow, the amplified instabilities peak from 50 to 80 kHz when increasing the Reynolds number. 
This trend continues until transition begins, when the peak broadens and begins to decrease in amplitude. At the two downstream locations considered here (x=0.701m and x=0.752m), for unit Reynolds number greater than 3.5 million, the spectra collapse and are generally broadband, this indicates a fully turbulent boundary-layer. 
Small peaks around 20 kHz are observed at lower Reynolds numbers, but no physical explanation is available to explain these.

The adverse pressure gradient has a destabilizing effect, a narrow frequency band is amplified and the boundary-layer turns turbulent at low Reynolds number.

%-----------------
\subsection{Cross-comparisons: STABL / BAM6QT in conventional noise}
%-----------------
Two runs are considered for the numerical / experimental comparison respectively at Re${/m}$ = 3.3x10$^{6}$ and 6.0x10$^{6}$. In conventional noise environment, the experimental Mach number is equal to 5.8 because of the larger turbulent boundary-layer thickness which develops on the nozzle walls in these conditions. The aerodynamic conditions of two selected runs are given in Table \ref{tab:aero_cond_noisy}.

These aerodynamic conditions are not exactly the ones corresponding to the numerical cases presented previously at Mach 6, so new STABL computations have been performed at Mach 5.8.

\begin{table}
\begin{center}
	\begin{tabular}{|l|c|c|c|c|c|c|c|c|c|}
        \hline
		Run & Mach & pi (bar/psia)& Ti (K)& Re${/m}$ & p$_{\infty}$ (Pa)  & T$_{\infty}$ (K)  & $\rho_{\infty}$ (kg/m$^3$) & $V_{\infty}$ (m/s) \\
        \hline
		Run25  &  5.8  & 2.85 /  41.3   & 430.  &    3.31x10$^{6}$  &  222.1 &  55.6 &  0.01391 &  867.3 \\
        \hline
		Run18  &  5.8  & 5.17 /  75.0   & 430.  &    6.00x10$^{6}$  &  403.0 &  55.6 &  0.02523 &  867.3 \\
        \hline
    \end{tabular}
	\caption{Aerodynamic conditions: some conventional runs}
    \label{tab:aero_cond_noisy}
    \end{center}
\end{table}

\begin{figure}
	\centering
	\includegraphics[width=0.86\linewidth]{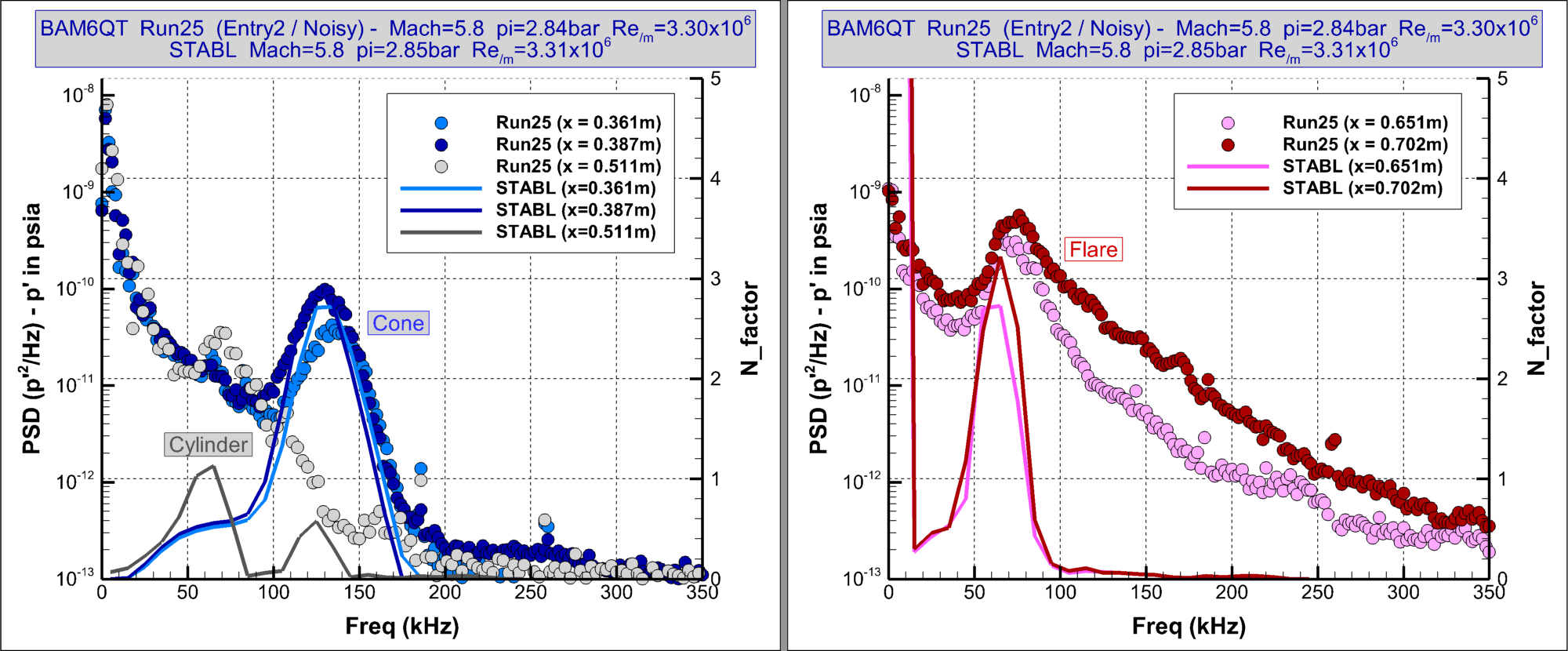}
	\caption{CCF3-5 - PSDs for conventional runs and STABL N-factors at Re${/m}$ = 3.3x10$^{6}$}
	\label{fig:CCF_Run25_vs_STABL}
\end{figure}

\begin{figure}
	\centering
	\includegraphics[width=0.86\linewidth]{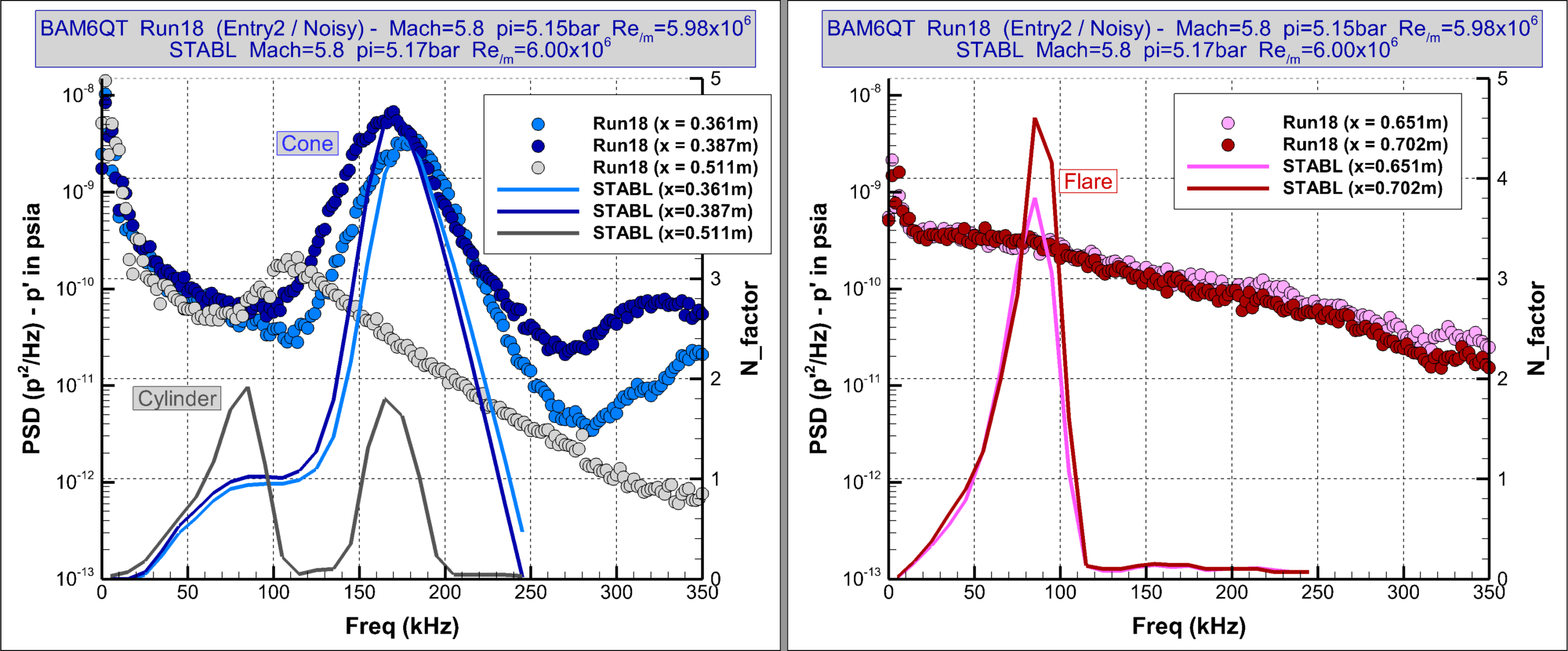}
	\caption{CCF3-5 - PSDs for conventional runs and STABL N-factors at Re${/m}$ = 6.0x10$^{6}$}
	\label{fig:CCF_Run18_vs_STABL}
\end{figure}

Figures \ref{fig:CCF_Run25_vs_STABL} and \ref{fig:CCF_Run18_vs_STABL} show comparisons of measured power spectra with computed N factors. Indeed, the N factor corresponds roughly with wave amplitude since it is a representation of how much a wave has grown along the body \cite{berridge2010Nfactor}. 
The initial amplitude of the wave is not known so this N factor cannot give an absolute wave amplitude. Nevertheless, an interesting numerical / experimental analogy can be made by adjusting manually the maximum N factors to fit the maximum power spectra value. This kind of plot has another advantage: if the stability computation is accurate and if second-mode waves are detected by the experiments, the peaks of amplified instability should be at the same frequencies. In this case, if the computational and experimental frequencies match, no doubt the detection of the waves is correct. Figures \ref{fig:CCF_Run25_vs_STABL} and \ref{fig:CCF_Run18_vs_STABL} display such N factor / power spectra comparison for the two noisy runs considered here. The overall agreement is correct, the following analysis can be made.

%-----------------
\subsubsection{Conical part}
%-----------------
On the cone, second-mode waves can be seen on each PCB sensor location (x=0.361m and x=0.387m). The large peaks are centered respectively at around 130-140 kHz for the lower Reynolds number and at around 160-170 kHz for the larger Reynolds number. 
A slightly larger amplitude is logically observed at the second PCB location. A correct agreement is obtained with the STABL N factors adjusted manually to fit with the experimental power spectra. At the second sensor position, the N factor value is equal to almost 3 at Re${/m}$ = 3.3x10$^{6}$ and to 4.6 at Re${/m}$ = 6.0x10$^{6}$. 

For these conventional runs, these cross-comparisons confirm that the boundary-layer remains laminar on the cone in these conditions even if some non-linear interactions have been observed experimentally from unit Reynolds numbers around 5 million (see figure \ref{fig:CCF10_Noisy_Cone_PCB_HighFreq}). 
%It can be stated that that the onset of transition occurs for N-factors between on the cone in conventional noise in BAM6QT.
%A new increase of the Reynolds number (above 6 millions) should lead to a turbulent spectrum knowing that the onset of transition can be estimated using N factor values around 4 to 5 for conventional noisy runs.

%-----------------
\subsubsection{Cylinder part}
\begin{figure}
	\centering
	\includegraphics[width=0.90\linewidth]{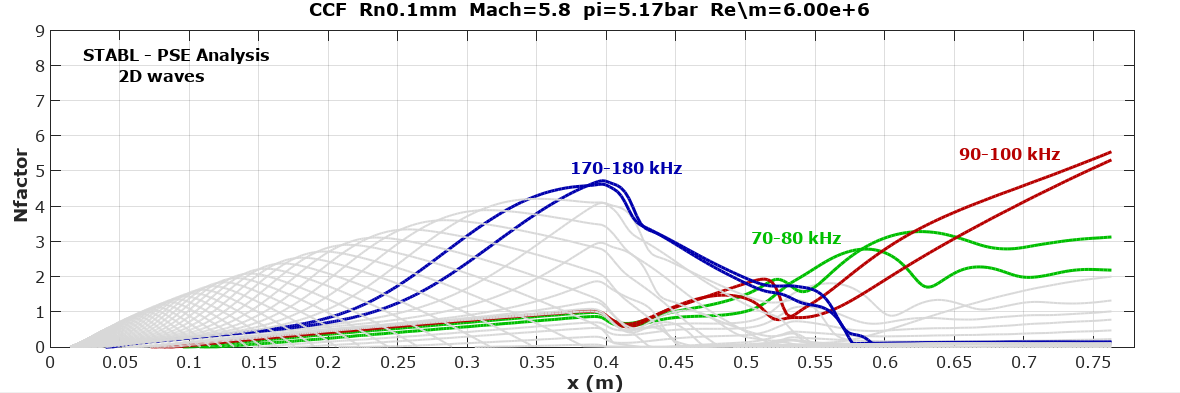}
	\caption{STABL N factors at Re${/m}$ = 6.0x10$^{6}$}
	\label{fig:CCF_Run18_STABL_Nfactor}
\end{figure}

%-----------------
On the cylinder part, the comparison between computation and experiment is not as good as on the cone. At the considered PCB location (x=0.511m) just before the end of the cylinder, two peaks can be observed on the N factor predicted numerically but only one on the experimental power spectra. Furthermore, the N factor and PSD levels do not fit. Figure \ref{fig:CCF_Run18_STABL_Nfactor} allows to explain the two peaks computed with STABL at Re${/m}$ = 6.0x10$^{6}$.
The first peak at around 170 kHz comes from the second-mode waves highly amplified on the cone and progressively damped on the cylinder part.
The second peak corresponds to lower frequencies at around 70-90 kHz coming from the conical part and amplified along the cylinder part. 
The peak detected in the experiment is at 110 kHz, at a higher frequency than the numerical prediction. The reason of this discrepancy is explained in \cite{benit2023-jfm}.

The measured / computed comparison is inconclusive on this cylinder part.
It seems that the damping effect produced at the cone-cylinder junction is "felt" slightly differently in the computation than in the experiment. Further numerical studies will be needed on the cylinder part, perhaps with more advanced tools.

%-----------------
\subsubsection{3.5° flare part}
%-----------------
On the flare, the boundary-layer is laminar at Re${/m}$ = 3.3x10$^{6}$ with experimental second-mode peaks observed at around 70 and 75 kHz. The computed frequencies are at a slightly lower frequency but second-mode waves are clearly identified here by the computed and experimental approaches. 
For the run at Re${/m}$ = 6.0x10$^{6}$, the PCB measurements indicate a turbulent spectrum on the flare while the stability analysis indicates N factors at around 4.5 for wave frequencies around 90 kHz for both sensors (x=0.651 m and x=0.702 m). In noisy conditions, such N factor values are indeed near to transition or turbulent. 

Here the flow has become turbulent on the flare even if the predicted N factor is slightly lower on the flare than on the cone at the considered x-locations - this fact is questionable and this situation is not fully explained at this moment. One possible explanation could come from the one discussed by \cite{marineau2015BLStab}: in experiments an increased tunnel noise exists at lower frequencies while a decreased tunnel noise is observed at high frequency. So the transition could appear on the lower frequency band on the flare rather than on the higher frequency band on the cone. 
The receptivity could also be different.\\

These cross-comparisons between STABL and BAM6QT are very satisfying: the agreement on the conical part and on the flare is very good but a little less conclusive on the damping cylinder region. 

As shown through all the previous analyzes and results, transition to turbulence on CCF3-5 is driven by the cone and the flare where both, computations and experiments, appear very reliable. This study confirms the efficiency of the combination local stability / wind tunnel experiments to provide physical and reliable extensive description of the development of boundary-layer instabilities on a given configuration.\\

At this point of the article, it appears very interesting to go further in the analysis by adding the major asset of BAM6QT: the capacity to be run in quiet mode with low noise levels comparable to flight.

%==================
\section{BAM6QT - Quiet runs on CCF3-5}
%==================
Surface pressure power spectral densities were obtained along each of the three sections of the model at an extensive range of freestream unit Reynolds number, up to Re${/m}$ = 12.0x10$^{6}$, in quiet flow.\\

Although the configuration considered here is CCF3-5, the results presented here on the conical part are the ones obtained on CCF10. Indeed, in the initial experimental campaign on CCF3-5 in quiet conditions, it appeared that the fluctuations along the flare seem to be transmitting their signal upstream to the cone and cylinder segments, potentially through model vibration from the longer CCF3-5 model. 
The conical part of the object is shared by the two configurations, so the measurements should be the same on this part of the object. So here, the quiet PSDs on the cone are the ones from CCF10 while the PSDs on the cylinder and the flare are the real ones from CCF3-5.

%-----------------
\subsection{Pressure fluctuation measurements (quiet flow)}
%-----------------

%-----------------
\subsubsection{Conical part}
\begin{figure}
	\centering
	\includegraphics[width=0.46\linewidth]{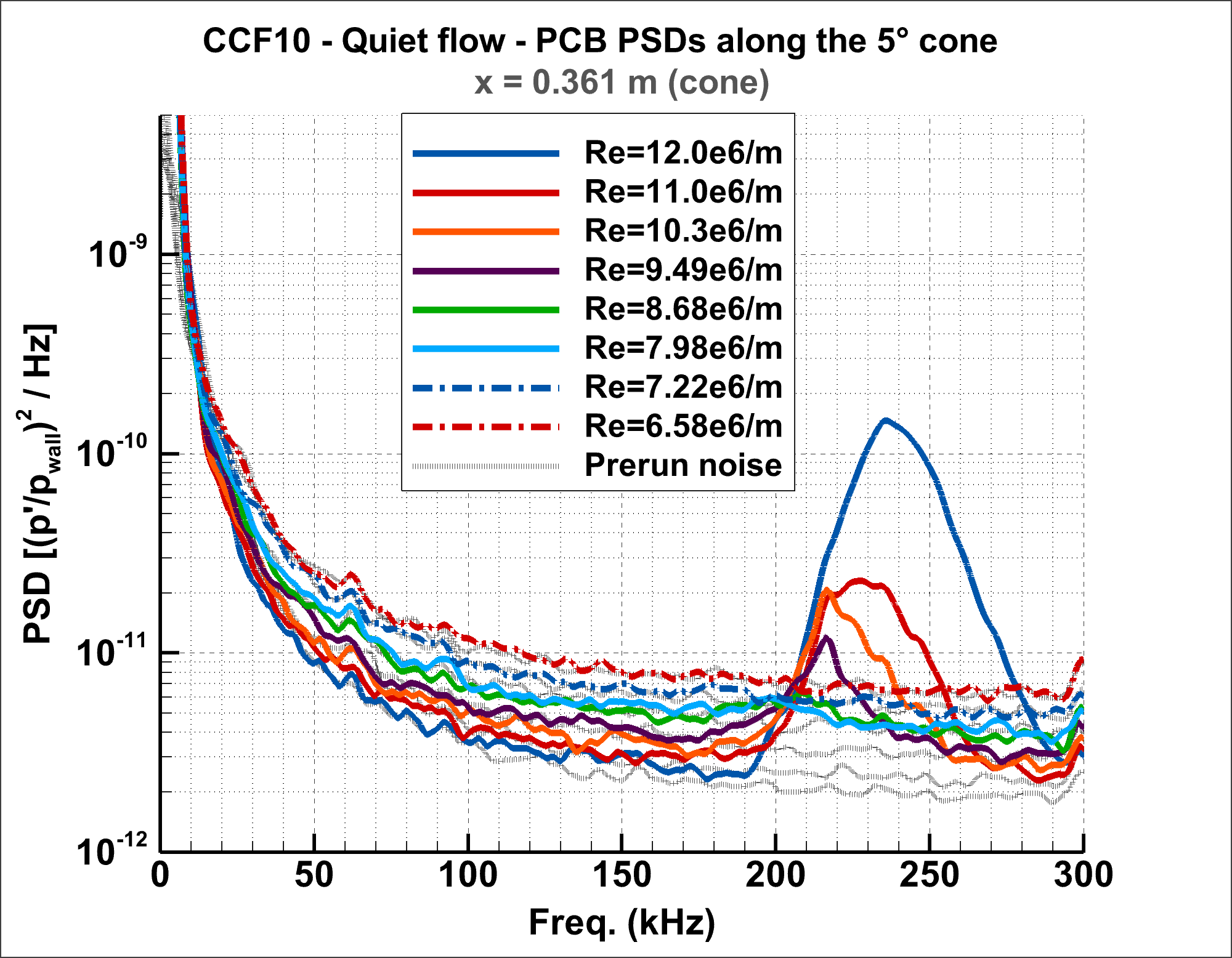} % Fig 5.10 Liz Thesis
	\includegraphics[width=0.46\linewidth]{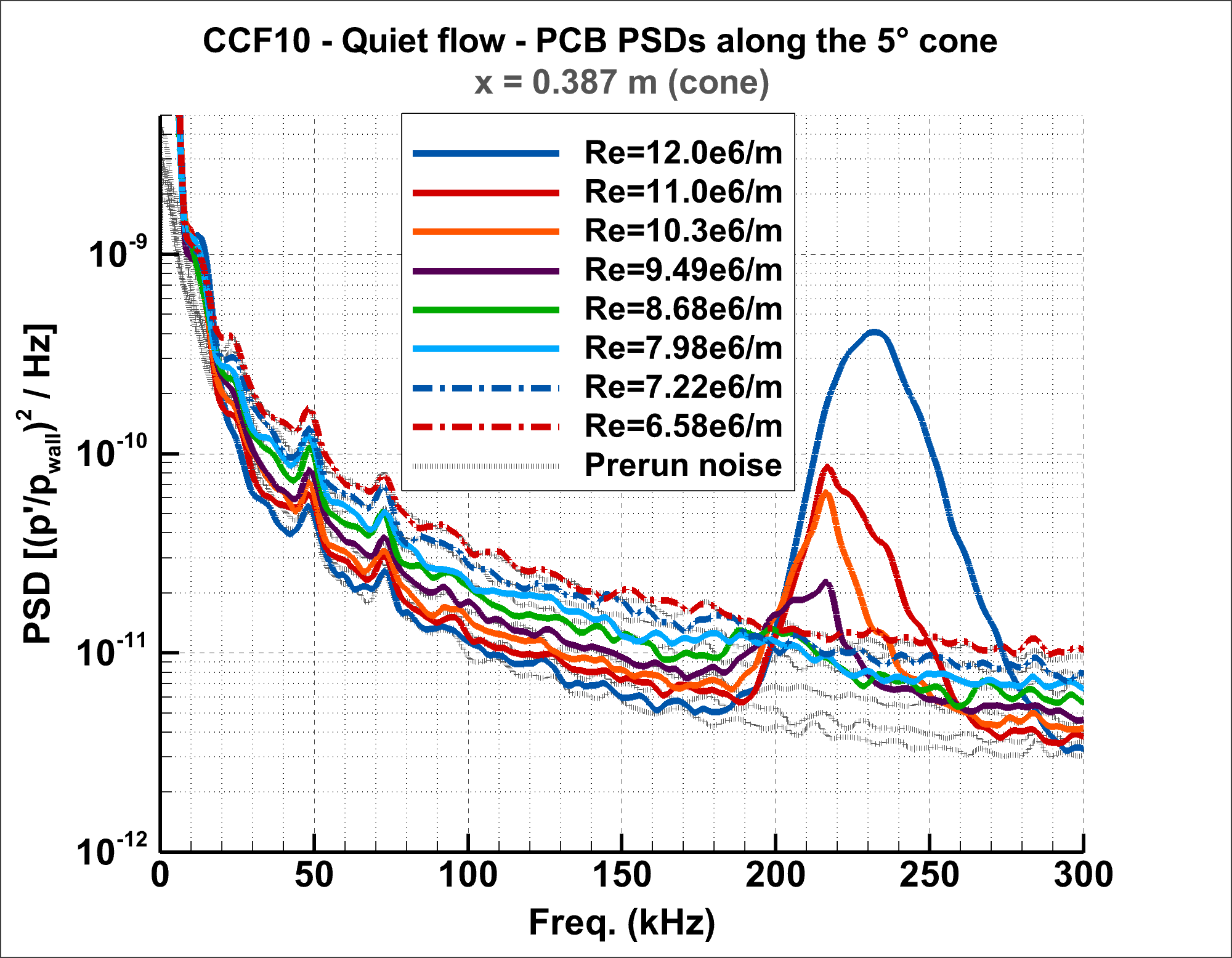} % Fig 5.10 Liz Thesis
	\caption{CCF3-5 / BAM6QT Quiet runs - PCB PSDs along the 5 degrees cone} 
	\label{fig:CCF10_Quiet_Cone_PCB}
\end{figure}

%-----------------
\label{sec.CCF10_Quiet_Cone_PSDs}
Figure \ref{fig:CCF10_Quiet_Cone_PCB} shows the PSDs from the sensors on the cone portion of the model normalized by the wall pressure computed with STABL at unit Reynolds numbers between 6 and 12 million. The second mode instability can be seen very clearly between 200 and 300 kHz. As expected with second mode fluctuations, the peak frequency and amplitude of the instability both increase with increasing unit Reynolds number. Expressed differently, the thinner the boundary layer, the higher the frequency and amplitude of the second mode at a given x-location, as long as no non-linear effect appears.

These experimental results show that the boundary-layer is laminar on the cone until very high Reynolds number in quiet conditions. Only the spectrum at the highest Reynolds number (Re${/m}$ = 12.0x10$^{6}$) starts to slightly broaden.
This can be the sign of the very start of non-linear interactions, a marker of the beginning of the transition to turbulence process.
Whatever, it can be guessed that the boundary-layer is not on the point to become turbulent and that higher Reynolds numbers (not achievable at the moment of this study in quiet conditions) could show additional second-mode amplifications for Reynolds numbers until 14-16 per million, perhaps more (a higher Reynolds number range is now possible in quiet mode in the BAM6QT). 

So here, in quiet conditions, the boundary-layer is still laminar at Re${/m}$ = 12.0x10$^{6}$ while it became turbulent at the end of the cone for unit Reynolds number between 8 and 9 million in conventional conditions. This confirms the importance of using quiet tunnels to be more representative of real flight conditions.

%-----------------
\subsubsection{Cylinder part}
\begin{figure}
	\centering
	\includegraphics[width=0.46\linewidth]{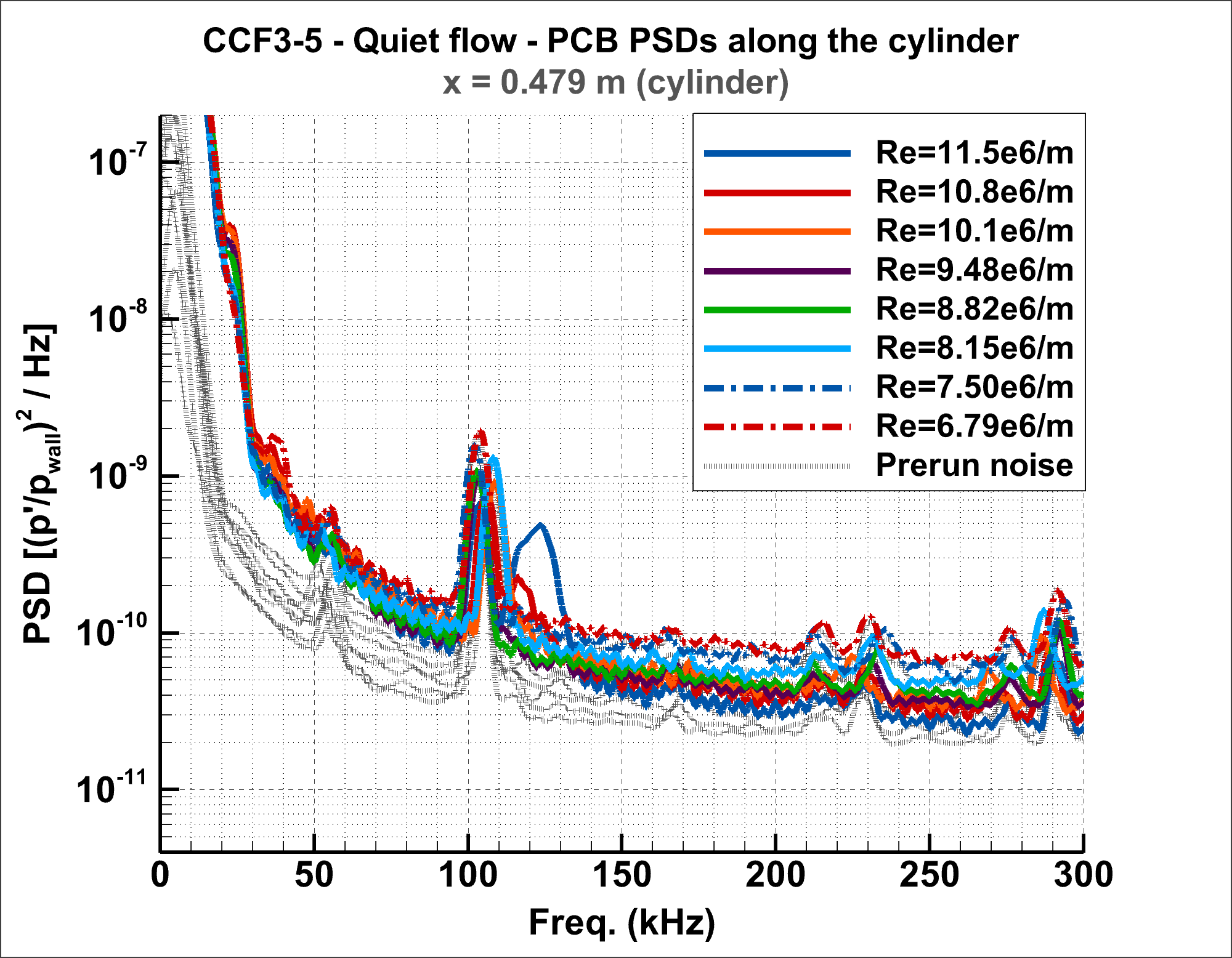} % Fig 4.4c Liz Thesis
	\includegraphics[width=0.46\linewidth]{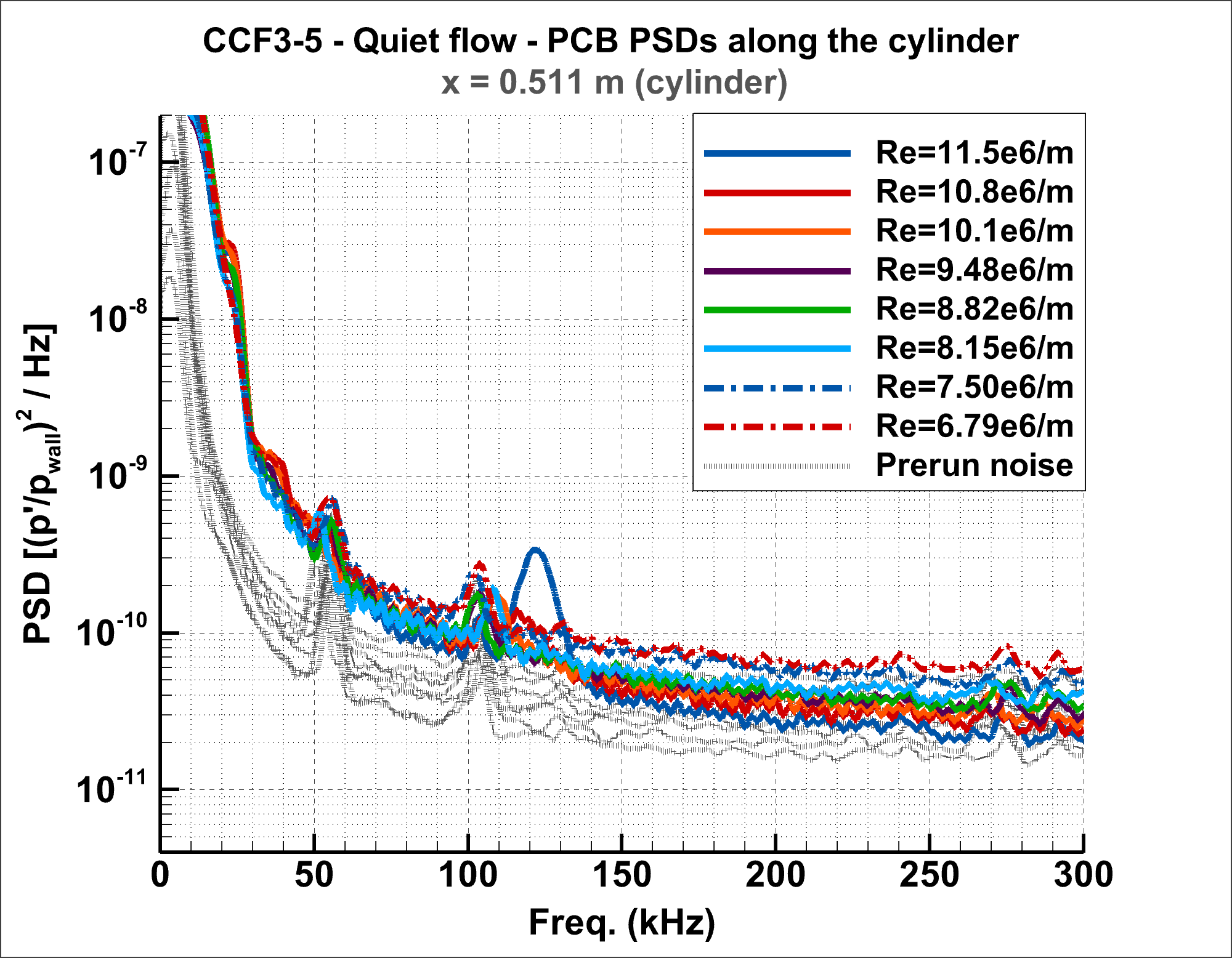} % Fig 4.4d Liz Thesis
	\caption{CCF3-5 / BAM6QT Quiet runs - PCB PSDs along the cylinder} 
	\label{fig:CCF3-5_PCB_Cyl_Quiet}
\end{figure}

%-----------------
Over the cylinder, the second-mode is damped following the flow expansion at the cone-cylinder junction (see figure \ref{fig:CCF3-5_PCB_Cyl_Quiet}). The amplitude of the waves is very small. The pressure fluctuations generated by the damped instability waves coming from the cone, as well as the ones directly amplified on the cylinder, are too small to be detected by the sensors. 

Sharper peaks at various frequencies are also present in the prerun voltages, and are therefore due to electronic noise.

%-----------------
\subsubsection{3.5° flare part}
\begin{figure}
	\centering
	\includegraphics[width=0.46\linewidth]{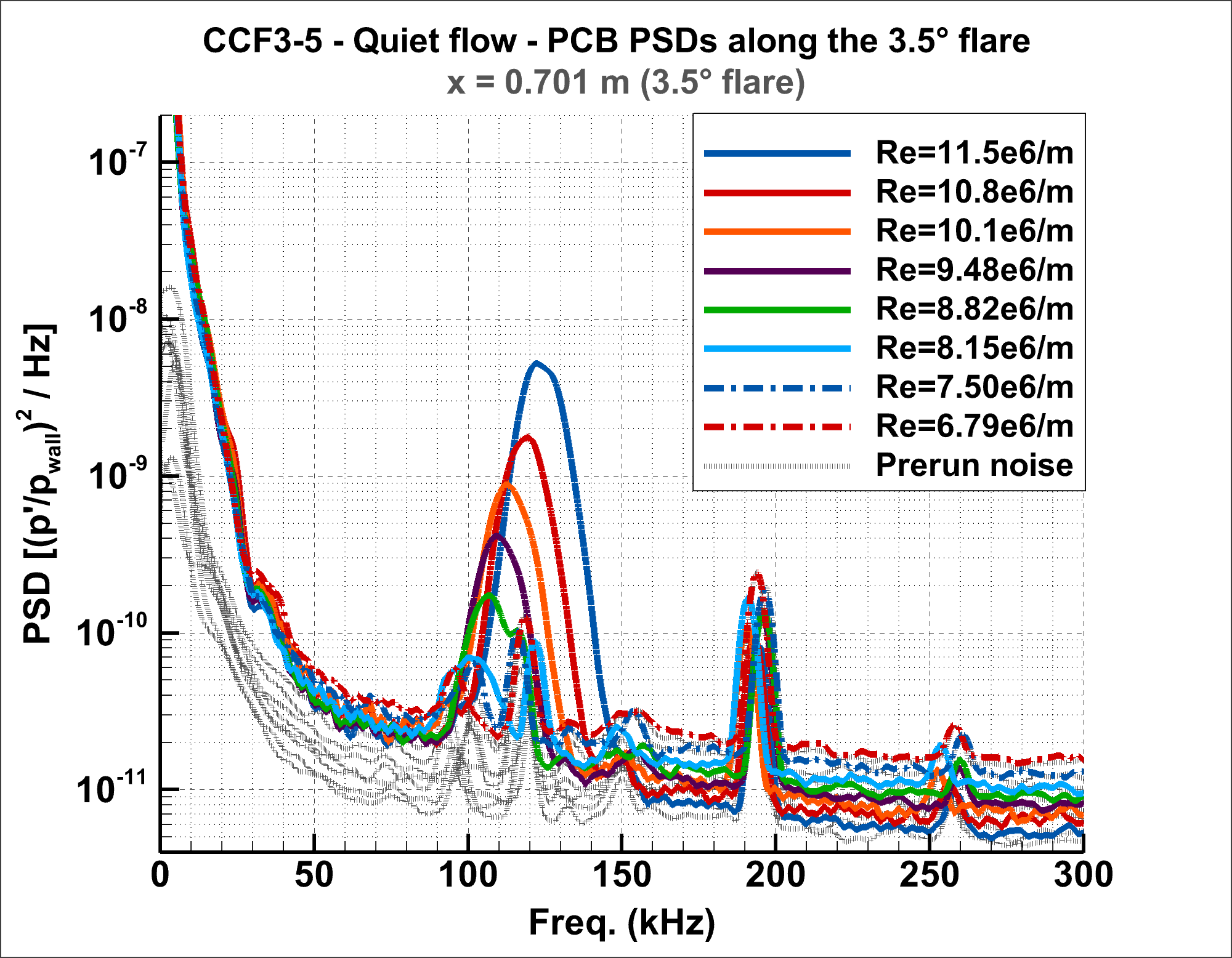} % Fig 4.5 Liz Thesis
	\includegraphics[width=0.46\linewidth]{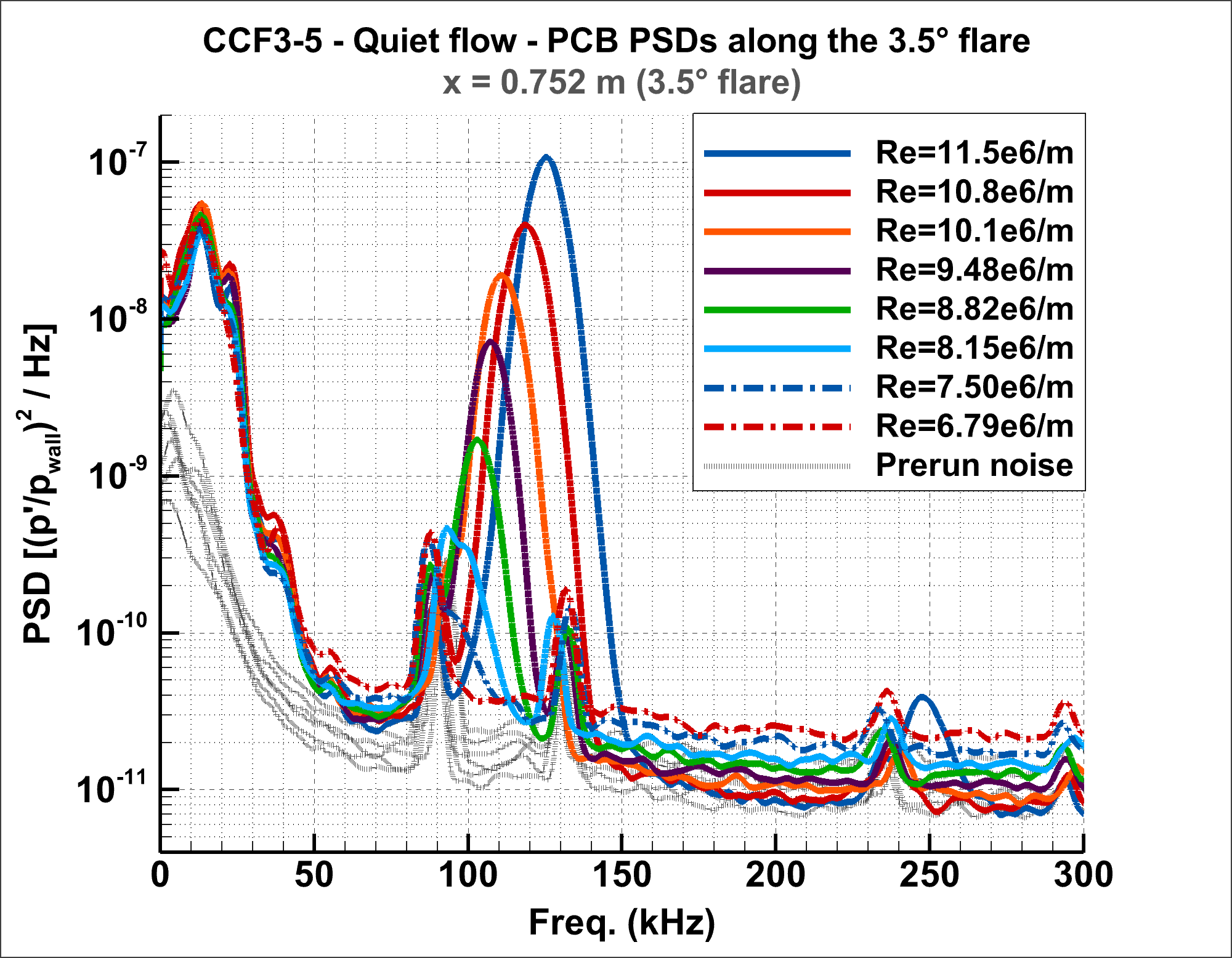} % Fig 4.5 Liz Thesis
	\caption{CCF3-5 / BAM6QT Quiet runs - PCB PSDs along the 3.5 degrees flare} 
	\label{fig:CCF3-5_PCB_Flare_Quiet}
\end{figure}

%-----------------
Figure \ref{fig:CCF3-5_PCB_Flare_Quiet} shows PSDs of pressure fluctuations along the flare. The second mode can be seen amplifying as it moves downstream. The instability peaks are detected at frequencies from 100 to 130 kHz in the considered Reynolds number range (from 6.8 to 11.5 million per meter). The two sensors are spaced 5 cm apart but the peak frequency remains almost constant, this can be related to the relatively constant boundary-layer thickness on the flare as shown in figure \ref{fig:CCF3-5_10delta}. The gain in amplitude is significant between the two locations, it surely would lead to transition breakdown on a longer flare.

As before, sharper spikes present in the prerun data (at around 100 kHz, 120 kHz and 200 kHz) are due to electrical noise. The frequencies of these noise spikes vary between the channels, similar results have been observed in other cases \cite{benit2021dissert}. %with the GN815 cards used to acquire these data.

%-----------------
\subsection{Cross-comparisons: STABL / BAM6QT in quiet flow}
%-----------------
For quiet flow conditions, one very interesting case is the STABL computation performed at high Reynolds number (Re${/m}$ = 11.2x10$^{6}$) presented in detail in section \ref{sec.stab_analysis}.
The computed N factors evaluated along the configuration are re-plotted in figure \ref{fig:CCF_Run28_STABL_Nfactor} with an emphasis on the largest wave amplitudes of the second-mode instabilities on each section of the geometry.
The maximum N factor values are above 8 on the cone and near 9 at the end of the flare. Only quiet runs can measure such amplified instabilities - this flow is turbulent at this Reynolds number in conventional wind tunnel conditions as confirmed by the previous analysis in conventional wind tunnel environment.
For comparisons, figure \ref{fig:CCF3-5_Cone_Cyl_Flare} sums up the experimental results obtained for two unit Reynolds numbers around 11 millions on each part of the configuration. 

%-----------------
\subsubsection{Conical part}
%-----------------
The PCB measurements peak around 215-235 kHz for two unit Reynolds number Re${/m}$ = 11.0x10$^{6}$ and 12.0x10$^{6}$ for the two instrumented locations (x=0.361m and x=0.387m) as visible on left illustration in figure \ref{fig:CCF3-5_Cone_Cyl_Flare}. The PSE analysis shows second-mode waves in the same frequency band as the experiments. So, the measured and computed second-mode frequencies compare very well.
The numerical N factor values peak at high levels, 7.3 and 7.7 respectively at Re${/m}$ = 11.2x10$^{6}$, at the two x-locations on the cone. 

%-----------------
\subsubsection{Cylinder part}
\begin{figure}
	\centering
	\includegraphics[width=0.40\linewidth]{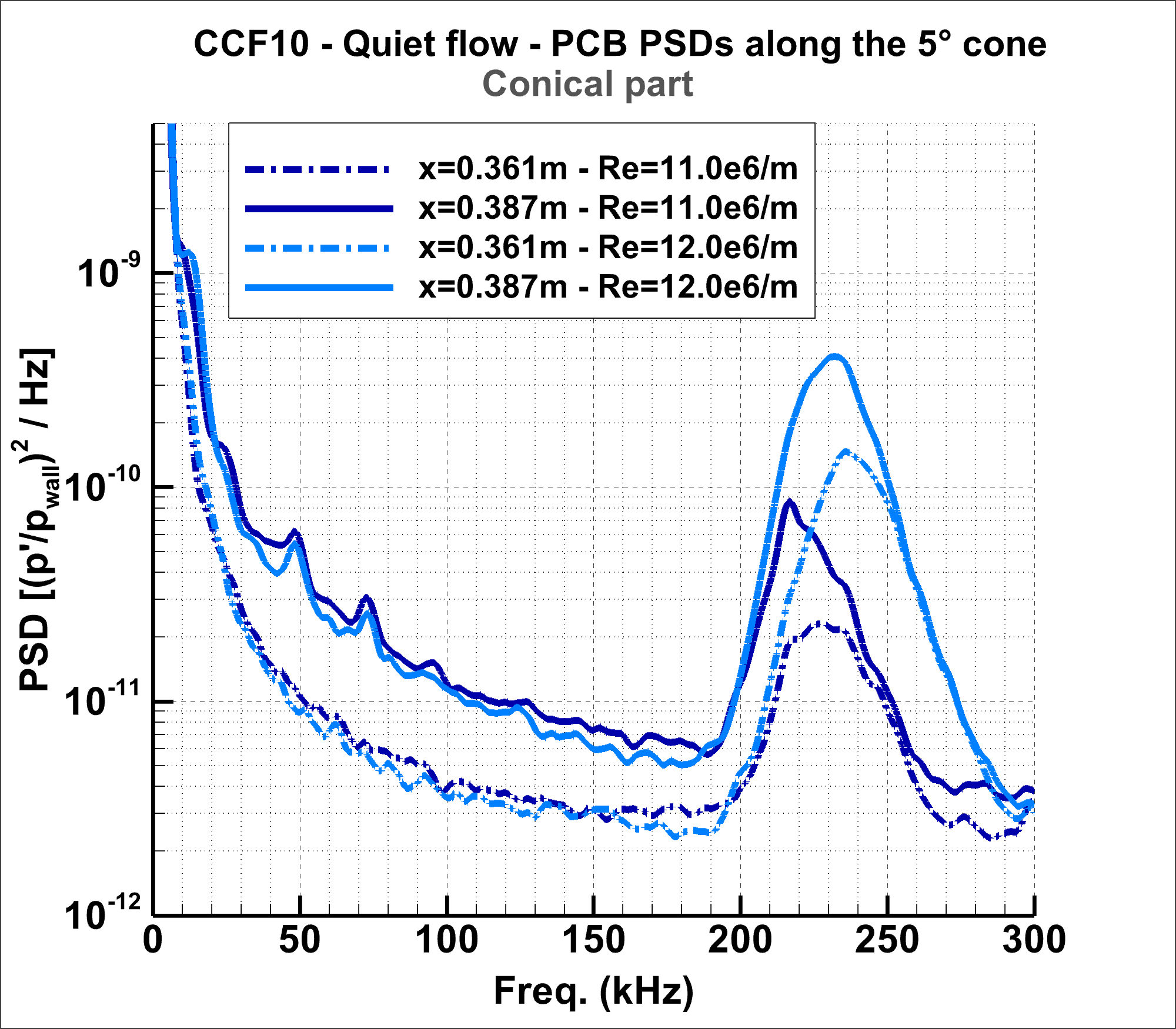}
	\includegraphics[width=0.40\linewidth]{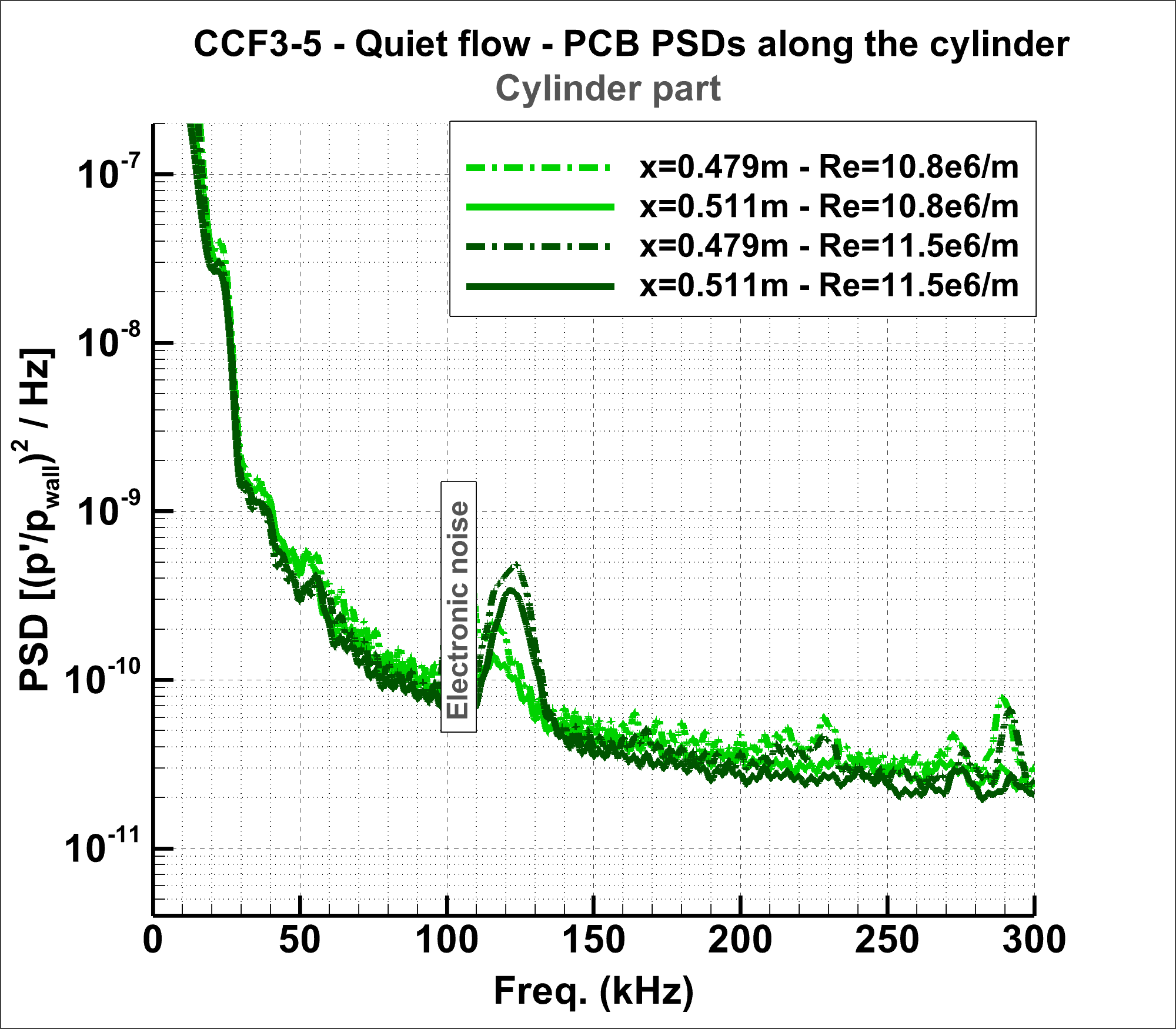}
        \includegraphics[width=0.40\linewidth]{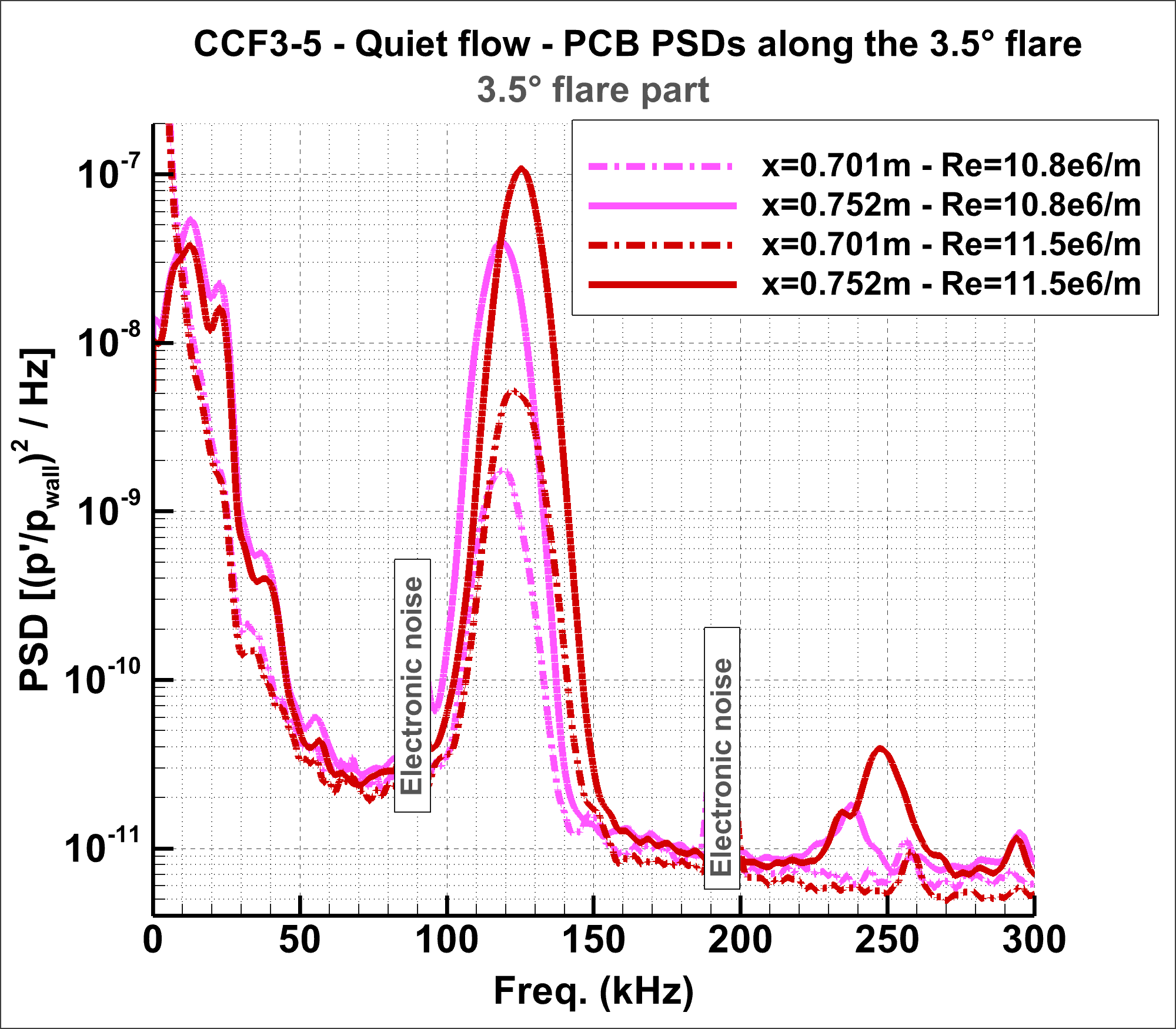}
	\caption{CCF3-5 / BAM6QT Quiet runs - PCB PSDs along the 3 sections} 
	\label{fig:CCF3-5_Cone_Cyl_Flare}
\end{figure}

%-----------------
On the cylinder part, as indicated previously, the stabilizing effect of the flow expansion leads to very limited wave amplitudes as shown experimentally in the central illustration of figure \ref{fig:CCF3-5_Cone_Cyl_Flare}.  The experimental waves peak at around 110 kHz. 
From a computational point of view, figure \ref{fig:CCF_Run28_STABL_Nfactor} shows that, after a first amplification on the cone, these 105-115 kHz instabilities are amplified again along the cylinder. The agreement on the second-mode frequency is very good and this analysis seems to explain the origin of this frequency range observed on the cylinder.
These second-mode instabilities reach limited amplitudes  corresponding to N factor around 3.2 at x=0.511m and Re${/m}$ = 11.2x10$^{6}$.

%-----------------
\subsubsection{3.5° flare part}
\begin{figure}
	\centering
	\includegraphics[width=0.88\linewidth]{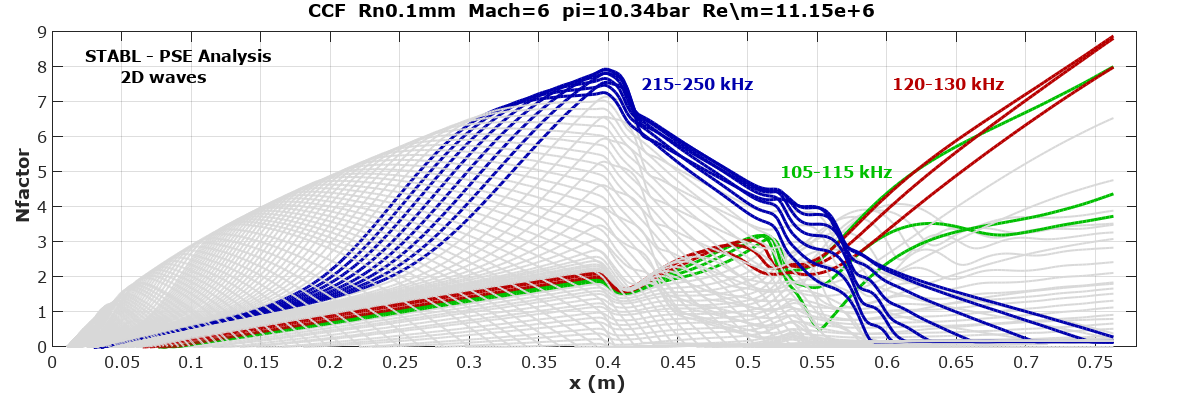}
	\caption{CCF3-5 - STABL N factors at Re${/m}$ = 11.2x10$^{6}$}
	\label{fig:CCF_Run28_STABL_Nfactor}
\end{figure}

%-----------------
Along the conical flare, the experimental second-mode peaks at around 125 kHz. This matches very well with the predicted second-mode fluctuations. Figure \ref{fig:CCF_Run28_STABL_Nfactor} displays three interesting discrete frequencies: 120, 125 and 130 kHz which clearly show the evolution of these instabilities along the configuration. After a slight amplification of the 2D waves on the cone and on the cylinder, a strong N factor increase is observed along the flare under the influence of the adverse pressure gradient. Knowing the relatively constant boundary-layer thickness and the limited evolution of the velocity profile in this region, only a limited frequency range is amplified along the flare.
The STABL results indicate second-mode N factor values equal respectively to 7.2 and 8.6 for the two considered PCB locations (x=0.701m and x=0.752m). 
Finally, the waves reach very high N factor values near 9 at the end of the configuration.
The numerical/experimental cross-comparison on the flare is also very good.

%===============
\section{Quiet flow v.s Conventional flow}
%===============
As shown in previous sections, the experimental noise level has a tremendous impact on the boundary-layer state. 
Because of very low freestream noise levels, quiet tunnels are a real asset for studying laminar-turbulent transition more reliably than conventional wind tunnels, and above all, make it possible to generate efficient data for estimating transition in free flight in the atmosphere.\\

Going back to the cone-cylinder-flare configuration, in quiet flow, the second-mode starts to emerge on the conical part only at Re${/m}$ = 9.5x10$^{6}$ (waves' amplitudes are too small to be detected below this Reynolds number).
In conventional noisy conditions, non-linear interactions are already detected from Re${/m}$ = 5.2x10$^{6}$ and the boundary-layer becomes turbulent at unit Reynolds number around 9 million - when the second-mode waves start to be detected in quiet conditions. 
This is a huge difference.
In quiet flow, the boundary-layer is still laminar even at Re${/m}$ = 12.0x10$^{6}$ with perhaps some non-linear interactions. It seems that the flow on the cone could stay laminar or transitional until higher unit Reynolds numbers of the order of 14-16 million.
For such high Reynolds numbers, the thin boundary-layer will lead to the amplification of higher frequency waves than the ones detected in noisy conditions, this is also an important difference. Not only is the noise level modified but also, as a consequence, the frequency and amplitude of instability waves.

On the cylinder part, the wave amplitudes are too small and more uncertain on this damping region to realize an efficient analysis of noisy / quiet influence on transition development. Nevertheless, it seems that the scenario of the second-mode waves evolution observed in quiet flow and numerically is coherent (see previous section), it was inconclusive in conventional noise.

On the 3.5 degree flare, in conventional noisy conditions, only a limited number of runs allow the study of the second-mode waves evolution. Starting at Re${/m}$ = 2.2x10$^{6}$, instabilities are clearly detected between 60 and 80 kHz before a broaden spectrum is observed at Re${/m}$ = 3.7x10$^{6}$ (a marker of non-linear-interactions). Finally, the boundary-layer turns turbulent at low unit Reynolds number, around Re${/m}$ = 4x10$^{6}$. The Reynolds number range to study boundary-layer stability is very limited in noisy environment on the flare.
In quiet conditions, only experimental runs for unit Reynolds number above 8 millions start to detect the second-mode instabilities. The waves have reached the minimum amplitudes to be measurable by the PCB sensors, but the boundary-layer is still fully laminar at this Reynolds number.   
Going further, very high amplitude levels appear and increase until the maximum Reynolds number achievable Re${/m}$ = 11.5x10$^{6}$ where a harmonic is detected at the location x=0.752m. The boundary-layer is still not fully turbulent at these high Reynolds number conditions.\\

Conventional noise experiments have proven to allow to obtain some information about laminar-turbulent transition but the extrapolation to flight and the prediction of flight performance remain uncertain as stated by \citet{chazot2019far}. Using low noise wind tunnels is the only way to be more representative of free-flight even if a full duplication of the flight conditions is also not possible (Mach number, wall temperature, scale model,...).

Nevertheless, the combination of very low freestream noise levels and limited cost of a quiet tunnel compared to expensive single one-shot hypersonic flight tests give quiet tunnels a real advantage for the extensive study of laminar-turbulent transition mechanisms.\\

As a conclusion, quiet tunnels are experimental facilities of primary importance in the perspective of free-flight transposition for the laminar-turbulent prediction. Indeed, boundary-layer transition brings problems that affect thermal protection risk and local high heat fluxes, aerodynamic coefficients dispersion, flight stability, trim capability and control authority as well as vehicle weight reduction. These complex issues must be addressed by the better suited ground-based experimental tool: the quiet wind tunnel.\\\\

\begin{figure}
 	\centering
 	\includegraphics[width=0.80\linewidth]{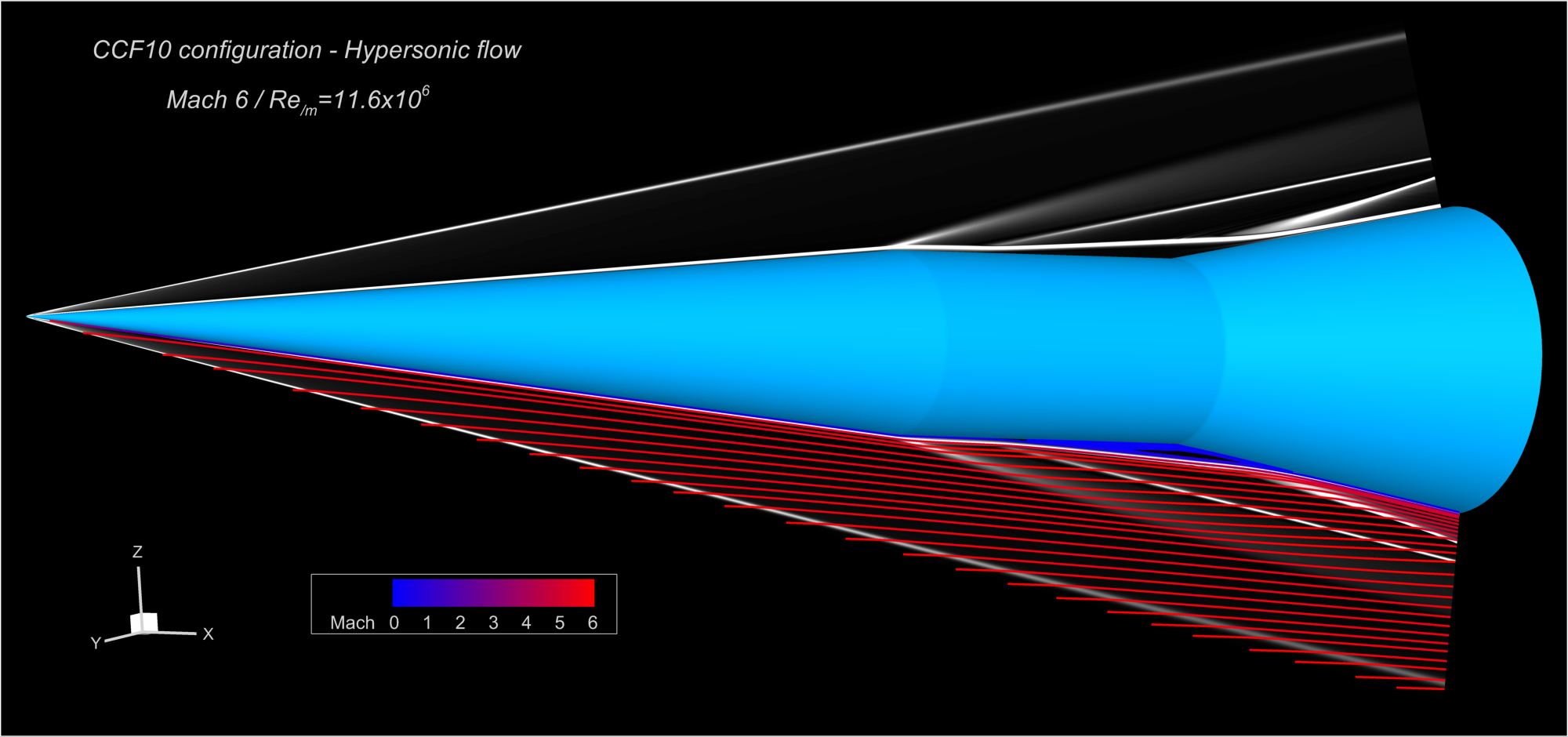}
 	\caption{CCF10 - Flow topology through density gradient magnitude and streamlines colored by Mach number at Re${/m}$ = 11.6x10$^{6}$ (DPLR2D laminar computation)}
 	\label{fig:CCF10_Flow_topology}
 \end{figure}

Before concluding, some short comments on CCF10. This configuration has been initially designed in order to generate a more complex aerodynamic flow than CCF3-5 (see CCF10 laminar computation in figure \ref{fig:CCF10_Flow_topology}). This incremental complexity can allow to address at the same time the three basic problems: boundary-layer transition, shock-boundary-layer interaction and correlation between numerical simulations and wind tunnel experiments (and also engage reflection about free flight transposition).
%Mastering these three themes is essential for successful hypersonic projects.

%===============
\section{On-going studies on CCF10 and CCF12 geometries}
%===============
%---------------
\noindent {\it{\textbf{The 10-degree flare case (CCF10)}}}\\
%---------------
As said before, the CCF10 configuration is the second step of the incremental process where the complex topic of shock-wave/boundary-layer interaction is added. 
Local stability theory cannot be applied, without asking questions about the reliability of the results, to cases with separation bubbles because the parallel flow assumption is not respected. Consequently, the numerical analysis is not performed on CCF10 but an extensive experimental study of this case can be found in \cite{benit2020-3072}.\\
%but, in a spirit of openness, below are some experimental results. The purpose of these experimental data is to highlight shortly the impact of flow separation on boundary-layer stability.

%===============
%\section{Perspectives and on-going studies on the third geometry: CCF12}
%===============
%---------------
\noindent {\it{\textbf{The 12-degree flare case (CCF12)}}}\\
%---------------
In order to promote the flow separation development and the amplification of laminar-turbulent transition instabilities, a third geometry called CCF12 has been defined based on the results obtained in \cite{esq2019-2115}, \cite{benit2020-3072} and \cite{esq2023-3af}. Indeed, Paredes indicated in \cite{paredJ061829} that a flare angle increase to 12° could promote shear-layer wave amplification above the separation bubble region.
CCF12 has a slightly longer cylinder (extended to x=546.61m) in order to contain the whole separation bubble on the cylinder part (without reaching the cone-cylinder junction). 

\begin{figure}
	\centering
	\includegraphics[width=0.76\linewidth]{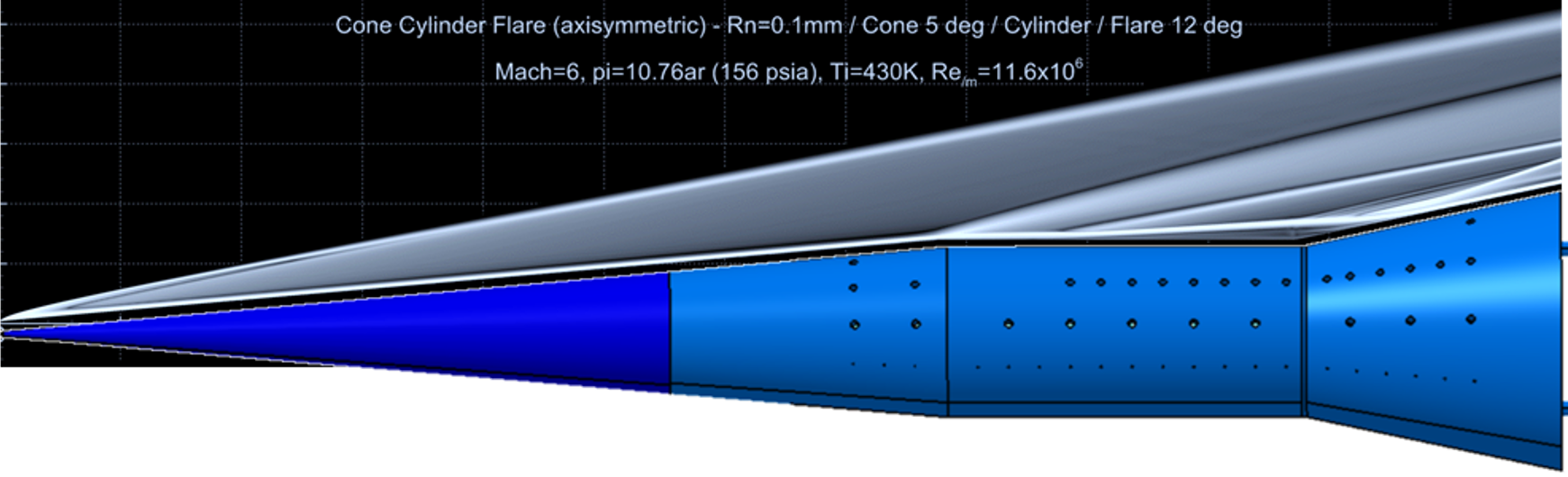}
	\caption{CCF12 model for R2Ch experiments (CEA-ONERA collaboration, 2023)}
	\label{fig:CCF12Model}
\end{figure}

Numerical studies and experiments on CCF12 have been presented in \cite{li2022-3855} and \cite{benit2023bluntcone}.
Wind tunnel tests have also been conducted recently at ONERA-R2Ch, in summer 2023, on the CCF12 model in a CEA-ONERA collaboration (see CCF12 laminar flow and model in figure \ref{fig:CCF12Model}). These experimental and numerical studies have been realized in the framework of the NATO-AVT-346 group. Papers from \cite{caillaud2025aiaaJ} and \cite{benit2025aiaaJ} summarize the results obtained within different wind tunnels (Purdue BAM6QT, AFRL M6LT and ONERA-R2Ch) and with different computational tools. 

%\begin{figure}
%	\centering
%	\includegraphics[width=0.96\linewidth]{IMG/CCF12_Rn0_1mm_VueXY_0_65m}
%	\caption{The CCF12 cone-cylinder-flare geometry}
%	\label{fig:CCF12geom}
%\end{figure}

%Very interesting elements of wind tunnel / flight tests can be found in \cite{Ref_HIFiRE} \cite{Wheaton} \cite{Guohua}\\\\

%===========
\section{Conclusion}
%===========
In order to progress in the understanding of hypersonic transition with pressure gradients, flow expansion and recompression, new axisymmetric configurations have been designed for experimental tests in BAM6QT.
The final shapes result in two slender cone-cylinder-flare configurations, CCF3-5 and CCF10.
The first configuration allows to keep an attached boundary-layer along the surface while the second geometry, by increasing the flare angle, generates a flow separation at the cylinder-flare junction.

The influence of the separation bubble on the flow has been quantified thanks to accurate laminar computations specifically prepared for local stability analyses. 
Following the parallel flow assumption, only CCF3-5 has been considered for stability analyses and comparisons with experimental results. For this case at zero-degree angle of attack, the dominant instability is the second-mode of Mack. Stability analyses have shown damped waves following the flow expansion and amplified second-mode instabilities, in a narrow frequency range, on the flare through the adverse pressure gradient. As a result, high second-mode amplitudes are observed at the end of the cone and on the flare with computed N-factor going from 5 to 9 for unit Reynolds number between 5 million to 11 million. 

Very instructive numerical / experimental cross-comparisons with conventional noisy and quiet flow measurements have confirmed the physical development of the second-mode waves.
The experimental noise level has a tremendous impact on the boundary-layer state. Second-mode starts to emerge only at high Reynolds number in quiet flow while the boundary-layer has become turbulent at limited Reynolds number in conventional noisy conditions. 
In low noise level environment, it seems that the flow on the cone could stay laminar or transitional until higher unit Reynolds numbers, of the order of 14-16 million, perhaps more.
For such high Reynolds numbers, the thin boundary-layer will lead to the amplification of higher frequency waves than the ones detected in noisy conditions. Not only is the noise level modified but also, as a consequence, the frequency and amplitude of instability waves. This confirms that quiet tunnels are a real asset for studying transition with noise levels comparable to flight and can generate efficient data for estimating transition in free flight in the atmosphere.

Just before finishing, it can be said that these axisymmetric shapes, CCF3-5 and CCF10, are very well suited for computational and experimental study on the second-mode waves in the presence of pressure gradients. A rewarding point is that these geometries have been the basis of the CCF12 configuration, with increased flare angle to 12-degree. This geometry has been thoroughly studied in the framework of the NATO-AVT-346 group, in order to explain the physical mechanisms responsible for hypersonic laminar-turbulent transition for flow with separation bubbles. 
 
%During re-entry or hypersonic flight, transition is an important issue and has a huge impact on vehicle design.

%================
\section{Acknowledgments}
%================
Sebastien Esquieu thanks Professor Steven P. Schneider for hosting him in the Aerospace Sciences Laboratory during the 2017-2018 internship (French DGA and CEA grant) and for the continuous exchanges in the NAVO-AVT-346 framework. Many thanks also to Heath Johnson from University of Minnesota for the access to the STABL suite and also to CEA and ONERA teams for the follow-on study on the CCF12 geometry.\\

\noindent \textbf{Fundings.} The work at Purdue was mostly funded by the U.S. Air Force Office of Scientific Research under Grants FA9550-17-1-0419 and FA9550-22-1-0110.  Any opinions, findings, and conclusions or recommendations expressed in this material are those of the authors and do not necessarily reflect the view of the United States Air Force.

% ===================================%
%%%%%%%%% Bibliography %%%%%%%%%
% ===================================%
\bibliographystyle{plain}
\bibliography{references}

\end{document}